\documentclass[fleqn,usenatbib]{mnras}

\usepackage{amssymb}
\usepackage{amsmath}

\usepackage{graphicx}
\usepackage{graphics}
\usepackage{color}
\usepackage{fancyhdr} 
\usepackage{graphicx}
\usepackage{float}
\usepackage{comment}
\usepackage{subfig}
\usepackage[utf8]{inputenc}
\usepackage[justification=centerlast, format=plain, labelfont=bf]{caption}
\usepackage[table,xcdraw,dvipsnames]{xcolor}

\def\VEV#1{\left\langle #1 \right\rangle}

    \newcommand{\be}{\begin{equation}}
  \newcommand{\ee}{\end{equation}}
    \newcommand{\ba}{\begin{align}}
  \newcommand{\ea}{\end{align}}

\newcommand{\Msun}{M_{\odot}}

\newcommand{\MUV}{M_{\rm UV}}
\newcommand{\LHa}{L_{\rm H\alpha}}

\newcommand{\sigmaUV}{\sigma_{M_{\rm UV}}}
\newcommand{\sigmaHa}{\sigma_{\Ha}}

\newcommand{\sigmaPSD}{\sigma_{\rm PS}}
\newcommand{\tauPSD}{\tau_{\rm PS}}
\newcommand{\Ha}{{\rm H \alpha}}
\newcommand{\etaHaUV}{\eta_{\rm H\alpha,UV}}
\newcommand{\tage}{t_{\rm age}}
\newcommand{\xiion}{\xi_{\rm ion}}
\newcommand{\AmpHa}{\mathcal {A_{\Ha/\rm UV}}}

\defcitealias{BC03}{BC03}
\defcitealias{BPASS}{BPASS}
\newcommand{\jbm}[1]{}

\title[Increased Burstiness in Smaller High-z Halos]{Relatively Fast and Reasonably Furious:  
Evidence for Increased Burstiness in Smaller Halos at Cosmic Dawn}

\author[J.~B.~Mu\~noz et al.]{
Julian B.~Mu\~noz,$^{1,2,3}$\thanks{E-mail: julianbmunoz@utexas.edu}
John Chisholm,$^{1,2}$
Guochao Sun,$^{4}$
Jenna Samuel,$^{1,2}$
Jordan Mirocha,$^{5,6}$
\newauthor
Emily Bregou,$^{1,2}$
Alessandra Venditti,$^{1,2}$
Mahdi Qezlou,$^{1}$
Charlotte Simmonds,$^{7,8}$
and Ryan Endsley$^{1,2}$
\\
$^{1}$Department of Astronomy, The University of Texas at Austin, 2515 Speedway, Stop C1400, Austin, TX 78712, USA\\
$^{2}$Cosmic Frontier Center, The University of Texas at Austin, Austin, TX 78712, USA\\
$^{3}$Texas Center for Cosmology \& Astroparticle Physics, Austin, TX 78712, USA\\
$^{4}$CIERA and Department of Physics and Astronomy, Northwestern University, 1800 Sherman Ave., Evanston, IL 60201, USA\\
$^{5}$California Institute of Technology, 1200 E. California Boulevard, Pasadena, CA 91125, USA\\
$^{6}$Jet Propulsion Laboratory, 4800 Oak Grove Drive, Pasadena, CA 91109, USA\\
$^{7}$Departamento de Astronom\'ia, Universidad de Chile, Camino El Observatorio 1515, Las Condes, Santiago, Chile\\
$^{8}$Kavli Institute for Cosmology and Cavendish Laboratory, University of Cambridge, Cambridge CB3 0HA, UK
}

\date{Accepted XXX. Received YYY; in original form ZZZ}

\pubyear{2026}

\begin{document}
\label{firstpage}
\pagerange{\pageref{firstpage}--\pageref{lastpage}}
\maketitle

\begin{abstract}
We introduce an effective framework to model star-formation burstiness and use it to jointly fit galaxy UV luminosity functions (UVLFs), clustering, and H$\alpha$/UV ratios, providing the first robust empirical evidence that early galaxies hosted in lower-mass halos are burstier. Using $z\sim 4-6$ observations, we find that galaxies show approximately $0.6$ dex of SFR variability if hosted in halos of $M_h = 10^{11}\, M_\odot$ (typical of $M_{\rm UV}\approx -19$ galaxies at $z = 6$). This translates into a scatter of $\sigma_{M_{\rm UV}}\approx 0.75$ mag in the UVLF, in line with past findings. Strikingly, we find that burstiness grows for galaxies hosted in smaller halos, reaching $\gtrsim 1$ dex for $M_h \leq 10^{9}\, M_\odot$ (corresponding to $\sigma_{M_{\rm UV}} \approx 1.5$ mag for faint $M_{\rm UV} \gtrsim -15$ galaxies). 
Extrapolating to higher redshifts, when small halos were more prevalent, the inferred mass-dependent burstiness can reproduce observed UVLFs up to $z\sim 17$ within 1$\sigma$, potentially alleviating the tension between pre- and post-JWST galaxy-formation models. Current observations allow us to constrain the main burst timescale to approximately $20$ Myr, consistent with expectations from supernova feedback, and suggest broad distributions of ionizing efficiencies at fixed $M_{\rm UV}$. Our results demonstrate that mass-dependent burstiness, as predicted by hydrodynamical simulations, is critical for understanding the mass assembly of early galaxies.  
\end{abstract}


\section{Introduction}

How the first galaxies assembled their stellar masses remains one of the key open questions in astrophysics. 
Over the last decades the community has converged on the idea of feedback regulation~\citep{Dekel:1986ehj,Faucher-Giguere:2013nkp,
Hopkins:2013vha, Somerville2014}, where star formation self-balances by disturbing the surrounding gas, reaching a star-forming ``equilibrium''. 
Despite its success explaining a broad array of galaxy observations, we do not know what drives feedback in the early Universe. 
We expect a plethora of physical processes to play an important role, including stellar winds~\citep{Vink2001_winds,Murray:2004dd}, supernovae~\citep{McKee1977}, and black holes~\citep{Silk:1997xw,DiMatteo:2005ttp}. 
These act on distinct timescales and with strengths that depend differently on halo mass. 
For instance, radiation and stellar winds act promptly following star formation~\citep[on $\sim$ Myr timescales,][]{Hopkins:2011xm}, whereas supernovae lag behind~\citep[by tens of Myr,][]{Leitherer_SB99}. 
Likewise, stellar feedback should strengthen for halos with lower masses~\citep[which have shallower potential wells,][]{Furlanetto22_feedback},
whereas black holes are expected to drive feedback in heavier ones. 
Therefore, understanding how feedback acts as a function of timescale and mass is key to decoding the mechanisms that shape galaxy formation.

Feedback gives rise to star-formation variability --- or {\it burstiness}  --- which we can probe through the spectral energy distributions (SEDs) of galaxies.  
As an example, Balmer-line emission is driven by strong ionizers like O and B type stars which live short ($\lesssim 10$ Myr) lives, whereas rest-frame UV light traces star formation on longer ($\sim 100$ Myr) timescales.
As a consequence, their ratio constrains the strength of recent star-formation bursts~\citep{Weisz12}. 
This insight has allowed observational studies to measure star-formation histories up to moderate redshifts $z\lesssim 3$, 
finding that high-mass galaxies tend to have roughly constant or decreasing star-formation histories, whereas low-mass ones are far more episodic~\citep[e.g.,][]{1912.06523,Faisst2019_Habursty}.
Higher-redshifts SFHs have, however, remained out of reach, as galaxies become too faint and redshifted~\citep{Stark_JWST_review}.

The James Webb Space Telescope (JWST) is now allowing us to study optical and UV tracers up to increasingly early times, and thus to constrain galaxy formation at higher redshifts than ever before. 
The first few years of JWST have revealed a bursty early universe, with a large diversity amongst the star-formation histories of different galaxies~\citep[e.g.,][]{Carnall2023_SMACS,Whitler2022:JWST_Ages,CurtisLakeJades_xiion,Hsiao23_xiion,Dressler2024_bursty, 2506.16510,Tang25_JWST_spectra}.
Moreover, the discovered overabundance of UV-bright galaxies at redshifts $z\gtrsim 9$~\citep{Eisenstein_JADES,Casey:2022amu,Castellano_GLASS_hiz,Adams_Conselice_JWST_2023,Rieke:2023tks,Harikane_UVLFs} may be due to larger burstiness than expected in the first billion years~\citep[][]{mirocha23, Sun23_FIRE_bursty,mason23,shen23,Munoz:2023cup}. 
Despite this evidence, the strength, timescale, and mass dependence of burstiness --- and thus of the feedback mechanisms shaping early galaxies --- remain elusive.

The challenge is in translating galaxy observations into population-level constraints.
Observational studies can recover star-formation histories (SFHs) for individual galaxies~\citep[]{2212.01915}, akin to performing a ``longitudinal'' study of each object over time.
While powerful, these SFHs can be prior dominated~\citep{Leja:2019} and difficult to translate into insights on the entire population (e.g., whether burstiness grows or decreases with halo mass due to feedback). 
Theoretical simulations, on the other hand, can produce a population of galaxies residing in cosmological structures~\citep[e.g.,][]{Somerville:1998bb,Benson2012_galacticus,Tacchella,Behroozi2019_UM}, enabling ``transverse'' studies that show the variation across galaxies at a fixed time. 
This allows them to model bursty galaxy SEDs, though in practice full simulations can be computationally expensive and parametrically expansive, restricting inference on data. 
Analytic work has been able to significantly speed up this process by bypassing simulations~\citep{Trenti10,Mason15,Sabti:2021xvh}, but abandons bursty star-formation histories in favor of effective parameters (such as the UV-scatter $\sigmaUV$), obscuring the feedback physics. 
Understanding the origin of burstiness requires bridging these longitudinal and transverse methods.

Here we present a framework designed to efficiently model the time-series burstiness of an entire galaxy population, and use it to fit multi-wavelength JWST+HST observations at $z\gtrsim 4$.  
Rather than forward-model the SED of each object in a simulation, which is computationally demanding, we assume SFHs have lognormal fluctuations (drawn from a power spectrum or PSD model; see~\citealt{1901.07556,2003.02146,2007.07916}) and analytically obtain the probability distribution functions (PDFs) of galaxy luminosities such as $L_{\rm UV}$ and $\LHa$, and their ratio as a function of halo mass, $M_h$. 
This allows us to predict observables for an entire galaxy population in seconds. 
We leverage this effective model to measure the amplitude, timescale, and mass behavior of star-formation burstiness by simultaneously fitting high-$z$ UV luminosity functions (UVLFs), clustering data, and $\Ha$/UV ratios. 
Our key result is an empirical measurement of increased burstiness for smaller halo masses, with typical burst timescales of $20$ Myr, which aligns with the expectations from supernova-driven feedback.

This paper is structured as follows. In Sec.~\ref{sec:themodel} we introduce the model and in Sec.~\ref{sec:compareobservations} show how we compare to observations. The rushed reader may want to skip to Sec.~\ref{sec:results}, where we present our results and  Sec.~\ref{sec:discussion} where we discuss them. We conclude in  Sec.~\ref{sec:conclusion}.  Throughout this paper we use AB magnitudes~\citep{OkeGunn} and a flat $\Lambda$CDM {\it Planck} 2018 cosmology~\citep{Planck:2018vyg}, unless otherwise specified.

\section{Efficiently modeling burstiness}
\label{sec:themodel}

We assume that dark-matter halos host galaxies that grow stochastically over time, with that growth decomposed into an average component (for all halos of a certain mass $M_h$) and fluctuations around it. 
More specifically, we take the fluctuations to be lognormal, as expected of a scale-free stochastic process~\citep[][see also Appendix~\ref{App:hydro}]{Pallottini:2023yqg}.
In that case a galaxy hosted in a halo of mass $M_h$ has a star-formation history
\be
\ln\dot M_\star (M_h, t) = \ln \overline {\dot M_\star}(M_h, t) + x(t),
\label{eq:defSFRx}
\ee
determined by the {\it median} (i.e., no-burst average) star-formation rate $\overline {\dot M_\star}$ (SFR, for halos of mass $M_h$ at time $t$) and a Gaussian random variable $x$  that captures the effect of burstiness, modulating the SFR of each individual galaxy. 
In this notation $\dot M_\star$ becomes a lognormal variable, and its time-series statistics can be derived from those of the Gaussian variable $x$, as we will describe below. 

In the rest of this Section we will first tackle the two separable components: the average SFR and the fluctuations on top, before moving to compute how burstiness translates into observables such as the UV magnitude $\MUV$, the $\Ha$ luminosity $L_{\Ha}$, and their ratio.

\subsection{The average star-formation rate}

The median SFR $\overline {\dot M_\star}(t)$ captures the average growth of structures in the universe, as well as the overall effect of feedback. 
In the absence of burstiness ($x\equiv 0$), this quantity can be determined through techniques such as abundance matching~\citep{Moster13_AM,Vale:2004yt}.
We, instead, model the average halo-galaxy connection analytically~\citep{Mason15,Sabti:2021xvh}, by assuming that halos accrete gas which then forms stars at a rate
\be
\overline {\dot M_\star}(M_h, t) = f_\star[M_h(t), t] f_b \dot M_h[M_h(t), t],
\ee
where $f_b = \Omega_b/\Omega_m$ is the baryon fraction, $f_\star$ is the star-formation efficiency (SFE), and $\dot M_h(t)$ the average halo mass-accretion rate (as its fluctuations will be included in $x$).
This expression has to be evaluated at the past halo mass $M_h(t)$, and for simplicity we assume exponential halo growth, so $M_h(z) \propto e^{-\alpha z}$ with $\alpha=0.79$, as found in  simulations~\citep[][but mismodeling of this term can be reabsorbed into the $f_\star$ parameters below, \citealt{Munoz:2023cup}]{Dekel:2013uaa}.
For the SFE we take a simple form
\be
f_\star (M_h,z) = \dfrac{2 \epsilon_\star(z)}{(M_h/M_p)^{\alpha_\star} + (M_h/M_p)^{\beta_\star}},
\ee
with four free parameters $(\epsilon_\star, M_p, \alpha_\star>0,$ and $\beta_\star<0)$ that are allowed to vary with redshift.
This double power-law form captures the mean effect of feedback, reducing the average SFE from its maximum at $\epsilon_\star$ both for lower  and higher masses than the pivot mass $M_p$. The two power-law indices $\alpha_\star$ and $\beta_\star$ regulate the strength of the suppression. Despite its simplicity, this SFE is physically motivated by energy/momentum-conserving feedback regulation~\citep{Furlanetto2017_feedback}, and is
flexible enough to fit luminosity functions both from HST/JWST observations and hydrodynamical simulations~\citep{Sabti:2021xvh, Tacchella}.

\begin{figure*}
    \centering
    \includegraphics[width = 0.61\linewidth]{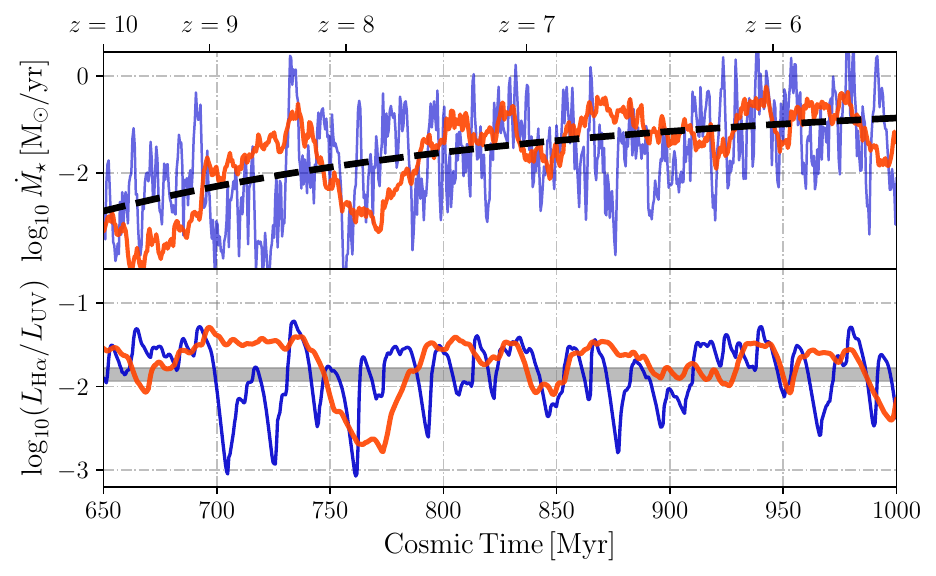}
    \includegraphics[width = 0.38\linewidth]{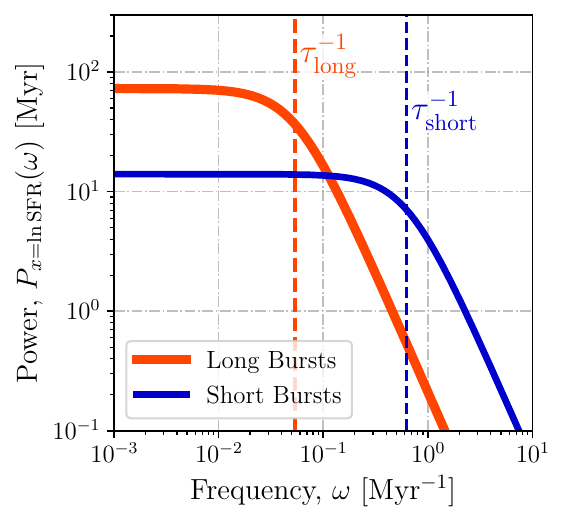}
    \caption{{\bf Left:} Example star-formation histories for galaxies hosted in ``low-mass'' halos, with $M_h=10^{10}\,\Msun$ at $z=5.5$. {\bf Top left} shows the average star-formation rate of all galaxies (black dashed, with no bursts) and two example bursty galaxies  drawn from the power spectrum models in the right panel (red and blue, corresponding to long vs short bursts). {\bf Bottom left} shows the $\Ha$/UV ratio that would be measured at each time, where the gray band represents the no-burst equilibrium value. 
    {\bf Right:} Power spectra of $x$ (the log-SFR fluctuation; see Eq.~\ref{eq:defSFRx}) for two illustrative bursty models as a function of frequency $\omega$. The red model has more power at low frequencies and a cutoff, corresponding to longer coherent star-formation bursts than its blue counterpart. 
    }
    \label{fig:SFHs_PS}
\end{figure*}

Using this parametrization, Fig.~\ref{fig:SFHs_PS} shows the average star-formation history of a ``low-mass'' halo with $M_h=10^{10} \Msun$ at $z\approx 5.5$, or 1 Gyr after the Big Bang. 
The average SFR rises exponentially fast in this model, from a mere $\overline {\dot M_\star}\approx 2\times 10^{-3} \,\Msun \, \rm yr^{-1}$ at $z \approx 10$ (650 Myr after the Big Bang), to $\overline {\dot M_\star}\approx 0.1 \,\Msun\, \rm yr^{-1}$ at $z\approx 6$, following the growth of structure in our Universe.
In this Figure, and until Sec.~\ref{sec:results} where we will vary all parameters, we take fiducial values of $\log_{10} \epsilon_\star = -1.5$, $\alpha_\star = 0.7$, $\beta_\star = -0.4$, and $\log_{10}M_p/\Msun = 11.1$, inspired by our HST+JWST fit in Sec.~\ref{sec:results}.

An important note is that the mean of a lognormal variable is larger than its median ($\VEV{e^x} > e^{\VEV{x}} = 1$), as fluctuations asymmetrically increase $e^x$ upwards. 
The {\it mean} SFR can be computed as
\be
\VEV{\dot M_\star} (M_h, t) = \overline {\dot M_\star}(M_h, t) \VEV{e^{x}} = \overline {\dot M_\star}(M_h, t) e^{\sigma_x^2/2},
\ee
which is larger than the median and grows with the variance $\sigma^2_x$ of $x(t)$.
Consequently, our fiducial peak SFE $\epsilon_\star\approx 0.03$ is lower than the $\epsilon_\star\approx 0.1$ commonly assumed at high $z$~\citep[e.g.,][see also Appendix B of \citealt{Nikolic:2024xxo}]{SunFurlanetto2016,Sipple:2023tgt}.

\subsection{Bursts from a power spectrum}

Burstiness enters our formalism through the $e^{x(t)}$ factor in Eq.~\eqref{eq:defSFRx}, which modulates the SFR of each individual galaxy around the average. This is a generic re-writing of the equation, as in principle $x(t)$ is a random variable with arbitrary properties for each galaxy.
However, in practice it is convenient (and often a good approximation) to assume that $x(t)$ is Gaussianly distributed, and that galaxies are instances ``drawn'' from the same distribution.
In that case, the statistics of $x(t)$ are fully determined by its correlation function. 
Following \citet{1901.07556}, we will assume the functional form of a damped random walk for $x$, which has zero mean ($\VEV{x}=0$) and a correlation function 
\be
\xi_x(\Delta t) \equiv \VEV{x(t) x(t+\Delta t)} =  \dfrac{\sigmaPSD^2}{2} e^{-|\Delta t|/\tauPSD},
\label{eq:corrfuncx}
\ee
where $\sigmaPSD$ and $\tauPSD$ parametrize the strength and correlation length of bursts, respectively, with the subscript PS indicating power spectrum.  
This choice encodes our understanding that the physical processes that affect SFHs are stochastic, so we quantify their overall amplitude and coherence timescale~\citep{bathtub,Dave_12_analytic}. 
In this model SFR bursts tend to be coherent on timescales $\tauPSD$, and their amplitude from one timestep to another is set by $\sigmaPSD$ (in fact by setting $\Delta t=0$ we see that the variance of $x$ is $\sigma_x^2 = \sigmaPSD^2/2$, where the factor of 2 is for notational convenience in the power spectrum below). For $\tauPSD \to 0$ we recover uncorrelated/white noise bursts, whereas for $\tauPSD\to \infty$ or $\sigmaPSD\to 0$ we obtain no variability in $x$, and thus no bursts.
For a pedagogical introduction to SFHs from this power-spectrum formalism we recommend the reader to consult \citet{2208.05938}.
Generically we can make both $\sigmaPSD$ and $\tauPSD$ depend on $M_h$ and $z$, and we will focus on the former here.

We can Fourier transform the correlation function to obtain the log-SFR power spectrum:
\be
P_x(\omega) = \int dt\, \xi(t) e^{i \omega t} = \dfrac{\sigmaPSD^2\, \tauPSD}{1 + (\tauPSD \, \omega)^2} ,
\label{eq:powspecx}
\ee
where $\omega$ is a frequency (with units of $1/$Myr), and $P_x$ has units of $1/\omega$ (or Myr). In past literature the $\tauPSD$ has been dropped in the numerator, implicitly substituting that factor for either 1 Myr or 1 Gyr, depending on the context, but we find that doing so leads to spurious correlations between $\sigmaPSD$ and $\tauPSD$.
Compared to past work in e.g.,~\citet{1901.07556} our power spectra can be related as $P_{x,\rm here}(\omega) = P_{x,\rm past}(\omega)\ln(10)^2$, as we work with ln(SFR), and the amplitudes can be translated through $\sigmaPSD^{\rm here} = \sigmaPSD^{\rm past} \ln(10)/\sqrt{\tauPSD}$.

To illustrate this formalism, throughout this Section and the next we will show results for two example galaxy-formation scenarios. 
The first is the ``long-bursts'' model, which has a burstiness amplitude $\sigmaPSD = 2$ (corresponding to 0.6 dex of SFR scatter at any time) and a relatively long timescale $\tauPSD= 20$ Myr. 
The second, or ``short-bursts'' model has $\sigmaPSD= 3$ (or 0.9 dex of SFR variation) and $\tauPSD= 2$ Myr, so galaxies burst strongly even at short timescales.
The top panel of Fig.~\ref{fig:SFHs_PS} shows an example SFH for each of these models, where we can see that larger $\tauPSD$ gives rise to longer, temporally coherent bursts. 
Fig.~\ref{fig:SFHs_PS} also shows the power spectra of the two models, which drops at a different frequency $\omega$ for each model, characterizing their typical burst timescale.  
The amplitude $\sigmaPSD$ and timescale $\tauPSD$ of these two example models are chosen to reproduce the same UVLFs, as we will show in Sec.~\ref{sec:compareobservations}, but note we will vary both of these parameters in our MCMCs.  
While both our lognormal approximation and the functional form for $P_x(\omega)$ are simplifying assumptions, they suffice to differentiate models with stronger/weaker burstiness and longer/shorter burst timescales. 
We leave improvements upon this model for future work, including multiple timescales~\citep{2006.09382} and duty cycles, but we note that in its current incarnation the model is able to reproduce the variability of SFHs seen in high-resolution hydrodynamical simulations that resolve the clustering of supernova feedback in the multiphase ISM \citep{Hu2023}, as we show in App.~\ref{App:hydro}.

The power spectrum $P_x(\omega)$ suffices to generate realizations of SFHs and from them synthesize observables.
This can, however, be prohibitively expensive for MCMC sampling over an unknown parameter space. A key advantage of our approach is analytically computing the observables in $<1$ s from the log-normal nature of the SFHs. 
The rest of this section shows how to do so. The reader uninterested in these calculations may want to skip to Sec.~\ref{sec:compareobservations}.

\begin{figure*}
    \centering
    \includegraphics[width=0.49\linewidth]{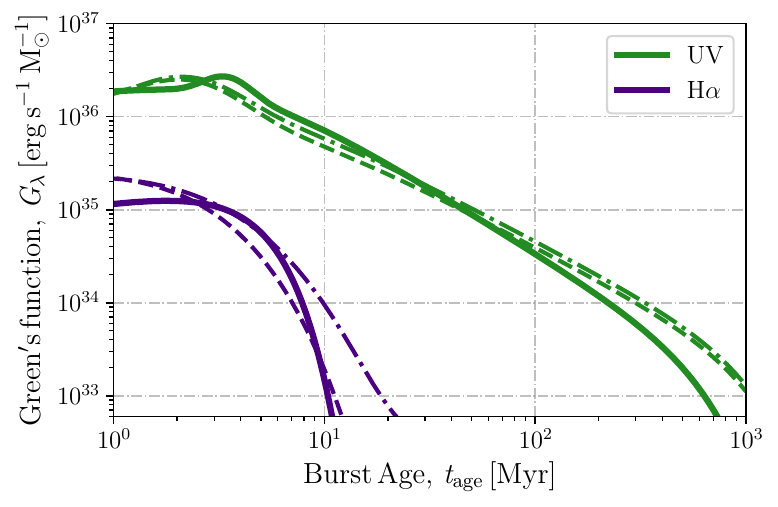}
    \includegraphics[width=0.48\linewidth]{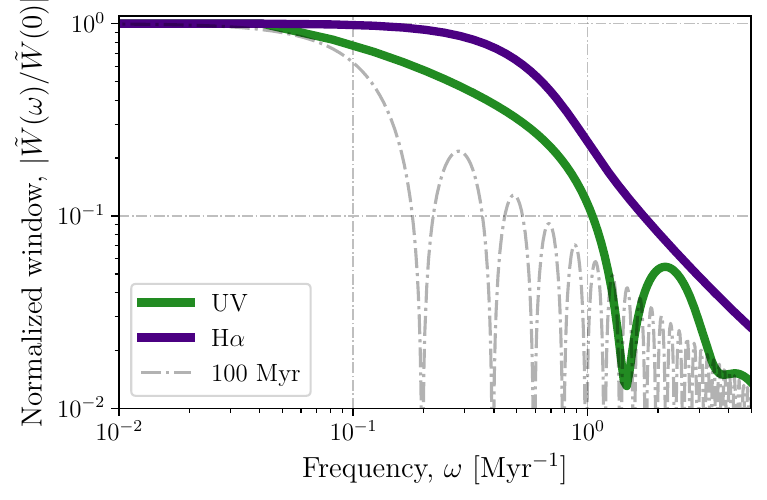}
    \caption{{\bf Left}: Green's function for UV (green) and $\Ha$ (purple) emission, which quantifies how much luminosity in each band is emitted by a 1$\Msun$ burst of star formation of age $t_{\rm age}$, so it can be understood as a mass-to-light ratio. 
    Our baseline is the \citetalias{BC03} model (solid), but we also show \citetalias{BPASS} for single stars (dashed) and with a default binary fraction (dot-dashed), which differ more in $\Ha$ than UV but give rise to very similar burstiness constraints, as we show in Appendix~\ref{app:extraposteriors}. 
    {\bf Right}: The normalized Fourier-space  ``window functions'' $\tilde W$ for UV and H$\alpha$, which are meant to be integrated over frequencies $\omega$ along with the SFR power spectrum (e.g., to obtain $\sigmaUV$). 
    The H$\alpha$ window extends to higher frequencies, as $\Ha$ light can capture fluctuations on shorter timescales. The UV light captures bursts over a broader frequency range than the typically assumed 100 Myr timescale (shown as a gray dot-dashed line for comparison).
    }
    \label{fig:windows}
\end{figure*} 

\subsection{Light from Mass: Green's functions}

When we observe a galaxy at a given redshift $z_{\rm obs}$, we are seeing the light from its stars of different ages combined and processed through dust and gas. 
In this sense, translating star-formation histories into light is an exercise in adding the SEDs of stars of different ages, requiring stellar population synthesis (SPS).  
If we take the simplifying assumption that the entire stellar population is processed in a similar way (e.g., they share dust attenuation) and has a similar origin (in terms of metallicities or initial mass function), as often done in high-$z$ SED fitting, we can find the SED of a galaxy by integrating over its past SFH.
That is, for a galaxy residing in a halo of mass $M_h$ we can write its luminosity at a certain band or wavelength $\lambda$ as an integral over ages:
\be
L_\lambda (M_h, t_{\rm obs}) = \int_0^{t_{\rm obs}} d \tage G_\lambda(\tage) \dot M_\star(M_h, \tage),
\label{eq:SEDintegraldef}
\ee
where $t_{\rm obs}$ is the age of the universe at the time of observation, $\dot M_\star(M_h, \tage)$ is the SFH (including the bursty component), and we call the SPS output $G_\lambda(\tage)$ a ``Green's function'', which translates the past SFH into light at the observed time, acting as an age-dependent mass-to-light ratio (see also Appendix B of \citealt{1901.07556}).
This Green's function $G_\lambda(\tage)$ quantifies how the luminosity $L_\lambda$ responds to stars that formed a certain time $\tage$ ago, so it will be useful for constraining burstiness. We take $G_\lambda(\tage)$ to only depend on age and not explicitly on the past SFH, but this assumption can be lifted to more accurately model the growth of metals, dust, and nebular gas properties as galaxies evolve.
This formalism clarifies how burstiness affects the observed light: the luminosity $L_\lambda$ is a weighted sum of the past SFH, including bursts, so it will inherit variability.

In this work we focus on two main observables: $L_{\rm UV}$ and $L_{\rm H\alpha}$, defined as the UV luminosity at rest-frame 1500 \AA \, and the continuum-subtracted $\Ha$ line-luminosity at 6563 \AA, respectively.
To obtain their Green's functions $G_\lambda$ we simply find the luminosity produced by $1\,\Msun$ of stars formed a time $\tage$ ago (assuming no dust, as we will add variable dust attenuation later). 
By default we use \textsc{Bagpipes}~\citep{bagpipes} and the 2016 version of the SPS model of \citet[][hereafter \citetalias{BC03}, with a \citealt{Kroupa2001_IMF} IMF]{BC03} with $Z=0.1 Z_\odot$ and $\log U = -2.5$ (though the results are not very sensitive to this choice, as the ionization parameter $U$ barely affects H$\alpha$ and increasing or decreasing the metallicity by a dex changes the $\Ha$/UV ratio by $\approx 20\%$, ~\citealt{Mehta23_uvcandels}, which we will re-absorb into a free $\AmpHa$ parameter). We will also show results for BPASS~\citep[][hereafter \citetalias{BPASS}]{BPASS} with the same assumed parameters.

Fig.~\ref{fig:windows} shows the Green's functions $G_\lambda$ for UV and H$\alpha$.  
With increasing age the most massive stars disappear, and with them the ionizing flux that gives rise to $\Ha$ photons, so by $\tage \approx 10$ Myr the $\Ha$ luminosity has dropped by two orders of magnitude. 
Comparatively, it takes $\approx 100$ Myr for the UV luminosity to decrease by the same amount.
In both cases, however, more recent star formation (with $\tage \lesssim$ few Myr) produces much higher luminosities, a phenomenon commonly referred to as outshining \citep{Papovich:2001bu,Harvey2025}.
For $L_{\rm UV}$ in particular there is a bump at $\sim 2-5$ Myr due to the Wolf-Rayet evolutionary stage, which high-$z$ studies are beginning to probe chemically~\citep{Berg25,Topping25}.
While the \citetalias{BC03} and \citetalias{BPASS} response functions $G_\lambda$ are not identical, and they can slightly shift by assuming different metallicities or gas conditions, they share the same physical features: the UV light tracks the SFH over longer timescales than H$\alpha$. This determines how each observable reacts to bursts on the SFH~\citep{Flores2021_indicators}.
We note that the overall amplitude of the Green's functions is fully degenerate with the peak SFE parameter $\epsilon_\star$.  
However, metallicity and the unknown IMF can change the relative amplitude of $G_{\Ha}$ and $G_{\rm UV}$, so we introduce a free parameter $\AmpHa$ (allowed to vary with halo mass) that rescales up and down the $\Ha$ Green's function to account for this uncertainty. Changes in the shape are tested separately in Appendix~\ref{app:extraposteriors}.

\subsection{PDFs of observables}

Our formalism makes it clear that burstiness in SFHs makes halos of a fixed mass $M_h$ host galaxies with different luminosities $L_\lambda$.
In order to build luminosity functions we will need 1D probability density functions (PDFs) $\mathcal P(L_\lambda | M_h)$, and to model observations of H$\alpha$/UV ratios we will additionally require the joint 2D PDF $\mathcal P(L_{\rm UV}, L_{\rm H\alpha} | M_h)$, both to be integrated against halo mass functions. 
Rather than taking effective burstiness parameters, such as $\sigmaUV$ or $\sigmaHa$ to be free and independent, as commonly done in analytic studies~\citep{shen23,Munoz:2023cup,Gelli24}, we will derive the PDFs of observables with our PS formalism.
We return to the luminosity from Eq.~\eqref{eq:SEDintegraldef} and rewrite it as
\ba
L_\lambda(M_h, t_{\rm obs}) &= \int_0^{t_{\rm obs}} d \tage G_\lambda(\tage) \overline{\dot M_\star}(M_h, \tage) e^{ x(\tage)} \nonumber \\ &= \int_0^{t_{\rm obs}} d \tage W_\lambda(M_h, \tage) y(\tage),
\label{eq:SEDintegraldef_y}
\end{align}
which is a linear combination of the SFR weighed by the Green's functions. In the last equality we have separated the (linear) SFR fluctuations normalized by their mean
\be
y = {\dot M_\star/\VEV{{\dot M_\star}}} = e^{x-\sigma_x^2/2},
\label{eq:y_definition}
\ee
and a multiplicative term that absorbs the mean SFR and Green's function,
\be
W_\lambda(M_h, \tage) = G_\lambda(\tage) \VEV{\dot M_\star}(M_h, \tage).
\ee
We call this $W_\lambda$ a ``window function'', as it will filter SFR fluctuations for each observable in Fourier space. 
With this expression, we can compute the average UV and H$\alpha$ luminosities simply as
\be
\VEV{L_\lambda}(M_h, t_{\rm obs}) =  \int_0^{t_{\rm obs}} \!\!\!  d \tage W_\lambda(M_h, \tage)  ,
\ee
where we have used our definition of $y$ from Eq.~\eqref{eq:y_definition}, which enforces $\VEV{y} = 1$.

Finding the full PDF of each observable is slightly more involved. 
We take advantage of the lognormal assumption 
of the SFH fluctuations, so each luminosity $L_\lambda$ will be a sum of lognormal random variables $y$ weighed by the window function $W_\lambda$, which allows us to compute their variance directly as
\be
\sigma^2_{L_\lambda} = \VEV{L_\lambda^2} - \VEV{L_\lambda}^2 = \int \dfrac{d\omega}{(2\pi)}  |\tilde W_\lambda(\omega)|^2 P_y(\omega),
\label{eq:varianceLum_PS}
\ee
where $P_y$ is the power spectrum\footnote{Observables depend on the SFR and not its log. 
Fortunately, for a (normalized) lognormal variable $y=e^{x-\sigma_x^2/2}$ the correlation function can be found from that of $x$ as $\xi_y(t) = e^{\xi_x(t)} - 1$~\citep{Xavier:2016elr}, and from it the power spectrum $P_y (\omega)$ through an FFT, as we detail in Appendix~\ref{app:powerspectrum_y}. } of $y$, and $\tilde W_\lambda(\omega)$ is the Fourier transform of $W_\lambda(\tage)$.
This $W_\lambda(\tage)$ now clearly acts as a window function, filtering how SFR fluctuations (i.e., bursts) at each frequency affect the variance of the luminosity $L_\lambda$. 
Our formalism mimics the calculation of, for instance, $\sigma_8$ in standard cosmology, with $P_y$ taking the place of the matter power spectrum and $\tilde W_\lambda$ that of the top-hat window function, and operating over frequencies (in Myr$^{-1}$) rather than wavenumbers (in Mpc$^{-1}$).

\begin{figure*}
    \centering
    \includegraphics[width = 0.31\linewidth]
    {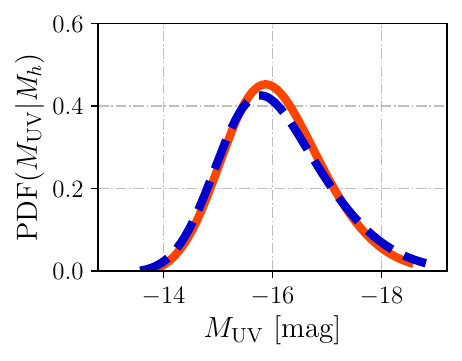}
    \includegraphics[width = 0.31\linewidth]{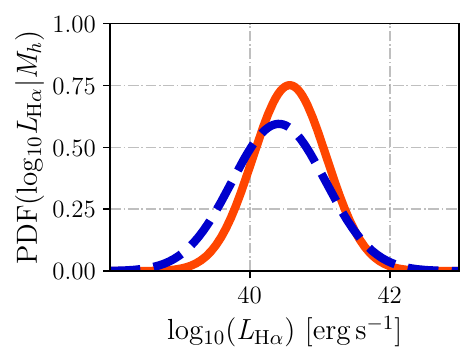}
    \includegraphics[width = 0.3\linewidth]{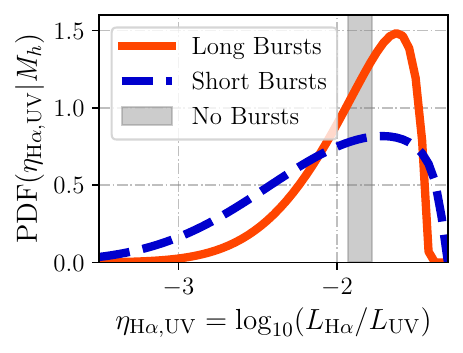}
    \caption{Predicted PDFs  of different observables for two illustrative models: one with long bursts (red) and another with short bursts (blue), in both cases for galaxies residing in halos of $M_h=10^{10}\,\Msun$ at $z=6$. 
    {\bf Left:} Both models predict nearly identical $\MUV$ distributions, as they have been calibrated to the same UVLFs.
    {\bf Center:} Their PDFs for $\log_{10}L_{\rm H\alpha}$ differ somewhat, though not enough to distinguish the burstiness timescale. 
    {\bf Right:} The H$\alpha$/UV ratios can efficiently differentiate between these two example models. The long-burst model (red) predicts a narrower PDF, closer to the no-burst expectation (gray), whereas in the short-burst model (blue) the H$\alpha$ and UV are less correlated and thus the distribution is broader.
    In both cases the PDF is skewed towards smaller values of $\etaHaUV$, representing ``off-mode'' galaxies. 
    }
    \label{fig:PDFs_at_Mh}
\end{figure*}

To build intuition, the right panel of Fig.~\ref{fig:windows} shows the normalized Fourier-space window functions $\tilde W_{\rm UV}$ and $\tilde W_{\Ha}$  as a function of frequency $\omega$. The $\Ha$ fluctuations pick up contributions from higher frequencies than the UV, as they are sensitive to shorter timescales. The UV window, on the other hand, starts to drop at lower frequencies $\omega\approx 0.1\,\rm Myr^{-1}$, as both young and old stars emit UV light so short and long bursts cannot be cleanly separated through UV alone (though the window function still has contributions at $\omega\approx 1 \,\rm Myr^{-1}$). 
As a point of comparison, we show a top-hat window function with a 100 Myr width, as commonly reported in SED fitting (e.g., SFR$_{100}$), which drops faster and oscillates rapidly, unlike either of the two observables. 
The window functions hold the information we need to extract burstiness, but they depend on both the Green's functions $G_\lambda$ and the average SFR $\overline {\dot M_\star}$, so we have to compute them on the fly at each $z$ and for each parameter set.

Given means and variances, we use the insight that a sum of lognormals is itself approximately lognormal~\citep{Lo2012} to find the PDF of each observable.
As we will see below, this is an excellent approximation for $L_{\rm H \alpha}$, whereas $L_{\rm UV}$ receives contributions from a broader set of timescales, which requires decomposing it into a short and a long-timescale component and finding the PDF of their sum directly (more details of this procedure are in Appendix~\ref{app:sum_of_lognormals}, and a comparison against a direct simulation in Appendix~\ref{app:comparison_simulation}).
We show in Fig.~\ref{fig:PDFs_at_Mh} the $\MUV$ and $L_{\Ha}$ PDFs for halos with $M_h=10^{10} \Msun$ at $z=6$ for our two example models with long and short bursts.
These two scenarios have very similar $\MUV$ PDFs as they have been calibrated to reproduce the observed UVLFs (which we will confirm soon). Both are very broad, with $\sigmaUV \approx 1.5$ due to the amount of burstiness in the SFHs.  
The $L_{\Ha}$ PDFs, however, differ for both models. The long-bursts model shows less variance at the smaller timescales where $\Ha$ light is produced, and thus its PDF is narrower. 
While the $L_\Ha$ PDFs are nearly exactly lognormal, the $\MUV$ ones are skewed, so UV-bright (more negative $\MUV$) galaxies are overrepresented. We will fit the 1D PDFs to its closest lognormal here, and leave for future work using the full-shape PDF and including a non-unity occupation fraction~\citep{Berlind:2001xk}.

In addition to the individual $\Ha$ and UV luminosities, a powerful observable to measure burstiness is their ratio, whose log$_{10}$ we define as
\be
\etaHaUV \equiv \log_{10}\left (L_{\Ha}/L_{\rm UV} \right)
\ee
for notational convenience. 
The lower-left panel of Fig.~\ref{fig:SFHs_PS} shows the temporal evolution of $\etaHaUV$ for two example galaxies, which moves up and down tracing their SFHs. 
Each value of $\etaHaUV$ can be obtained under many different SFH configurations, so this quantity may not be a reliable indicator of SFR$_{10}$/SFR$_{100}$ for individual galaxies~\citep{Fisher25_REBELS}. 
Yet, the distribution of $\etaHaUV$ values of an entire galaxy population encodes key information on the burstiness amplitude and timescale. As an example, the short-burst example galaxy shown in Fig.~\ref{fig:SFHs_PS} spends more time at low $\etaHaUV$ values than its long-bursts counterpart, so we expect a broader distribution of $\Ha$/UV ratios for that model.

Computing the PDF of $\etaHaUV$ requires not just those of $L_{\Ha},$ and $L_{\rm UV}$, but also their cross correlation, which we find as (see also e.g., \citealt{Sun2023})
\ba
\sigma^2_{L_{\lambda_1} L_{\lambda_2}} &= \VEV{L_{\lambda_1} L_{\lambda_2}} - \VEV{L_{\lambda_1}} \VEV{L_{\lambda_2}} \nonumber \\ &= \int \dfrac{d\omega}{(2\pi)} \tilde W^*_{\lambda_1}(\omega)\tilde W_{\lambda_2}(\omega) P_y(\omega).
\end{align}
Note that dust attenuation affects the UV and $\Ha$ luminosities differently, shifting the observed $\etaHaUV$. In this work we will do inference with dust-corrected $\Ha$/UV ratios, so we leave for future work modeling the impact of dust on this observable. 
The right panel of Fig.~\ref{fig:PDFs_at_Mh} shows the PDF of $\etaHaUV$ for halos of $M_h=10^{10}\,\Msun$ at $z=6$ (we invite the reader to visit Appendix~\ref{app:eta_PDF} for the mathematical derivation on how this PDF is computed from the cross-variances).
While both the $L_{\rm UV}$ and $L_\Ha$ PDFs are approximately lognormal, that of their ratio $\etaHaUV$ is not.
Both models show a broad --- and skewed --- distribution of $\Ha$/UV ratios at fixed $M_h$, peaking near the ``equilibrium'' or no-burst prediction of $10^{\etaHaUV} = 1/60-1/85$~\citep[][where the range corresponds to a 2 dex change in metallicity]{asada}, but extending towards negative values of $\etaHaUV$.
Physically, this PDF can be interpreted as the fraction of galaxies that have experienced a recent burst (and thus $\Ha$ emission) at each $\MUV$ and $M_h$, so the broad tail towards negative $\etaHaUV$ values signifies that an important fraction of galaxies are expected to be ``off'' in $\Ha$ ($\sim10$ Myr timescale) but ``on'' in the UV (or $\sim 100$ Myr).
This is a telltale sign of burstiness. 
The short-bursts model (with $\sigmaPSD=3$ and $\tauPSD = 2$ Myr), shows a broader PDF than its long-bursts counterpart, as the $\Ha$ and UV are less correlated if there are shorter on/off cycles versus long steady periods of star formation.

\begin{figure}
    \centering
    \includegraphics[width = 1.01\linewidth]
    {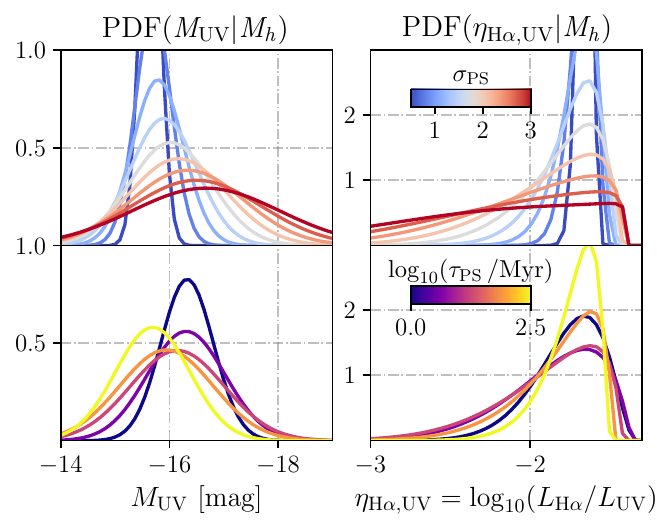}

    \caption{Predicted PDFs for $\MUV$ ({\bf left}) and the $\Ha$/UV ratio ({\bf right}) for galaxies hosted in halos with $M_h=10^{10}$ at $z=6$ as a function of the bursty power spectrum parameters. In the {\bf top} panels we vary the amplitude $\sigmaPSD$ and in the {\bf bottom} the timescale $\tauPSD$. The former has a stronger impact on the width of the PDFs, so it will be easier to constrain. Still, the timescale $\tauPSD$ affects the shape of the PDFs, so combining UV and $\Ha$ information will allow us to measure it. 
    }
    \label{fig:Change_PDFs_eta_sigmatau}
\end{figure}

In order to build further intuition, Fig.~\ref{fig:Change_PDFs_eta_sigmatau} shows how the PDFs $\mathcal P(\MUV|M_h)$ and $\mathcal P(\etaHaUV|M_h)$ change as a function of the PS amplitude and timescale.
Taking as starting point the fiducial long-bursts model (with $\sigmaPSD=2$ and $\tauPSD=20$ Myr), we vary one of these parameters at a time. 
The amplitude $\sigmaPSD$ has the biggest influence.  Reducing $\sigmaPSD$ returns narrower distributions of both $\MUV$ and $\etaHaUV$ (around the equilibrium value), whereas large $\sigmaPSD$ produce high scatter in the two observables and thus more bright (negative $\MUV$) and ``off-mode'' (low $\etaHaUV$) galaxies. 
The timescale $\tauPSD$ has a more modest effect on the shape. 
Yet, very long burstiness timescales ($\tauPSD\gtrsim 100$ Myr) produce a sharper $\etaHaUV$ distribution, as well as fainter $\MUV$. 
Interestingly, small $\tauPSD\approx 1$ Myr produce relatively narrow $\MUV$ distributions (as this observable is averaged over long timescales) but broad PDFs for $\etaHaUV$ because star formation over shorter ($\Ha$) and longer (UV) timescales are no longer correlated.  
The distribution of $\etaHaUV$, therefore, holds extremely valuable information on the amplitude and timescales of bursts in the early universe, and by combining $\etaHaUV$ with UVLFs we will be able to measure both $\sigmaPSD$ and $\tauPSD$.

\section{Comparing to observations}
\label{sec:compareobservations}

We have built a model to find the PDF of observables ($\MUV,\, L_{\rm H\alpha},$ and the $\Ha$/UV ratio $\etaHaUV$) for halos of a certain mass $M_h$, which takes into account their star-formation variability.
We don't observe galaxies as a function of $M_h$, however, so in order to compare to data we need to combine those PDFs with halo mass functions to weigh by the abundance of each halo.  
In order to build intuition we will continue showing results for our two example (long- and short-bursts) models through this Section, deferring a full parameter search to Section~\ref{sec:results}.

\subsection{Luminosity functions}

The UVLF is defined as the comoving number density of galaxies per unit UV magnitude. 
As such, it can be obtained by summing over all halos with a weight given by how likely each is to host a galaxy with magnitude $\MUV$. That is,
\be
\Phi_{\rm UV}(\MUV) = \int dM_h \dfrac{dn}{dM_h} \mathcal P(\MUV|M_h),
\ee
where $dn/dM_h$ is the halo mass function, for which we use the parameterization based on $N$-body simulations from~\citet{Yung}.

In addition to burstiness, which enters the UVLFs through $\mathcal P(\MUV|M_h)$ (see Fig~\ref{fig:Change_PDFs_eta_sigmatau}), we have to model dust attenuation. 
Dust will both dim and redden galaxies, affecting the UV more than $\Ha$, and if unmodeled can give rise to incorrect inferences on star-formation parameters~\citep{Narayanan_18_dust}.
Following \citet{astro-ph/9903054}, we model the mean dust attenuation of galaxies of magnitude $\MUV$ as
\be
\VEV{A_{\rm UV}}= C_0 + C_1 \VEV{\beta} + 0.2 \ln (10) C_1^2 \sigma_\beta^2,
\ee
where $C_0$, $C_1$ are free parameters, and use the relation between the (mean) UV slope and magnitude ($\VEV{\beta}-\MUV$) from \citet[][see also~\citealt{Cullen23_jwst_UVSlopes,Topping24_UVslopes,Austin25_betaslopes,Jecmen_beta26}]{Bouwens2014} with a fixed scatter $\sigma_\beta = 0.34$~\citep[][which could be reabsorbed into the definition of $C_0$ and $C_1$ along with the dust-attenuation law, which can be different at high $z$, see~\citealt{McKinney:2025htm}]{Tacchella}.
Past determinations using this IRX-$\beta$ relation have found a broad range of dust-attenuation fits covering $C_0=2.5-4.5$ and $C_1=1.1-2.1$~\citep{Overzier11_dust,1204.3626,Casey:2014cqa,Bouwens16xiion}, so we will vary both parameters over this range in our analyses. 
Moreover, the dust content and geometry can change dramatically from one galaxy to another, in effect producing $\MUV$ variability that could be confused with burstiness~\citep{Cochrane,1904.07238,Carniani2018_dust_distr}.
To account for this dust stochasticity we introduce another free parameter, the dust variability $\sigma_{\rm dust}$, which gives rise to a $\MUV$ scatter
\be
\sigma_{\MUV, \rm dust} = \VEV{A_{\rm UV}} \times \sigma_{\rm dust}.
\ee
which we add in quadrature to that due to burstiness:
\be
\sigma_{\MUV, \rm total}^2 = \sigma_{\MUV, \rm burst}^2 + \sigma_{\MUV, \rm dust}^2.
\ee
In our MCMCs we will vary this parameter with a Gaussian prior of $\sigma_{\rm dust} = 0.5\pm 0.5$, always keeping it positive.
Through this section we will fix the dust parameters to $C_0=3.1$, $C_1=1.8$, and $\sigma_{\rm dust}=0.1$ when computing UVLFs, and we note that we always set a minimum $\sigmaUV=0.2$.

\begin{figure}
    \centering
    \includegraphics[width = 1.0\linewidth]
    {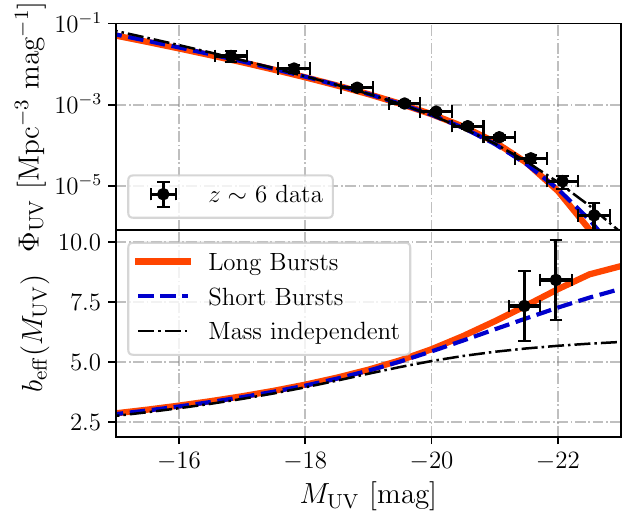}
    \caption{
    Predicted UVLF ({\bf top}) and bias ({\bf bottom}) at $z\sim 6$ for the two example models (with long and short bursts in red and blue, respectively), as well as one with mass-independent burstiness (black dotdashed).
    The three models predict very similar UVLFs, but the mass-independent case differs in bias at the bright end (where measurements lie). 
    The long- and short-bursts models predict the same PDF $\mathcal P(\MUV|M_h)$, as shown in Fig.~\ref{fig:PDFs_at_Mh}, so even with clustering we cannot distinguish them. 
    Black points show observations from HST~\citep{Bouwens21} and HSC~\citep{Harikane}.
    }
    \label{fig:UVLFz6}
\end{figure}

Fig.~\ref{fig:UVLFz6} shows predicted UVLFs at $z\sim 6$ for three scenarios. First, our two example models with long and short bursts (with have nearly identical $\MUV$ PDFs and thus UVLFs) both assume that galaxies become burstier if hosted in smaller halos, with $d\sigmaPSD/d\log_{10}M_h= -0.4$ (as the data prefers, more below in Sec.~\ref{sec:results}). For comparison, we show a mass-independent burstiness model, where $\sigmaPSD$ does not change with $M_h$.
Despite the very different galaxy-formation physics of these three models (long vs short bursts and mass-dependent vs independent $\sigmaUV$), they all predict very similar UVLFs. That is because the SFE $f_\star$ can be recalibrated in each of these models, showcasing the degeneracy between star-formation efficiency and burstiness. 
For context, Fig.~\ref{fig:UVLFz6} also shows the UVLF measured at $z\sim 6$ from \citet[][see also~\citealt{FinkelsteinBagley22_UVLF}]{Bouwens21}.  
Those data were obtained assuming a different cosmology ($\Omega_m=0.3$, $h=0.7$), so we correct the magnitudes and cosmological volumes and assume a minimum uncertainty of 20\% to account for cosmic variance and any unmodeled systematics~\citep{Sabti:2021xvh}.
Additionally, when comparing theory to observations we compute the UVLF at several redshifts and integrate them over a normalized redshift window that corresponds to each observed $z$ selection function.

\subsection{Clustering}

The spatial distribution or {\it clustering} of galaxies provides additional information on their formation. 
For instance, burstiness broadens the distribution $\mathcal P(\MUV|M_h)$, allowing smaller-mass halos, which are more homogeneously distributed, to populate the brighter $\MUV$ bins.
Thus, the clustering of galaxies as a function of luminosity depends sensitively on the amount of UV scatter, which will help in breaking key degeneracies. 
Here we will model galaxy clustering through the effective bias~\citep{Munoz:2023cup}
\be
b_{\rm eff} (\MUV) = \Phi_{\rm UV}^{-1} \int dM_h \dfrac{dn}{dM_h} \mathcal P(\MUV|M_h) b_h(M_h),
\ee
where we take the halo bias $b_h(M_h)$ from \citet{Tinker10_bias}. 
In essence, the effective bias quantifies how correlated galaxies are, with $b_{\rm eff}\gg 1$ implying galaxies tend to reside close to each other, as expected of very massive halos.  
There is additional information on the full correlation function of galaxies~\citep{Shuntov,Paquereau25_clustering}, which we will consider in future work.

\begin{figure*}
    \centering
    \includegraphics[width = 0.8\linewidth]
    {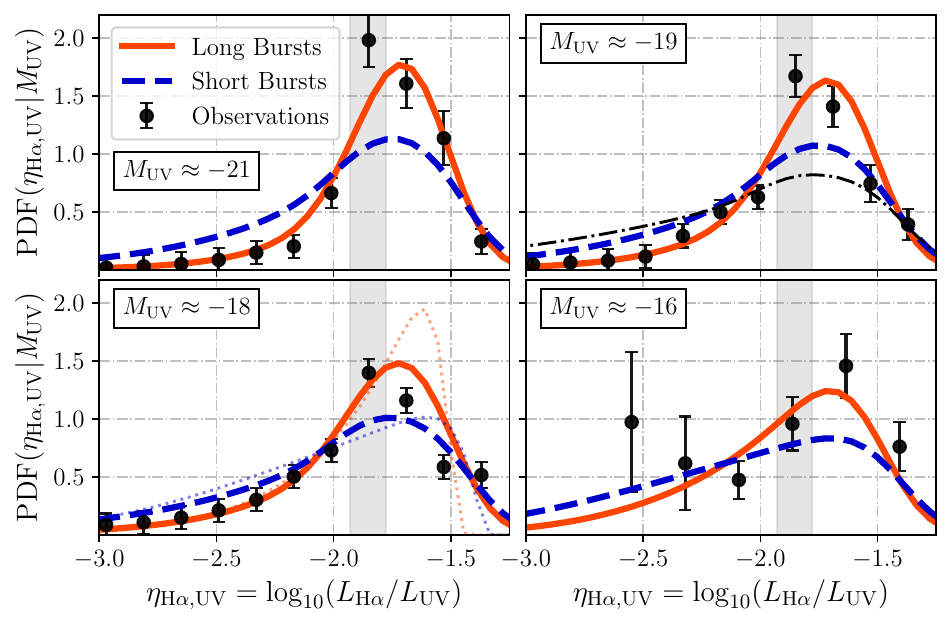}
    \caption{
    Histograms of the H$\alpha$/UV ratios predicted by the two example models at $z\sim 6$ (with long $\tauPSD=20$ Myr bursts in red and short $\tauPSD=2$ Myr in blue) compared to observations of UV-selected galaxies from~\citet[][dust-corrected and ordered from bright to faint $\MUV$ bins]{Endsley24_bursty,Chisholm}.
    This observable traces the variation in star-formation histories, and as such is sensitive to the amplitude and timescale of the SFR power spectrum.
    For instance, we see the PDFs broaden in the fainter $\MUV$ bins, which we will interpret as enhanced burstiness for smaller halo masses. 
    Likewise, the shape of the PDF is sensitive to the burstiness timescale, which appears better fit by longer bursts (red curves). 
    Sec.~\ref{sec:results} will confirm these intuitions through an MCMC search.
    For reference, the vertical gray bands show the no-burstiness expectation, much narrower than the observations. 
    In the second panel ($\MUV\sim -19$) we show the mass-independent bursty model from Fig.~\ref{fig:UVLFz6} as the black dot-dashed line, which predicts a PDF far too broad.
    All predictions have been smoothed by the observational uncertainty on $\etaHaUV$, and in the third panel ($\MUV\sim -18$) we show the un-smoothed curves as dotted lines for comparison.
    }
    \label{fig:HaUVratios_z6}
\end{figure*}

The bottom panel of Fig.~\ref{fig:UVLFz6} shows clustering predictions for three models along with the bias measurements from ~\citet[][where we translate their measurements from thresholds to $\MUV$ bins in Appendix~\ref{app:bias_threshold}]{Harikane}. 
The bias grows for brighter galaxies, as they tend to live in heavier halos, which cluster more strongly. 
This trend is less marked for the mass-independent case, where burstiness allows the brightest galaxies to be hosted by relatively light halos. This lowers the predicted bias compared to the other two models (and to the measurements, though error-bars remain sizeable). 
However, the long- and short-bursts models give rise to the same bias, so UV data alone cannot distinguish the timescales of star-formation variability. 
Let us now show how adding H$\alpha$ data can provide new, key information on burstiness.

\subsection{H$\alpha$/UV ratios}

Through JWST we have access to information on high-$z$ galaxies beyond their $\MUV$.  
In particular, H$\alpha$/UV ratios provide an economical way to quantify the burstiness of the high-$z$ Lyman-break galaxy (LBG) samples, like the ones used to build the UVLF, as $\Ha$ can be measured through medium bands without spectroscopic selection functions~\citep[e.g.,][]{Endsley23_reionization}. The $\Ha$/UV ratios are sensitive to both the timescale and amplitude of burstiness (as seen in Fig.~\ref{fig:Change_PDFs_eta_sigmatau}), but less so to parameters such as the star-formation efficiency, providing complementary information to the UVLF. 
In this work we will compare our predictions against JWST observations at $z\sim 4-6$ from two different sources, in all cases dust corrected (though these samples show little dust attenuation towards the faint end, with a median $A_V \leq 0.05$ for $\MUV\geq -19$~\citealt{Endsley24_bursty}). 

First, we combine the $z\sim 6$ data from \citet{Endsley24_bursty} and \citet{Chisholm}. 
The former is obtained from $N = 368$ galaxies in the GOODS and Abell 2744 fields, where the H$\alpha$/UV ratios are SED-fitted assuming a two-component SFH model in BEAGLE~\citep{Chevallard16_BEAGLE}. 
The latter data are obtained directly through the medium-band excess from $N=95$ galaxies in the GLIMPSE program~\citep{AtekChisholm25_glimpse}, which reaches fainter magnitudes.  We use GLIMPSE to construct a faint bin ($\MUV = -15$ to $-17$), and keep the $\MUV\leq -17$ galaxies of \citet{Endsley24_bursty}, split in three bins. 
Fig.~\ref{fig:HaUVratios_z6} shows the PDFs $\mathcal P(\etaHaUV | \MUV)$ of these $z\sim 6$ data.

\begin{figure}
    \centering
    \includegraphics[width=1.02\linewidth]{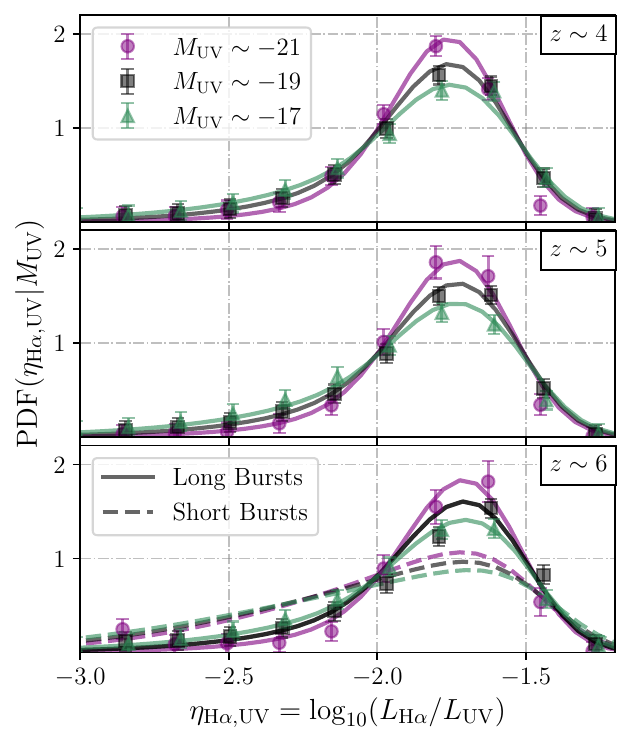}
    \caption{
    Histograms of $\Ha$/UV ratios observed in \citet{simmonds24_masscompl}, along with predictions from a fiducial long-bursts model (solid, with the short-bursts counterpart as dashed only in the last panel). The increased burstiness towards small halo masses translates into broader and more skewed PDFs towards the faint end at all redshifts. The $x$ position of each $\MUV$ bin has been shifted slightly in this plot for visualization purposes.
    }
    \label{fig:HaUV_ratios_allz}
\end{figure}

Second, we use the $z\sim 4, 5$ and 6 ratios from ~\citet[][which continue to higher $z$, albeit not through $\Ha$ necessarily]{simmonds24_masscompl}, obtained from SED-fitting $N\sim 10^4$ JADES galaxies with {\tt Prospector}~\citep{2012.01426}. 
Fig.~\ref{fig:HaUV_ratios_allz} shows the PDFs of $\etaHaUV$ split in three $\MUV$ bins centered at $\MUV=\{-21,-19,-17\}$ with width $\Delta \MUV=2$ (only reaching down to $\MUV=-16$, where the sample is considered fairly complete. See Appendix~\ref{app:HaUVhistograms} for alternative cuts.)  
These $\Ha$/UV observations cover a broader redshift range, and are obtained from a different photometric procedure.
However, at $z\sim 6$ (where they overlap) they agree well with those in Fig.~\ref{fig:HaUVratios_z6}, and differences between them can be thought of as modeling error in extracting $\Ha$/UV ratios from medium-band photometry.
As we will see, they give rise to similar burstiness parameters, and we encourage the reader to visit Appendix~\ref{app:HaUVhistograms}
for a detailed comparison.

The PDFs in Figs.~\ref{fig:HaUVratios_z6} and \ref{fig:HaUV_ratios_allz} all follow the same trend: fainter galaxies have broader distributions of H$\alpha$/UV, hinting at increased burstiness (larger $\sigmaPSD$) towards smaller halo masses. 
The uncertainties are smaller in the \citet{simmonds24_masscompl} dataset as it contains more objects, but are still somewhat dominated by the extraction of $\eta_{\rm H\alpha,UV}$ from photometric measurements of each object\footnote{This produces non-zero off-diagonal elements of the covariance matrix for small $\etaHaUV$ values, which we however disregard in this first analysis.}.
These PDFs and their uncertainties are found through a Monte-Carlo approach, as detailed in appendix~\ref{app:HaUVhistograms}.

Along with the measured PDFs, Figure~\ref{fig:HaUVratios_z6} shows the predictions\footnote{When comparing to data, we smooth theoretical predictions with a kernel that reproduces the observed errors on $\etaHaUV$, which grow towards smaller values as those are harder to differentiate in photometry. We find that the error scales as $\sigma(\etaHaUV) = a + b \times \etaHaUV $ with $a=-0.6$, $b=0.4$, and we set a minimum error of 0.15 dex to account both for modeling uncertainty and scatter in the $\Ha$/UV due to non-burstiness factors such as metallicity~\citep{Shivaie18}.
The third panel in Fig.~\ref{fig:HaUVratios_z6} shows the effect of smoothing the input curves, which is slight but noticeable.} for the two example models with long and short bursts. 
While these models agreed on the UVLF and its clustering, they predict different $\Ha$/UV ratios. 
Looking at the brightest bin ($\MUV\sim -21$), the long-bursts model is more peaked than the short-bursts counterpart, which shows a heavier tail towards negative values of $\etaHaUV$ (or off-mode galaxies). 
This mirrors the behavior seen for $\mathcal P(\etaHaUV|M_h)$ in Fig.~\ref{fig:PDFs_at_Mh}, where shorter bursts ``decorrelate'' the $\Ha$ and UV light, broadening the PDFs.
Moving towards fainter bins, we see both models predict broader PDFs, as we have assumed that burstiness grows towards small halo masses ($d\sigmaPSD/d\log_{10}M_h <0$), which increases variability and thus translates into wider distributions at faint $\MUV$ bins (populated by smaller $M_h$ halos).
The long-bursts model fits observations well in all $\MUV$ bins, whereas the short-burst model has a PDF that is too flat, overshooting the ``off-mode'' low $\etaHaUV$ values and underproducing the peak near $\etaHaUV\sim -1.5$ to $-2$ in the data.
This trend continues at other redshifts, as Figure~\ref{fig:HaUV_ratios_allz} shows. 
In addition, a mass-independent bursty model far overpredicts the width of the $\etaHaUV$ PDF at the bright end, as highlighted by the second panel of Fig.~\ref{fig:HaUVratios_z6}.
We emphasize these are just example models shown to build intuition, and in the next Section we will carry out an MCMC analysis to find the burstiness parameters preferred by data.

These methods are implemented in the publicly available code {\tt Zeus21}\footnote{\url{https://github.com/julianbmunoz/Zeus21}}, which now can take PS parameters as inputs (rather than, say, $\sigmaUV$) and compute PDFs as described in Sec.~\ref{sec:themodel}. These are then convolved with halo mass functions to produce observables. A full run takes  $\lesssim 1$ s, allowing efficient MCMC exploration of the parameter space.

We note, in passing, that we could co-add $\Ha$ luminosity functions.
The advent of JWST spectroscopy has allowed for precise determinations of the abundance of H$\alpha$-emitting galaxies, and thus to build $\Ha$LFs and clustering in a similar manner to UVLFs~\citep{CoveloPaz25_haLF, Korber25_HbLF}.
This can potentially break degeneracies in the timescale of burstiness, though current $\Ha$LF data are not powerful enough to fully disentangle burstiness parameters~\citep{2410.21409}.
Additionally, differences in the selection functions for galaxies in UV and $\Ha$ may hinder a combined analysis. 
As such, we will not directly include $\Ha$LFs in our likelihoods, but we show in Appendix~\ref{app:HaLF} that our model (calibrated to UV and $\Ha$/UV data) fits well the $z\sim 4-6$ observations from \citet{CoveloPaz25_haLF}.

\section{Results}
\label{sec:results}

We now fit different data sets with our model in order to constrain the amount of burstiness, its mass dependence, and the timescales involved.
We will first focus on the $z\sim 6$ data from Fig.~\ref{fig:HaUVratios_z6}, and then combine information from redshifts $z\sim 4-6$ (Fig.~\ref{fig:HaUV_ratios_allz}, as for higher $z$ we lose direct access to $\Ha$ emission through JWST/NIRCam). 
This serves a double purpose.
First, it will showcase how co-adding different observations is required to break degeneracies and extract the physics of burstiness. Second, it allows us to test whether our results depend on the dataset, prior set, and photometric procedure taken to obtain $\Ha$/UV ratios.

\subsection{Fitting at a single redshift}

We begin by fitting $z\sim 6$ observations. 
We build a likelihood $\mathcal L$ for each of our three observables by assuming Gaussian and uncorrelated measurements:
\be
\ln \mathcal L = - \sum_i \dfrac{(d_i-m_i)^2}{2 \sigma_i^2},
\ee
where $d_i$ is the data (UVLFs, bias, or $\Ha$/UV ratios), $\sigma_i$ the reported uncertainty, and $m_i$ is our model prediction.
In this equation $i$ will be $\MUV$ bins for the UVLF [$d_i = \Phi_{\rm UV}(\MUV)$], and clustering [$d_i = b_{\rm eff}(\MUV)$], and bins of $\etaHaUV$ for the $\Ha$/UV ratios [$d_i =\mathcal P (\etaHaUV | \MUV)$].
For simplicity, we assume that the different observables are uncorrelated, so we can just multiply their likelihoods when co-adding data (leaving for future work the use of synthetic observations based on state-of-the-art simulations to properly account for the full covariance). We leverage our analytic approach to MCMC search over parameter space, in all cases varying the 4 parameters that control the average halo-galaxy connection (i.e., the star-formation efficiency parameters $\log_{10}\epsilon_\star$, $\log_{10} M_p/\Msun$, $\alpha_\star$, and $\beta_\star$; where we keep their redshift evolution fixed when fitting at a single $z$), the three dust parameters ($C_0, C_1$, and $\sigma_{\rm dust}$), two parameters that control the mean $\Ha$ luminosity and its mass dependence ($\AmpHa$ and $d\AmpHa/d\log_{10}M_h$, to account for changes in, e.g., metallicity and IMF), and the four parameters that define burstiness ($\sigmaPSD$, $\tauPSD$, and their derivatives $d\sigmaPSD/d\log_{10}M_h$, $d\tauPSD/d\log_{10}M_h$ against halo mass).
While this may appear a very broad set of parameters (13 at a single $z$, 15 in total, see Table~\ref{tab:parameters} for a summary with their prior ranges), we will show that current data is able to constrain most of them well, barring $\beta_\star,C_0$, and $C_1$, which we vary within their prior ranges as a source of uncertainty. 
We run our MCMC and show the most relevant posteriors below. For the full corner plots of all parameters we encourage the reader to visit Appendix~\ref{app:extraposteriors}.

\begin{table}
\centering 
\begin{tabular}{|l|cc|}
\hline
Parameter                                            & Range         & Prior                 \\ \hline
$\log_{10} \epsilon_*$                               & $[-3,0]$      & Flat                  \\
$\log_{10}M_p\,[M_\odot]$                               & $[10,15]$     & Flat                  \\
$\alpha_*$                                           & $[0,1.5]$     & Flat                  \\
$\beta_*$                                            & $[-3, 0]$     & Flat                  \\
$d\log_{10} \epsilon_*/dz$                           & $[-1.5, 1.5]$ & $\mathcal N(0,0.2)$   \\
$d\log_{10} M_p/dz$                          & $[-1.5, 1.5]$ & $\mathcal N(0,0.2)$   \\ \hline
$\sigmaPSD$                              & $[0.1, 5.0]$  & Flat                  \\
$\log_{10}\tauPSD$ [Myr]                   & $[0,2.5]$     & Flat                  \\
$d\sigmaPSD/d\log_{10}M_h$               & $[-1.5, 1.5]$ & Flat                  \\
$d \log_{10}\tauPSD$$/d\log_{10}M_h$ & $[-1.5, 1.5]$ & $\mathcal N(0,0.2)$   \\ \hline
$\AmpHa$                                   & $[0.5, 2.0]$  & $\mathcal N(1,1)$     \\
$d\AmpHa/d\log_{10}M_h$                    & $[-1.5, 1.5]$ & $\mathcal N(0,0.2)$   \\ \hline
$C_{0,\rm dust}$                                     & $[2.5,4.5]$   & $\mathcal N(3.5,1.0)$ \\
$C_{1,\rm dust}$                                     & $[1.1,2.1]$   & $\mathcal N(1.6,0.5)$ \\
$\sigma_{\rm dust}$                                  & $[0, 3.0]$    & $\mathcal N(0, 1.0)$ \\ 
\hline
\end{tabular}
\caption{Parameters we vary in our MCMC analyses, their range, and assumed prior  (where $\mathcal N(\mu, \sigma)$ is a Gaussian of mean $\mu$, and standard deviation $\sigma$). The first box shows the parameters that model star-formation efficiency (anchored at $z=8$), second box burstiness (anchored at $M_h=10^{10}\,\Msun$), third box the IMF/metallicity re-scaling of $L_{\Ha}$, and last box dust attenuation.
For their definitions see Secs.~\ref{sec:themodel} and~\ref{sec:compareobservations}.
}
\label{tab:parameters}
\end{table}

\subsubsection*{Bursty or efficient?}

\begin{figure}
    \centering
    \includegraphics[width = 0.554\linewidth]{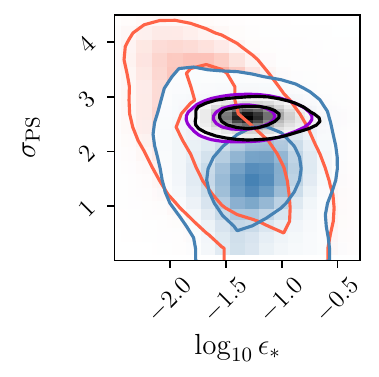}
    \includegraphics[width = 0.428\linewidth]{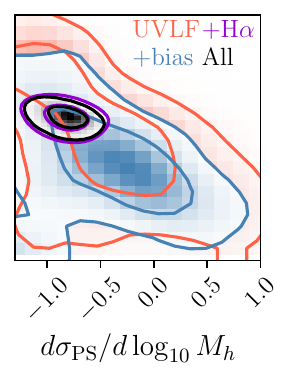}
    
    \caption{Posteriors (1 and 2$\sigma$ in light and dark colors, respectively) when fitting different combinations of $z\sim 6$ data, with UVLF alone in red, UVLF+clustering in blue, UVLF+H$\alpha$/UV ratios in purple, and with all data in black.
    {\bf Left:} Amplitude of burstiness ($\sigmaPSD$ at $M_h=10^{10}\,\Msun$) against star-formation efficiency ($\epsilon_\star$, at the peak halo mass), where UVLF data alone (red) shows a clear degeneracy, partially broken through clustering (blue) which disfavors high burstiness.
    Including H$\alpha$/UV ratios (purple) strongly breaks this degeneracy, preferring bursty models, which is confirmed when co-adding all datasets (black). 
    {\bf Right:} Amplitude of burstiness vs its slope against (log$_{10}$) halo mass, where the UV  data (red and blue) are fairly degenerate. Including H$\alpha$ data reveals a strong preference for growing burstiness towards smaller masses (and thus towards the faint end of the UVLF).
    These posteriors are obtained by varying our entire parameter suite,  and the full corner plot is shown in Appendix~\ref{app:extraposteriors} for reference.
    }
    \label{fig:cornerssigmaPS_z6}
\end{figure}

Burstiness and star-formation efficiency are degenerate in UVLF observations. 
Fig.~\ref{fig:cornerssigmaPS_z6} (left panel) illustrates this point through the posterior of a UVLF-only MCMC. Increasing the burstiness amplitude $\sigmaPSD$ is allowed as long as the peak star-formation efficiency $\epsilon_\star$ decreases. Burstier galaxies can be less efficient on average but produce the same UVLFs, as the more-abundant low-$M_h$ objects can ``scatter'' up in luminosity and populate the bright end of the UVLF.

This is a pervasive issue with one-point statistics, such as luminosity functions. Clustering information can break this degeneracy, as for instance the galaxy bias can pinpoint the halo masses in which galaxies of a certain $\MUV$ reside~\citep{Munoz:2023cup,Giavalisco01_clustering}. 
The difference between the red and blue distributions in
Fig.~\ref{fig:cornerssigmaPS_z6} shows how adding UV clustering measurements to the likelihood reduces the uncertainty, disfavoring the burstiest models (with $\sigmaPSD\gtrsim 3$, similar to the mass-independent model in Fig.~\ref{fig:UVLFz6}). 
Precise bias measurements of UV-selected galaxies are, however, limited to the brightest bins, so they only disfavor high burstiness in the heaviest halos. Upcoming JWST and future {\it Roman} observations will greatly improve these measurements, and can thus break the degeneracy between efficiency and burstiness at high $z$~\citep{Munoz:2023cup}.

Another avenue to break this degeneracy is using the temporal information encoded in the SEDs of galaxies. Burstiness gives rise to larger variability between the H$\alpha$ and UV, which broadens the PDFs of H$\alpha$/UV ratios of galaxies (as shown in Fig.~\ref{fig:Change_PDFs_eta_sigmatau}).
Fig.~\ref{fig:cornerssigmaPS_z6} shows that adding the H$\alpha$/UV observations to the likelihood efficiently breaks the degeneracy and constrains burstiness. 
We find a strong preference for bursty models, with an amplitude of fluctuations of $\sigmaPSD \approx 2-3$. 
As a reminder, the variance of $x=\ln(\dot M_\star)$ is $\sigma_x^2 = \sigmaPSD^2/2$, so our constraint is equivalent to a $0.6-0.8$ dex scatter on the SFR for $M_h=10^{10}\,\Msun$. 
Clustering information does not improve the constraint significantly over the $\Ha$/UV ratios at $z\sim 6$.

\subsubsection*{Heavy or light?}

A powerful indicator of the feedback mechanisms that shape SFHs is the mass dependence of burstiness.
So does burstiness grow or decrease with host-halo mass $M_h$?  
Our model parametrizes this through the derivative $d\sigmaPSD/d\log_{10}M_h$ of the burst strength against halo mass. 
Fig.~\ref{fig:cornerssigmaPS_z6} (right panel) shows the  posterior for this quantity from fitting UVLFs alone, which cannot fully distinguish the behavior of burstiness against halo masses, given the bursty/efficient degeneracy.
UV clustering disfavors strong burstiness at the bright end, where bias measurements reside.
These data prefer $d\sigmaPSD/d\log_{10}M_h<0$, but they are not sufficient to measure .  
Adding H$\alpha$/UV PDF data critically improves the posteriors, showing a strong preference for increased burstiness towards small halo masses (which tend to populate the faint end).  
This is not unexpected; we saw in Fig.~\ref{fig:HaUVratios_z6} that the observed H$\alpha$/UV PDF broadens for fainter $\MUV$ bins, implying more on/off mode star formation.  
Through our model we can physically connect $\MUV$ and $M_h$.  
Including clustering on top of the $\Ha$/UV ratios confirms this result, though the constraint is dominated by the ratios.

\begin{figure}
    \centering
    \includegraphics[width=0.99\linewidth]{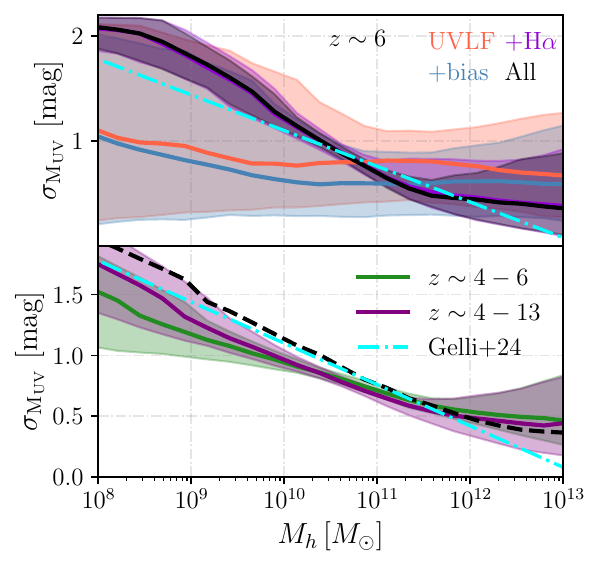}
    \caption{Inferred halo-mass dependence of the UV scatter $\sigmaUV$. 
    {\bf Top} shows the 1$\sigma$ contours from fitting different combinations of UV and H$\alpha$ observations at $z\sim 6$, which together allow us to measure the amplitude and mass dependence of $\sigmaUV$. We find increased scatter for lower masses, as predicted by hydrodynamical simulations (compiled in \citealt[][cyan dot-dashed line]{Gelli24}). 
    For heavy halos, with $M_h\gtrsim 3\times 10^{11}\,\Msun$, the UV scatter flattens due to the contribution from dust variability. 
    {\bf Bottom} combines all data at $z\sim 4-6$ (in green, from UVLFs, clustering, and the $\Ha$/UV ratios from~\citealt{simmonds24_masscompl} instead of from~\citealt{Endsley24_bursty} and~\citealt{Chisholm}) and at $z\sim 4-13$ (adding UVLFs for $z>6$, purple), which agree with each other and sharpen the measurement of increased burstiness for smaller $M_h$.
    Black-dashed line shows the $\sigmaUV$ predicted at $z = 10$, which is higher as earlier halos grow faster.  
    }
    \label{fig:sigmaUVHavsMh}
    \label{fig:sigmaUV_allz}
\end{figure}

To build intuition, in Fig.~\ref{fig:sigmaUVHavsMh} we project our results into the more familiar $\MUV$ scatter $\sigmaUV$ (in mag, defined as the width in $\MUV$ of galaxies hosted in halos of mass $M_h$). 
As expected, the UVLF alone cannot measure $\sigmaUV$, allowing for a broad range of values ($\sigmaUV\sim 0.4-1.4$) at $M_h\sim 10^{11}\,\Msun$, where it is best constrained. Adding bias measurements only rules out the burstiest models.
The situation is dramatically improved when including H$\alpha$/UV observations, which allow us to both constrain $\sigmaUV$ and find a clear preference for increased scatter towards smaller masses.  
We infer a UV scatter of $\sigmaUV\approx 0.75$ at $M_h=10^{11}\, \Msun$, similar to the results of \citet{Shuntov} using H$\alpha$ and OIII emitter clustering. We do not co-add these clustering data as line-emission-selected samples may not represent the overall LBG population, which includes non-line-emitter ``off-mode'' galaxies.

In addition to star-formation burstiness, the observed UV variability shown in Fig.~\ref{fig:sigmaUVHavsMh} includes dust attenuation, controlled by the free parameter $\sigma (A_{\rm UV})$ in our model. Our results suggest that dust variability dominates $\sigmaUV$ for the heaviest halos (with $M_h\gtrsim 3\times 10^{11}\,\Msun$, where the $\sigmaUV$ curves deviate from a simple power law in Fig.~\ref{fig:sigmaUVHavsMh}).
The overall dearth of massive halos at high $z$ impedes us from robustly constraining $\sigmaUV$ at the heavy end, so our results should not be overinterpreted in this regime (or at later times, $z\ll 4$). 
As a byproduct of our analysis we measure the $\Ha$ burstiness $\sigmaHa$, finding $\sigmaHa \approx 0.4$ dex at $M_h=10^{11}\,\Msun$ that grows to a larger $\sigmaHa \approx 1$ dex at $M_h=10^{9}\,\Msun$, mirroring the $\sigmaUV$ trend.

In short, our posteriors in Fig.~\ref{fig:sigmaUVHavsMh} show that current $z\sim 6$ observations prefer galaxies hosted in smaller-mass halos to be burstier.
For reference, we show the fit to $\sigmaUV$ from hydrodynamical simulations found in \citet[][see also~\citealt{2405.04578, Shen+25,Katz25_megatron, Jenna25,Spice26_UVLFscatter}]{Gelli24}, which predicts increased UV scatter towards smaller masses, which overlaps nicely with our $z\sim 6$ measurement. 
We will test this trend when co-adding different redshifts below.

\subsubsection*{Fast or slow?}

The last question we would like to answer is whether star-formation burstiness happens on fast or slow timescales. 
This measurement is more complex, as $\sigmaPSD$ is the main parameter that regulates the amount of scatter in observables, with $\tauPSD$ acting as a ``decorrelation timescale'' that cuts off shorter bursts (see Fig.~\ref{fig:Change_PDFs_eta_sigmatau}).

\begin{figure}
    \centering
\includegraphics[width = 0.99\linewidth]{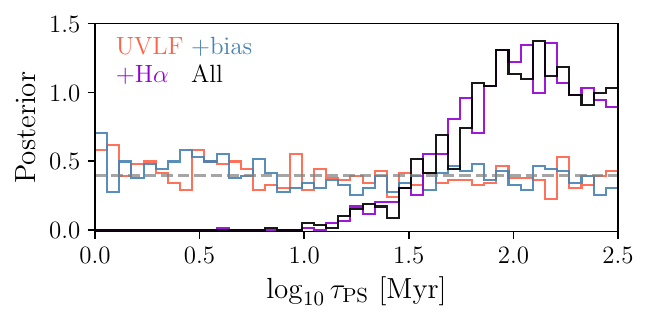}
    \caption{Posterior for the ($\log_{10}$) burst timescale $\tauPSD$ when fitting $z\sim 6$ Lyman-break galaxies. UV data alone (red and blue) are unable to constrain this quantity, and their posteriors overlap the prior (flat, gray dashed line). By combining UV with H$\alpha$ (purple and black) we find a clear preference for moderately long bursts, with $\tauPSD\gtrsim 20$ Myr at 95\% CL. 
    }
    \label{fig:cornerssigmaPS_z6_tau}
\end{figure}

Fig.~\ref{fig:cornerssigmaPS_z6_tau} shows the posterior of $\tauPSD$, where UVLFs alone or added to clustering cannot distinguish the timescales of burstiness, essentially allowing all values of $\tauPSD$.
This was expected, as any one tracer has no hope of measuring both of these parameters.
Co-adding information from UV and $\Ha$ can break the degeneracy and begin to measure $\tauPSD$. 
With these single-$z$ data we do not find sharp constraints, though our posteriors are peaked towards medium-to-long timescales ($\tauPSD\gtrsim 20$ Myr at 95\%CL). 
This disfavors very short (white-noise-like) $\tauPSD\sim$ a few Myr, as expected from prompt star formation and feedback (e.g., from stellar winds and radiation).  
We will return to measuring the timescale with data over multiple $z$ below, but we note that our findings so far align with the claims in \citet{Weisz12} and \citet{Endsley24_bursty} of burstiness on $\sim 30$ Myr timescales at $z\sim 0$ and 6, respectively, as well as with the cosmic noon study of \citet{Ciesla2024_bursty}, which found long $\sim 100$ Myr timescales for variability through the star-forming main sequence. The timescales may be different in line-emitting samples~\citep{RobertsBorsani25_burstiness} and in the LBG-selected sample we focus on.

\subsection{Fitting $z\sim 4-6$}

Let us now extend the analysis to data at other redshifts.
We remain at $z\lesssim 6$ where $\Ha$ can be directly measured in NIRCam medium bands, and fit the UVLFs from \citet{Bouwens21}, biases from \citet{Harikane}, and H$\alpha$/UV ratios from \citet[][in Fig.~\ref{fig:HaUV_ratios_allz}]{simmonds24_masscompl} at $z\sim 4-6$, where again we assume they are uncorrelated so we multiply their likelihoods.

\begin{figure}
    \centering
    \includegraphics[width=0.98\linewidth]{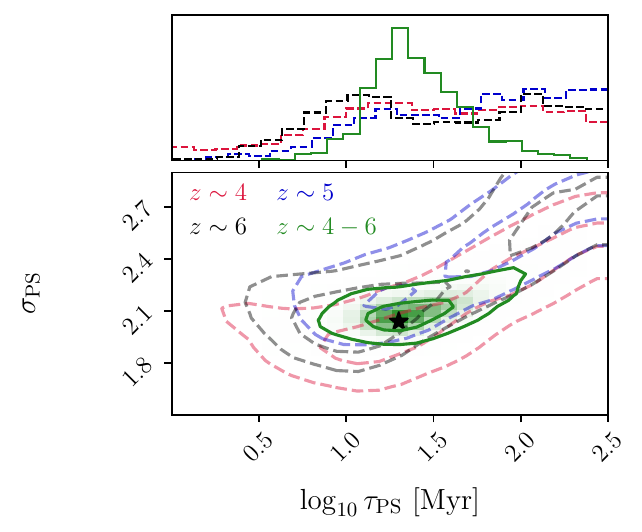}
    \caption{Posteriors for the burstiness parameters (amplitude $\sigmaPSD$ vs timescale $\tauPSD$).
    At each $z$ (dashed lines) the two parameters are degenerate, but combining the $z\sim 4-6$ observations (solid green) we see a clear preference for intermediate timescales $\tau\approx 20\,\rm$ Myr and moderate burstiness $\sigmaPSD\approx 2$. 
    Black star represents the long-burst model shown as an example through the text (the short-burst model lays above this plot, so it is ruled out).
    } 
    \label{fig:posteriossigmatau_z456}
\end{figure}

Before combining all redshifts $z\sim 4-6$, let us fit one at a time to ensure they are consistent.  
Fig.~\ref{fig:posteriossigmatau_z456} shows the posteriors for the PS parameters $\sigmaPSD$ and $\tauPSD$ at each $z$. These all overlap, showing consistent results between them (as well as with the $z\sim 6$ results from Fig.~\ref{fig:cornerssigmaPS_z6_tau}, though with slightly weaker burstiness $\sigmaPSD$). 
That is, all $z\sim 4-6$ data separately prefer similar burstiness parameters, with no obvious redshift evolution on either $\sigmaPSD$ or $\tauPSD$. 
We find $\sigmaPSD\approx 1.8-2.6$ at all $z$, with a clear degeneracy in  the $\sigmaPSD-\tauPSD$ plane, and a large amount of overlap in the ellipses despite the different evolution of dust, metals, stellar masses, and even cosmic age between $z=4$ and 6. 
From comparing the posteriors of the PS parameters at each redshift we find that $-0.1<d\sigmaPSD/dz<0.3$ and $|d\log_{10}\tauPSD/dz|<0.4$ within 68\% CL, both consistent with zero.  
We also find that all three redshift separately prefer growing burstiness towards the faint end, with very similar slopes (full posterior in Appendix~\ref{app:extraposteriors}).

\begin{figure*}
    \centering
    \includegraphics[width = 0.98\linewidth]
    {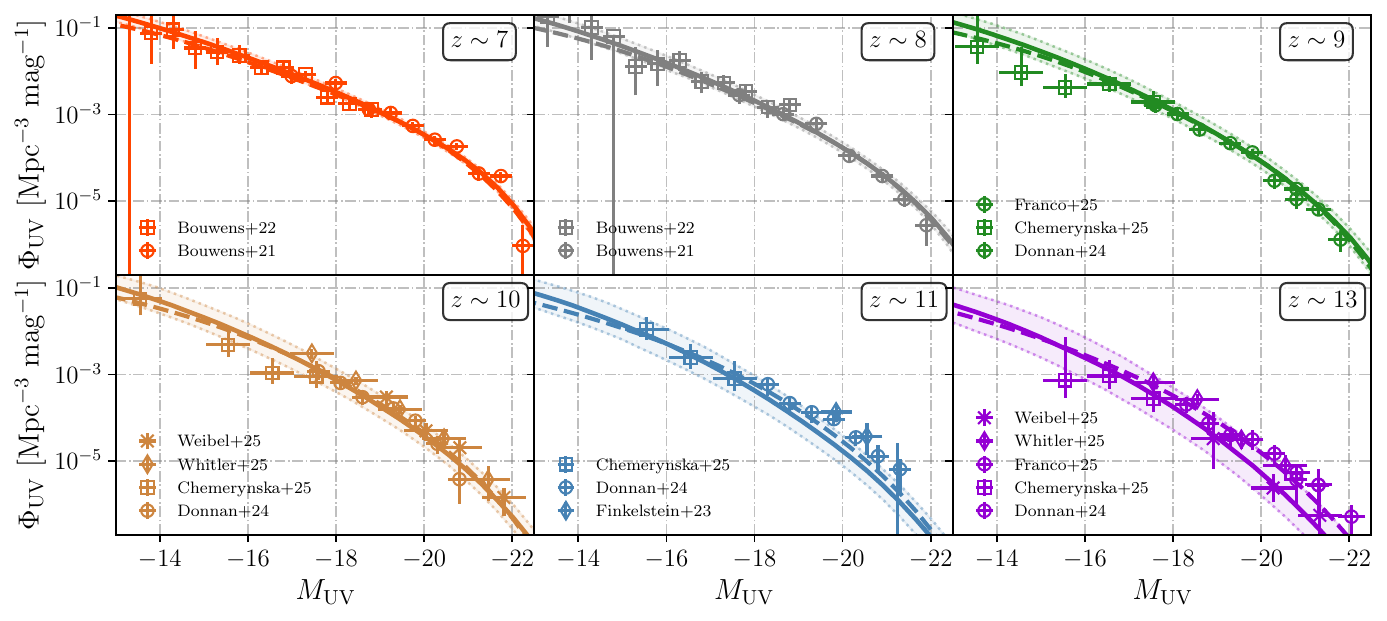}
    
    \caption{
    Predicted UVLFs for the model calibrated at $z\sim4-6$ when extrapolated to higher $z$ (solid lines). At each $z$ we show the median and 1$\sigma$ posteriors, which agree well with observations (empty symbols, which have not been used to calibrate this model). For reference, the dashed line shows the same model but calibrated to all $z\sim 4-13$, which provides a slightly better fit. 
    }
    \label{fig:UVLFallz}
\end{figure*}

As there is no apparent evolution on the PS parameters in Fig.~\ref{fig:posteriossigmatau_z456} we also perform an analysis that combines the information on the entire $z\sim 4-6$ data (augmenting our model by allowing the main SFE parameters $\log_{10}\epsilon_\star$ and $\log_{10} M_p/\Msun$ to vary smoothly with redshift through derivatives).
The additional power of co-adding redshifts breaks parameter degeneracies (especially between $\sigmaPSD$ and its derivative against halo mass) allowing us to measure the amplitude and timescale of burstiness simultaneously.
We find an amplitude of burstiness 
$\sigmaPSD = 2.11^{+0.11}_{-0.08}$ and a correlation timescale   
$\log_{10}\tauPSD /{\rm Myr}  = 1.40^{+0.32}_{-0.24}$. 
Variability in the SFH is built as a superposition of bursts (drawn as a Gaussian random field) with a characteristic coherence length $\tauPSD$. As such, our results indicate that bursts are coherent over $\tauPSD= 25^{+27}_{-11}$ Myr timescales.
We find little to no preference for changing timescales with halo mass ($d\log_{10} \tauPSD/d\log_{10} M_h= 0.04 \pm 0.26$ is consistent with zero, given the large uncertainties).

Our analysis of the $z\sim 4-6$ data also cements a strong preference for increased burstiness towards smaller halo masses, as we find $d \sigmaPSD/d\log_{10} M_h= -0.50 \pm 0.13$, more than 4$\sigma$ below zero. 
The lower panel of 
Fig.~\ref{fig:sigmaUV_allz} illustrates what these parameters imply in terms of the UV scatter $\sigmaUV$ evaluated at $z = 6$ and $z = 10$. 
The same PS parameters give rise to a higher UV scatter $\sigmaUV$ at early times ($z\sim 10$), since earlier halos grow faster. 
The $\sigmaUV$ posterior shows a marked increase in UV stochasticity towards small halo masses, with an overall value of $\sigmaUV\gtrsim 1$ at $M_h\lesssim 10^{10}\,\Msun$ (that decreases to $\sigmaUV\approx 0.5$ for the heaviest halos).
This confirms the single-$z$ results we found above, and shows our results are robust to the photometric procedure used to extract the $\Ha$/UV ratios.

\subsection{Extrapolated UVLFs}

Given that our model can predict the $z\sim4-6$ observations of UVLFs, clustering, and H$\alpha$/UV ratios (as well as $\Ha$LFs, see Appendix~\ref{app:HaLF}) without explicit time evolution in the burstiness parameters, we now venture to extrapolate its predictions to earlier times.

First, we show the predicted UVLFs at $z\sim 7-13$ in Fig.~\ref{fig:UVLFallz} and compare them against observed compilations from HST and JWST~\citep{Bouwens21, Bouwens22, Finkelstein_CEERS, Donnan24_UVLF, Whitler25_UVLF,Weibel,Chemerynska25_UVLF,Franco_25_COSMOSW}.
We focus on this $z$ range where spectroscopic follow-ups have been able to robustly confirm photometric redshifts, which are far more rare at $z > 13$~\citep{GSz14, Momz40}.
While the model has not been calibrated to any of these data, it is clear that it can reproduce them well, extending the results of past work in e.g.,~\citet{Tacchella, Sabti:2021xvh} in using lower-$z$ calibrations to reproduce higher-$z$ observations.
A critical piece of this successful extrapolation is the mass dependence of the burstiness strength. 
The $z\sim 4-6$ data prefers increased burstiness towards small halo masses (cf.~Fig.~\ref{fig:sigmaUVHavsMh}), which implies more variability at early times when halos tend to be smaller (as advocated by \citealt{Gelli24}). 
As a consequence, our model can explain the bulk of the UV-bright early JWST galaxies~\citep{Finkelstein_CEERS2}, subject of much debate.
While our extrapolation loses predictive power the farther away it is from $z\sim 4-6$, where it is calibrated, it performs well up to $z\sim 13$. It slightly underpredicts the bright end of the $z\gtrsim 11$ UVLFs, perhaps pointing to new ingredients active in that regime, such as feedback-free starbursts~\citep{Dekel23_FFB}, differences in dust~\citep{Ferrara2022,McKinney:2025htm}, more efficient UV emission (e.g., through a top-heavy IMF~\citealt{Inayoshi22_SF,Hutter:2024cvr,Yung24_brightUV}), or AGN contamination~\citep{Hegde:2024kph}.
Yet, our $z\sim4-6$ calibrated model predicts the correct amplitude and faint-end slopes of the observed UVLFs up to $z\sim 13$ with no new ingredients.

Having built confidence that the model can reproduce the $z\sim 4-13$ observations, we now calibrate it to the data over that entire redshift region (i.e., our previous $z\sim 4-6$ data plus the UVLFs in Fig.~\ref{fig:UVLFallz}), and show in the bottom panel of Fig.~\ref{fig:sigmaUV_allz} the predicted UV variability $\sigmaUV$ as a function of halo mass.
The posteriors from $z\sim 4-6$ and $z\sim 4-13$ data overlap, in both cases showing agreement with the predictions of hydrodynamical simulations and with the burstiness inferred using the \citet{Endsley24_bursty} and \citet{Chisholm} observations (though with slightly lower $\sigmaPSD$).
As a cross check, we compute a simple Gaussian tension metric $|\theta_1-\theta_2|/\sqrt{\sigma_1^2 + \sigma_2^2}$ between the $z\sim 4-6$ and $z\sim 4-13$ datasets for each parameter $\theta$ (with uncertainty $\sigma$). We find that all parameters shift by $\leq 1\sigma$ (as expected given the two datasets are correlated), and in particular the burstiness slope $d\sigmaPSD/d\log_{10}M_h$ steepens by 0.7$\sigma$ (from $-0.50\pm0.13$ to $-0.63\pm 0.14$). We conclude that the fits including the high-redshift UVLFs are statistically consistent with our $z\sim 4-6$ calibration.

\begin{figure}
    \centering
    \includegraphics[width = 0.98\linewidth]
    {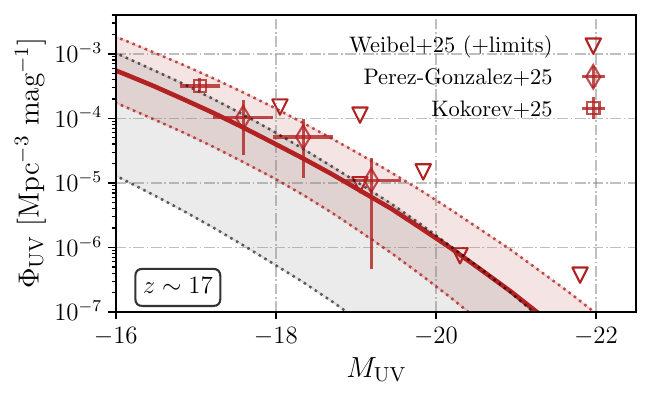}
    
    \caption{Predicted UVLFs at $z\sim 17$, compared against observations. 
    The gray region is the posterior of the predicted UVLF using only $z\sim 4-6$ data, whereas the red one uses all data up to $z\sim 13$. 
    Upper triangles show all limits, and  for visual clarity we have shifted the $\MUV\sim -19$ measurement from \citet{PG25} by 0.1 mag so it is visible against the \citet{Weibel} limit  at the same $\MUV$. 
    }
    \label{fig:UVLFz17}
\end{figure}

We can use the $z\sim 4-13$ calibrated model to predict even higher redshifts, where recent surveys have been reporting galaxy candidates. Fig.~\ref{fig:UVLFz17} shows the predicted UVLF at $z\sim 17$, which is in agreement with the candidates reported in \citet{Kokorev25_glimpse, PG25}, as well as the the upper limits from the those works and \citet{Weibel}.
We warn against over-interpretation of these ultra-high-$z$ predictions, as we have not included a Pop III component in our modeling~\citep{Venditti:2025mgi, Bromm, hegde, Cruz} or any galaxy formation below the atomic-cooling threshold~\citep{Oh}.
Nevertheless, the overall success of the low-$z$ calibrated model in predicting higher-$z$ UVLFs indicates that the physics of galaxy formation may not change dramatically at $z\sim 10$, but instead evolve smoothly as dark-matter halos and galaxies grow over cosmic age.

\subsection{Other high-$z$ observables}

\begin{figure}
    \centering
    \includegraphics[width = 0.98\linewidth]
    {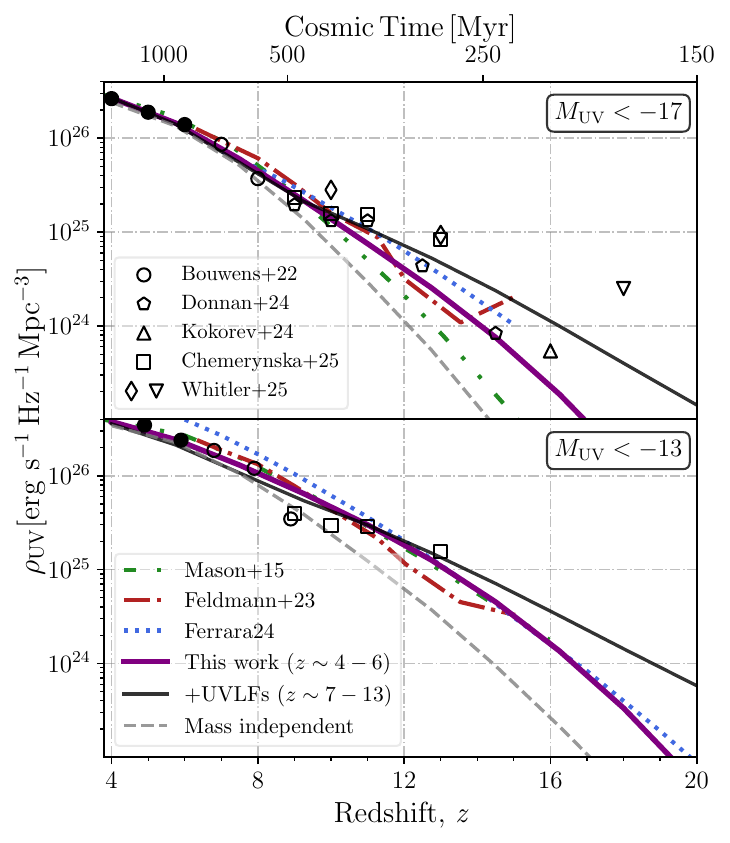}
    
    \caption{UV luminosity density as a function of redshift $z$ (and cosmic time, top $x$ axis), when integrated down to galaxies with $\MUV = -17$ ({\bf top}) and $\MUV = -13$ ({\bf bottom}). 
    Different symbols show observations (from HST and JWST, with upper/lower limits as triangles), and lines are theoretical models. 
    Pre-JWST models (e.g.,~\citealt{Mason15} in green) fit data well for $z\lesssim 10$, but fall at higher $z$. The post-JWST models of~\citet[][FIREbox, red dot-dashed]{Feldmann:2022qvd} and \citet[blue dotted]{Ferrara24} explain the JWST observations through different physical mechanisms.
    Our bursty model (purple line) is fit to data at $z\sim4-6$ (filled circles), and when extrapolated to higher $z$ it overlaps nicely with current observations (empty symbols),  as the mass-dependent burstiness flattens $\rho_{\rm UV}$ against $z$. 
    The agreement improves when also calibrating to UVLFs at $z\sim 7-13$ (black line), as expected. 
    For reference, the prediction of a model with mass-independent burstiness (gray dashed) falls far below observations at $z\gtrsim 10$. 
    }
    \label{fig:rhoUV}
\end{figure}

We can contextualize our results through the UV luminosity density $\rho_{\rm UV}$, defined as the integral of the observed (i.e., dust-attenuated) UVLF above two cutoff magnitudes $M_{\rm UV}=-17$ and $-13$. 
This quantity tracks the star-formation-rate density of the Universe~\citep{Madau:2014bja}. 
Fig.~\ref{fig:rhoUV} shows the prediction of our model, calibrated at $z\sim 4-6$, which agrees well with $\rho_{\rm UV}$ measurements as early as $\sim200$ Myr after the Big Bang ($z\sim 17$).  
For comparison, we also show a model with fixed (mass-independent) stochasticity, which has $\rho_{\rm UV}$ drop dramatically for $z\gtrsim 9$, falling far below observations. 
This highlights how the growing burstiness towards smaller $M_h$ that we infer at lower $z$ is key to matching high-$z$ observations. 
The model calibrated to JWST observations shows a much shallower decline of the UV density with redshift. 
For $z\lesssim 10$ our results are comparable to the pre-JWST predictions of~\citet{Mason15}, and for higher $z$ they agree well with those of \citet{Feldmann:2022qvd, Ferrara24}, as well as \citet{Robertson:2015uda}. 
Increased burstiness at early times therefore implies fairly active star formation, which improves the detection prospects of upcoming line-intensity mapping observations~\citep{Pritchard2012_review21cm,Bernal:2022jap}.
For instance, we can translate UV density into a star-formation rate density (SFRD) with a constant $\kappa_{\rm UV} = 1.1\times 10^{-28}$ $\Msun\, \rm yr^{-1}  s\,Hz \, \,erg^{-1}$~\citep{Madau:2014bja}, to find that our model predicts SFRD $\approx 7.3\times 10^{-4}\,\Msun \, \rm yr^{-1}\,Mpc^{-3}$ at $z\sim 15$, and  $\approx 1.5\times 10^{-4}\,\Msun \, \rm yr^{-1}\,Mpc^{-3}$ at $z\sim 20$ (in both cases down to $\MUV=-13$).

\begin{figure}
    \centering
    \includegraphics[width = 0.98\linewidth]
    {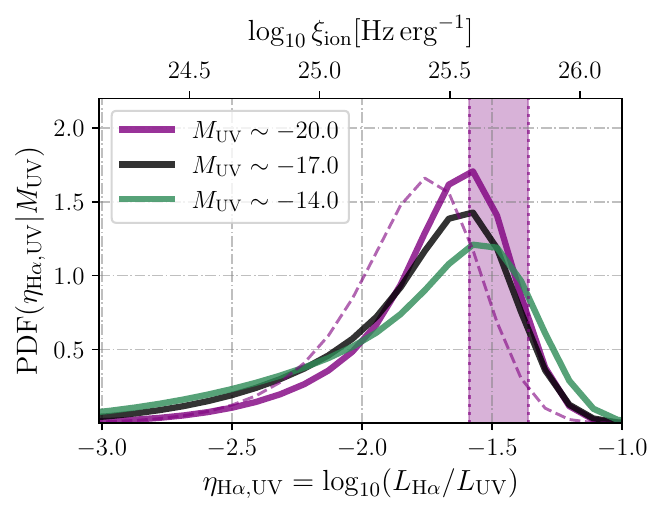}
    
    \caption{
    Predicted $\Ha$/UV ratios at $z\sim 10$ for our best-fit model, dust-corrected and calibrated at $z\sim 4-6$. For reference, we show the PDF at $z\sim 5$ as a thin dashed line (for the brightest bin, $\MUV\sim -20$). Increased scatter towards small halo masses translates into more burstiness at higher $z$, and thus broader and less Gaussian PDFs.
    We have assumed an observational/intrinsic scatter of 0.15 dex, independently of $\eta_{\Ha,\rm UV}$.
    The vertical band shows the measurement of one object at $z\sim 10$ from~\citet{Kokorev25_capers}.
    We show as the top $x$ axis the ionizing efficiency $\log_{10}\xi_{\rm ion}$, and note that we have not evolved the IMF or metallicity with redshift, so any shifts are entirely due to burstiness.
    }
    \label{fig:HaUVratios_highz}
\end{figure}

Beyond UV light, we can also predict distributions of $\Ha$/UV ratios at different redshifts and $\MUV$ to understand the burstiness of different galaxy populations. 
Fig.~\ref{fig:HaUVratios_highz} shows the PDF  $\mathcal P(\etaHaUV|\MUV)$ for $z\sim 10$ galaxies with $\MUV\sim -14, -17$, and $-20$, as predicted by our model calibrated at $z\sim 4-6$.
At this higher-$z$ the PDFs will be broader and less Gaussian, a tell-tale sign of burstiness.
In particular, all the $\MUV$ bins show a significant tail towards small $\Ha$/UV ratios, or off-mode galaxies, far broader than at $z\sim 5$ (also shown for comparison).
The physical reasons are clear: at earlier times the HMF is dominated by smaller halos, which makes all galaxies burstier. 
Extrapolating towards lower $z$, therefore, we expect more Gaussian PDFs, as heavier halos show less burstiness, though the faintest galaxies will still show skewed, bursty distributions of $\Ha$/UV~\citep{1809.06380}.
It is increasingly difficult to observe these PDFs at higher redshifts, as $\Ha$ emission falls out of JWST/NIRCam coverage. 
Yet, other star-formation tracers such as H$\beta$, H$\gamma$, or metal lines follow similar star-formation physics. As an example, we show the measurement of a H$\gamma$/UV ratio for a $\MUV\sim -20$ galaxy at $z\sim 10$ from \citet[][translated to $\Ha$ by  assuming a ratio $\rm H \gamma/H\alpha = 0.164$]{Kokorev25_capers}, which is near the peak of the predicted PDF.
The other galaxy in that work has an upper limit of $\etaHaUV\leq -1.7$, which constrains the object to be in the lower half of the PDF, also in agreement with our predictions.

\begin{figure}
    \centering
    \includegraphics[width = 0.98\linewidth]
    {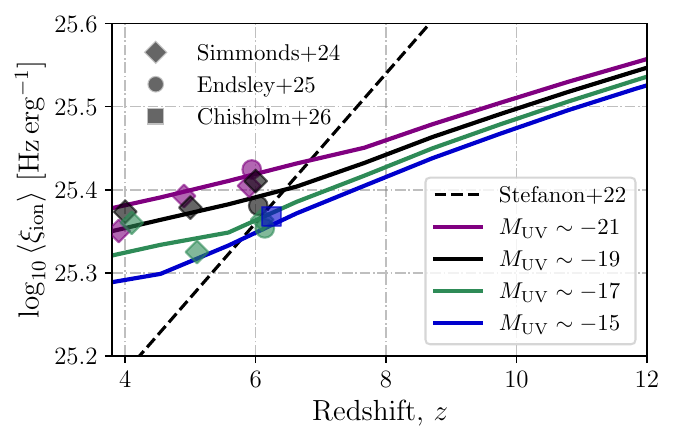}
    
    \caption{Log$_{10}$ of the mean ionizing efficiency for galaxies of different $\MUV$ predicted by our best-fit model as a function of $z$. This model is calibrated on $z\sim 4-6$ data (whose means are shown as filled symbols, slightly offset for clarity). Increased burstiness at lower masses translates into a higher mean ionizing efficiency at earlier times. For comparison, we show the pre-JWST estimate of this quantity from \citet{Stefanon22_xiion}. 
    }
    \label{fig:xiionvsz}
\end{figure}

Burstiness likewise affects the ionizing efficiencies $\xi_{\rm ion}$ of galaxies. 
There is ongoing debate on the average value of $\xi_{\rm ion}$ at high redshifts and faint magnitudes~\citep{Simmonds24_xiion,Pahl, Endsley23_reionization, Prieto,Begley25_OIIIHb}, as elevated values would imply faster reionization than expected~\citep{Munoz:2024fas}. 
Fig.~\ref{fig:HaUVratios_highz} shows $\xi_{\rm ion}$ at $z\sim 10$ for different $\MUV$, obtained by assuming case B recombination~\citep{Osterbrokferland2006}\footnote{We note that case B may not always be justified~\citep{scarlata2024}, and other factors that evolve with cosmic age such as the IMF or dust can affect $\xi_{\rm ion}$~\citep{Shivaie18}, shifting the mean of this distribution with $z$. 
We do not vary the IMF or metallicity against redshift in this work, so any reported change on $\Ha$/UV or $\xi_{\rm ion}$ is solely due to burstiness.}.
Under this assumption, $\xi_{\rm ion}$ follows $\etaHaUV$ so its PDF is broad and its average is shifted to higher $\xi_{\rm ion}$ values. Importantly, the width and non-Gaussian tail of the predicted $\xiion$ PDF separates the values of the median, mean, and log-mean of this quantity.
For instance, for $\MUV\approx -17$ galaxies at $z\sim 10$ we predict a log-mean $\VEV{\log_{10} \xi_{\rm ion}} \approx 25.3$ Hz erg$^{-1}$, but a mean $\log_{10} \VEV{\xi_{\rm ion}} \approx 25.5$ Hz erg$^{-1}$ (comparable to the median), which is 0.2 dex higher. 
More generally, we use our predicted PDFs to compute the mean $\VEV{\xiion}$ as a function of redshift and $\MUV$ and show it in 
Fig.~\ref{fig:xiionvsz}.
Our prediction is calibrated to $z\sim 4-6$ data from \citet[][]{simmonds24_masscompl, Endsley24_bursty, Chisholm}, whose means it reproduces well, and extrapolated to higher $z$.
As in past work, we see a moderate decrease of $\xiion$ towards fainter galaxies, which tend to have more ``off-mode'' objects~\citep{Pahl, Endsley23_reionization}. 
Interestingly, we see a rise of the ionizing efficiency towards higher $z$ (despite assuming no redshift evolution for the IMF or metallicity), due to the increased burstiness at early times, which broadens the $\Ha$/UV distributions and shifts the mean. We find a simple linear fit for the mean ionizing efficiency
\be
\log_{10} \VEV{\xiion} = 25.38 - 0.01 (\MUV + 17) + 0.02 (z-6)
\ee
in ${\rm Hz\,erg^{-1}}$, though we caution that galaxies appear to converge at high $z$ where they all reside in bursty small-mass halos, which this simple fit does not capture. Our predicted trend has the same sign but significantly shallower slope than found on \citet{Stefanon22_xiion}. In future work we will examine this result when varying metallicity and IMF with redshift as well.

\section{Discussion}
\label{sec:discussion}

Our results demonstrate that HST+JWST observations require mass-dependent burstiness at $z\sim 4-6$, with burst timescales of $\approx 20$ Myr. 
Below we discuss caveats and assumptions in our modeling, and explore the implications for understanding the high-redshift galaxy population and its feedback mechanisms.

\subsection{Caveats}

Our model is designed to be as simple as possible while still capturing the diversity of early-universe galaxies. 
As such, we have made a number of simplifications. 
For instance, we only include central galaxies residing in halos (i.e., we ignore satellites and a full halo-occupation distribution model), do not extract information from the one-halo term of the correlation function, and do not account for photoheating feedback from reionization~\citep{Shapiro:2003gxa,Borrow:2022djx}. 
Moreover, we have ascribed variability in $\Ha$/UV ratios to either burstiness in SFHs or changes in metallicity from one system to another, while not including variation arising from the finite sampling of the IMF.
This sampling can partially mimic burstiness by broadening the $\Ha$/UV PDFs, but using \textsc{slug} \citep{slug,krumholz_slug_2015} we have verified that stochastic IMF sampling leaves the distributions largely symmetric for a given IMF and affects much fainter objects ($\MUV\approx -13$) than we focus on here~\citep[see also][]{Fumagalli_slug_2011}. 
Regardless, we plan to explicitly include these modeling improvements in future work.

Our assumed lognormal approximation and PS functional form are relatively rigid, and could be enhanced.
Multi-scale feedback processes in the first galaxies may produce a different power-law index at high PS frequencies~\citep{2410.21409}, or give rise to several timescales $\tauPSD$~\citep{2006.09382}. 
Additionally, for very large burstiness amplitudes ($\sigmaPSD\gtrsim 4$) a finite number of galaxies does not adequately sample the tail of the log-normal distribution, so care must be exercised when interpreting observations (with $\sim 10^3$ objects). 
Yet, the PS model is flexible enough to represent a variety of physical scenarios.
For instance, it can mimic a duty cycle where galaxies only form stars some fraction $f_{\rm duty}\leq 1$ of the time. We can compute the equivalent $f_{\rm duty}$ of our PS model by asking when the bursty SFR is above a threshold $\dot M_\star = \overline{\dot M_\star}/f_{\rm duty}$, assuming it is approximately zero elsewhere. Doing this calculation we find that our $\sigmaPSD\approx 2$ model translates into a duty cycle $f_{\rm duty}\approx 10-30\%$ (and a larger burstiness amplitude $\sigmaPSD = 3$ would correspond to a very small $f_{\rm duty}\approx 1-3\%$).

All of our inferences on star-formation  burstiness apply to high-$z$ UV-selected galaxies and should not be extrapolated to other galaxy populations, for instance selected in mm/submm, stellar mass, or as line emitters. 
While it is possible that the same model encompasses those (see, e.g., Appendix~\ref{app:HaLF} for $\Ha$-selected galaxies), or that it can be  extended to accommodate them, modeling the disparate selection functions of each population is beyond the scope of this work. 
In the same vein, we have focused on the power of joint UV and $\Ha$ observations to constrain burstiness, but other star-formation tracers can provide further insights at high $z$, including Balmer breaks, UV slopes, stellar masses, and other line tracers like H$\beta$+OIII.
As an example, Appendix~\ref{app:BalmerBreaks} highlights how our best-fit model reproduces the distribution of Balmer breaks from the JADES galaxy sample of \citet{Endsley24_bursty}, showing consistency across star-formation tracers.  
Spectroscopic determinations will significantly strengthen the quality of the datasets, and allow us to constrain burstiness without an SED prior. 
Moreover, spatially resolved SED fitting may provide additional information through the morphology of star-forming clumps.
We leave a detailed modeling of these for future work.

Finally, while our approach does not independently fit the SED of each object, it still relies on stellar population synthesis models to convert mass to light (e.g., in Fig.~\ref{fig:windows}).
Through the text we have assumed the SPS model from \citetalias{BC03},  but enhanced it with two free parameters ($\AmpHa$ and $d\AmpHa/d\log_{10}M_h$) that rescale the $\Ha$ luminosity with respect to UV (as they are both rescaled by the free SFE parameter $\epsilon_\star$).
To quantify how our results depend on the specific choice of SPS model, we have repeated the entire analysis with \citetalias{BPASS}, both with and without binary stars.
We find highly consistent results, with overlap in the burstiness amplitude $\sigmaPSD \approx 2$, mass behavior $d\sigmaPSD/d\log_{10}M_h \approx -0.5$, and timescale $\tauPSD\approx 20$ Myr (as differences in mean $\Ha$/UV luminosity are absorbed into the nuisance parameter $\AmpHa$). 
A full posterior comparison can be found in Appendix~\ref{app:extraposteriors}.

\begin{figure}
    \centering
    \includegraphics[width = 0.98\linewidth]
    {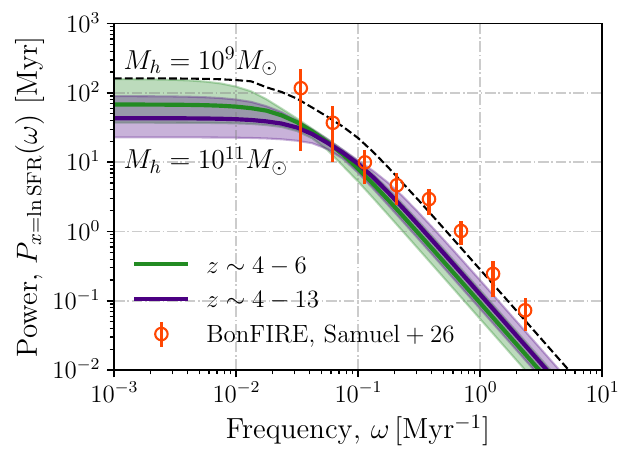}    
    \caption{
    Posterior for the power spectrum of $x=\ln$(SFR) for galaxies residing in halos of $M_h=10^{11}\,\Msun$, obtained from our empirical measurement of $\sigmaPSD$ and $\tauPSD$. The green region shows the median and uncertainties when calibrated for $z\sim 4-6$ data, and purple for $z\sim 4-13$. 
    We show, for reference, the power spectrum measured in the BonFIRE hydrodynamical simulations of \citet{Jenna25}, which is in good agreement with our empirical determinations.
    The dashed line represents our prediction for halos of $M_h\sim 10^9\,\Msun$, which are burstier. 
    }
    \label{fig:PS_vsomega_posterior}
\end{figure}

\subsection{Insights into galaxy formation}

While our model does not allow the burstiness parameters to evolve with redshift (only through the halo mass and accretion rates), it can still explain observations at earlier times than it was calibrated on. 
Our interpretation is that the physical mechanisms that are at play at $z\sim 4-6$ are sufficient to generate a large amount of burstiness at $z\gtrsim 10$, explaining the abundance of UV-bright galaxies during cosmic dawn (shown through the UVLFs in Fig.~\ref{fig:UVLFallz} and $\rho_{\rm UV}$ in Fig.~\ref{fig:rhoUV}). 
That is due to the mass dependence of burstiness. 
We find that galaxies hosted in halos with $M_h\sim 10^{11}\,\Msun$ (typical at $z\sim 6$) vary in their SFR by $\approx 0.6$ dex, whereas those hosted in smaller halos $M_h\sim 10^{9}\,\Msun$ (typical at $z\gtrsim 9$) vary by a larger $\approx 1$ dex. 
These results hinge on the deep $z\sim 4-6$ observations, which give us access to galaxies that are sufficiently faint to be hosted in low-mass halos, thus recreating the conditions of an earlier universe. 
This mass-dependent stochasticity boosts the clustering of galaxies at the bright end, where it is most readily measurable~\citep{Munoz:2023cup}.
We note that our results do not exclude other physical processes shaping high-$z$ galaxy formation, including the possibility that halo mass acts as a proxy for other quantities such as virial velocity or acceleration~\citep{Boylan-Kolchin:2024roq}.

The power-spectrum formalism that we employ here allows for a direct statistical comparison of burstiness between observations and simulations. 
To perform such comparison, we combine our posteriors for the burstiness parameters $\sigmaPSD$ and $\tauPSD$ to obtain a measurement of the (log) SFR power spectrum, which we show in Fig.~\ref{fig:PS_vsomega_posterior} (for galaxies hosted in halos of $M_h=10^{11}\,\Msun$). 
We find a power spectrum amplitude $\approx 10^2\,\rm Myr$ at the peak, which grows for halos with $M_h=10^9\,\Msun$, indicating increased burstiness. 
This posterior resembles our long-bursts model shown through the main text, and is in broad agreement with simulation results in \citet{2405.04578}.
In order to extend this comparison, we compute the power spectrum of the star-formation histories of galaxies in the BonFIRE suite of simulations of \citet{Jenna25}, and show it in Fig.~\ref{fig:PS_vsomega_posterior} for $N=14$ galaxies residing in halos of $M_h=10^{10.5} \,\Msun$ at $z=9$ (as close to $M_h=10^{11}\,\Msun$ as found in the box).  
The overall amplitude and slope of the power agree, showing that the amount of burstiness in FIRE-3 can explain the high-$z$ UVLFs, as advocated in \citet{Sun23_FIRE_bursty}.
The FIRE SFHs we use have support over cosmic ages $200-550$ Myr, limiting the frequency range accessible, but we find similar results with the smaller, higher-resolution CampFIRE box that is run to $z\sim 6$ (see Appendix~\ref{App:hydro}).

The power spectrum we measure in Fig.~\ref{fig:PS_vsomega_posterior} shows a cutoff at frequencies $\omega\gtrsim 0.05\,\rm Myr^{-1}$. 
Physically, this cutoff encodes a suppression of short-timescale fluctuations (high $\omega$) in the SFHs of galaxies. 
This hints at a burstiness in early galaxies dominated by processes with a coherence length of $\tauPSD\approx 20$ Myr, comparable to the timescales of supernova feedback~\citep{Faucher-Giguere:2017lgp} and giant molecular clouds~\citep{2203.09570,2403.19843}, rather than long-term processes such as mergers~\citep{0707.2960}, or short-term massive stellar winds~\citep{1110.4638}. 
Yet, we note that the PS model allows for bursts longer than $\tauPSD$, so some objects will have coherent up/downturns over $\sim 100$ Myr timescales, perhaps resembling quenched galaxies at lower redshifts~\citep{1809.00722}.
We only find a mild preference for $\tauPSD$ to grow with halo mass, but we note that lower-$z$ studies argue for longer timescales in heavier galaxies~\citep{1809.06380}.
In our single-$z$ fits we did not find $\tauPSD$ to evolve for $z\sim 4-6$ (see Fig.~\ref{fig:posteriossigmatau_z456}), whereas galaxy dynamical times naively scale as $H^{-1}(z)$, and should therefore change by $\sim 60\%$.

As for the mass behavior of the burstiness amplitude, our model does not allow us to directly compare with the supernova-feedback prediction from \citet{Faucher-Giguere:2017lgp} of increased burstiness as Poisson noise on stellar mass ($1/\sqrt{M_\star}$). Yet, our fit does prefer increased variability towards smaller $M_h$ (as $\sigma(\ln \dot M_\star)\propto -0.5 \, \log_{10} M_h$).
This aligns with expectations from burstiness dominated by stellar feedback, given that lighter halos have shallower potentials and smaller virial velocities, so their gas reservoirs are easier to disrupt. 
Our inferred slope for $\sigmaUV$ vs $M_h$ is comparable to that assumed in \citet{Gelli24}, though we emphasize that here it is a free parameter found through UV+$\Ha$ observations. 
Our $z\sim 4-6$ calibrated model fits $z\lesssim 11$ UVLFs similarly to that work (overlapping the observations within 1$\sigma$ theoretical uncertainties), and by adding higher-$z$ UVLFs to the likelihood the model parameters---both burstiness and SFE---slightly shift to better fit those observations (up to $z\sim 17$).

We can translate our results from the abstract space of PS to burstiness in observables beyond $\Ha$ and UV. For instance, we can use our formalism to predict the scatter in the star-forming main sequence (SFMS), which was the original motivation for the PS formalism~\citep{1901.07556}. Our analytic model cannot yet produce full PDFs for $M_\star(M_h)$, so we turn to a simulation for this exercise, which we describe in detail in Appendix~\ref{app:SFMS}. 
Burstiness gives rise to enhanced scatter on the SFMS, and observational studies do not yet agree on whether it grows or decreases with $M_\star$~\citep{Rinaldi24,Popesso22, 2212.02890,Clarke24,1706.07059,2508.04410}. 
Our model predicts enhanced scatter towards small $M_\star$, which follows from the model preference for increased variability towards lighter halos. In particular, using $\Ha$ as a SFR tracer we find $\sigma_{\rm log_{10} SFR} \approx 0.4$ dex at $M_\star =10^{11} \Msun$, comparable to the scatter in~\citet{Cole24_burstSFMS}, growing linearly to reach $\sigma_{\rm log_{10} SFR} \approx 0.7$ dex at $M_\star =10^{7} \Msun$.
The scatter is predicted to be smaller when using UV as a SFR tracer, due to its longer timescale. 
Over time halos will grow and thus become less bursty, so we expect less scatter in the SFMS towards $z\sim 0$~\citep{astro-ph/0311060,1405.2041}.

\section{Conclusions}
\label{sec:conclusion}

In this work we have developed a self-consistent, efficient framework to model burstiness in galaxy formation during the first billion years, publicly available through the {\tt Zeus21} 
software package.
Our model assumes dark-matter halos host galaxies that form stars stochastically around a parametrized mean, with log-normal fluctuations---bursts---drawn from a power spectrum with an amplitude $\sigmaPSD$ and timescale $\tauPSD$.
While simple, this model is able to reproduce the burstiness observed in hydrodynamical simulations and can predict observables in $<1$ s, enabling MCMC searches over parameter space. 
By jointly fitting the ``average'' and ``bursty'' parameters we are able to reproduce current UV, clustering, and $\Ha$/UV data at $z\sim 4-6$, and when extrapolating to earlier times we find agreement with current observations up to $z\sim 17$. 
Our main results are:

$\bullet$ A measurement of the burstiness of early-universe galaxies.
Combining observations at $z\sim 4-6$ we break the degeneracy between star-formation efficiency and burstiness, inferring an rms $\sigma (\log_{10} \dot M_\star) \approx 0.6$ dex or $\sigmaUV \approx 0.75$ mag for galaxies hosted in halos with $M_h=10^{11}\, \Msun$ (typical of the observed UVLF at these redshifts).

$\bullet$ The first robust determination of increased burstiness for early galaxies hosted in smaller dark-matter halos. 
We interpret $\Ha$/UV ratios vs $\MUV$ to find that galaxies get burstier by $0.15$ dex per decade in (smaller) $M_h$. For instance, we infer $\sigma (\log_{10} \dot M_\star) \approx 1$ dex for galaxies hosted in halos with $M_h=10^{9}\, \Msun$. 
We extrapolate this trend to earlier times, when halo masses are smaller, and find that it can explain the bulk of the observed UVLFs up to $z\sim 17$ with no new ingredients (though predictions fall 1$\sigma$ below measurements at the bright end).

$\bullet$ A tentative measurement of the timescale of burst cycles. The $z\sim 4-6$ data combined show a preference for bursts coherent on $\tauPSD\approx 20$ Myr timescales, 
constraining the physics that give rise to star-formation burstiness in the first billion years. 
Importantly, fully incoherent bursts, as expected of short $\tauPSD\sim$ Myr feedback processes, are disfavored.

Altogether, we constrain the first galaxies to be reasonably bursty on relatively fast timescales and find strong evidence for increased star-formation variability towards smaller masses, in agreement with models of supernova feedback. 
Building on these techniques, upcoming observations will be able to more precisely measure the timescales involved and their dependence on halo mass, putting tight bounds on the feedback mechanisms that shaped galaxy formation from the Big Bang to today.

\section*{Acknowledgements}
We thank Vasily Kokorev, Iryna Chemerynska, Alba Covelo-Paz, and Andrea Ferrara for sharing and documenting their data; as well as Jared Barron, Danielle Berg, Mike Boylan-Kolchin, Steve Finkelstein, Changhoon Hahn, Charlotte Mason, Jed McKinney, Jacob Shen, Dan Stark, and Yonatan Sklansky for useful discussions that improved the quality of this manuscript. 
This work has been possible thanks to public software: {\tt numpy} \citep{numpy}, {\tt scipy} \citep{scipy}, {\tt Zeus21} \citep{Munoz:2023kkg}, {\tt powerbox}~\citep{Murray_powerbox}, and {\tt emcee}~\citep{emcee}. 
JBM acknowledges support from NSF Grants AST-2307354 and AST-2408637, and the CosmicAI institute AST-2421782.
This research was supported in part by grant NSF PHY-2309135 to the Kavli Institute for Theoretical Physics (KITP). GS was supported by a CIERA Postdoctoral Fellowship, with additional support provided by NSF through grants AST-2108230 and AST-2307327; by NASA through grants 21-ATP21-0036 and 23-ATP23-0008; and by STScI through grant JWST-AR-03252.001-A. JS acknowledges support from program JWST-AR-06278 by NASA through a grant from the Space Telescope Science Institute, which is operated by the Association of Universities for Research in Astronomy, Inc., under NASA contract NAS 5-03127.

\section*{Data Availability}

The code used to compute the bursty galaxy model predictions will be made available on GitHub upon publication, and all data will be shared upon reasonable request. The observational data used in this work are publicly available from the referenced publications.

\bibliography{biblio}{}

\begin{thebibliography}{}
\makeatletter
\relax
\def\mn@urlcharsother{\let\do\@makeother \do\$\do\&\do\#\do\^\do\_\do\%\do\~}
\def\mn@doi{\begingroup\mn@urlcharsother \@ifnextchar [ {\mn@doi@}
  {\mn@doi@[]}}
\def\mn@doi@[#1]#2{\def\@tempa{#1}\ifx\@tempa\@empty \href
  {http://dx.doi.org/#2} {doi:#2}\else \href {http://dx.doi.org/#2} {#1}\fi
  \endgroup}
\def\mn@eprint#1#2{\mn@eprint@#1:#2::\@nil}
\def\mn@eprint@arXiv#1{\href {http://arxiv.org/abs/#1} {{\tt arXiv:#1}}}
\def\mn@eprint@dblp#1{\href {http://dblp.uni-trier.de/rec/bibtex/#1.xml}
  {dblp:#1}}
\def\mn@eprint@#1:#2:#3:#4\@nil{\def\@tempa {#1}\def\@tempb {#2}\def\@tempc
  {#3}\ifx \@tempc \@empty \let \@tempc \@tempb \let \@tempb \@tempa \fi \ifx
  \@tempb \@empty \def\@tempb {arXiv}\fi \@ifundefined
  {mn@eprint@\@tempb}{\@tempb:\@tempc}{\expandafter \expandafter \csname
  mn@eprint@\@tempb\endcsname \expandafter{\@tempc}}}

\bibitem[\protect\citeauthoryear{{Adams} et~al.,}{{Adams}
  et~al.}{2023}]{Adams_Conselice_JWST_2023}
{Adams} N.~J.,  et~al., 2023, \mn@doi [\mnras] {10.1093/mnras/stac3347}, \href
  {https://ui.adsabs.harvard.edu/abs/2023MNRAS.518.4755A} {518, 4755}

\bibitem[\protect\citeauthoryear{Aghanim et~al.}{Aghanim
  et~al.}{2020}]{Planck:2018vyg}
Aghanim N.,  et~al., 2020, \mn@doi [Astron. Astrophys.]
  {10.1051/0004-6361/201833910}, 641, A6

\bibitem[\protect\citeauthoryear{{Asada} et~al.,}{{Asada} et~al.}{2024}]{asada}
{Asada} Y.,  et~al., 2024, \mn@doi [\mnras] {10.1093/mnras/stad3902}, \href
  {https://ui.adsabs.harvard.edu/abs/2024MNRAS.52711372A} {527, 11372}

\bibitem[\protect\citeauthoryear{{Atek} et~al.,}{{Atek}
  et~al.}{2025}]{AtekChisholm25_glimpse}
{Atek} H.,  et~al., 2025, \mn@doi [arXiv e-prints] {10.48550/arXiv.2511.07542},
  \href {https://ui.adsabs.harvard.edu/abs/2025arXiv251107542A} {p.
  arXiv:2511.07542}

\bibitem[\protect\citeauthoryear{{Austin} et~al.,}{{Austin}
  et~al.}{2025}]{Austin25_betaslopes}
{Austin} D.,  et~al., 2025, \mn@doi [\apj] {10.3847/1538-4357/ae07db}, \href
  {https://ui.adsabs.harvard.edu/abs/2025ApJ...995...43A} {995, 43}

\bibitem[\protect\citeauthoryear{{Basu}, {Bhagwat}, {Ciardi}  \&
  {Costa}}{{Basu} et~al.}{2026}]{Spice26_UVLFscatter}
{Basu} A.,  {Bhagwat} A.,  {Ciardi} B.,   {Costa} T.,  2026, \mn@doi [\mnras]
  {10.1093/mnras/staf2240}, \href
  {https://ui.adsabs.harvard.edu/abs/2026MNRAS.545S2240B} {545, staf2240}

\bibitem[\protect\citeauthoryear{{Begley} et~al.,}{{Begley}
  et~al.}{2025}]{Begley25_OIIIHb}
{Begley} R.,  et~al., 2025, \mn@doi [\mnras] {10.1093/mnras/staf211}, \href
  {https://ui.adsabs.harvard.edu/abs/2025MNRAS.537.3245B} {537, 3245}

\bibitem[\protect\citeauthoryear{{Behroozi}, {Wechsler}, {Hearin}  \&
  {Conroy}}{{Behroozi} et~al.}{2019}]{Behroozi2019_UM}
{Behroozi} P.,  {Wechsler} R.~H.,  {Hearin} A.~P.,   {Conroy} C.,  2019,
  \mn@doi [\mnras] {10.1093/mnras/stz1182}, \href
  {https://ui.adsabs.harvard.edu/abs/2019MNRAS.488.3143B} {488, 3143}

\bibitem[\protect\citeauthoryear{{Benson}}{{Benson}}{2012}]{Benson2012_galacticus}
{Benson} A.~J.,  2012, \mn@doi [\na] {10.1016/j.newast.2011.07.004}, \href
  {https://ui.adsabs.harvard.edu/abs/2012NewA...17..175B} {17, 175}

\bibitem[\protect\citeauthoryear{{Berg} et~al.,}{{Berg} et~al.}{2025}]{Berg25}
{Berg} D.~A.,  et~al., 2025, \mn@doi [arXiv e-prints]
  {10.48550/arXiv.2511.13591}, \href
  {https://ui.adsabs.harvard.edu/abs/2025arXiv251113591B} {p. arXiv:2511.13591}

\bibitem[\protect\citeauthoryear{Berlind \& Weinberg}{Berlind \&
  Weinberg}{2002}]{Berlind:2001xk}
Berlind A.~A.,  Weinberg D.~H.,  2002, \mn@doi [Astrophys. J.]
  {10.1086/341469}, 575, 587

\bibitem[\protect\citeauthoryear{Bernal \& Kovetz}{Bernal \&
  Kovetz}{2022}]{Bernal:2022jap}
Bernal J.~L.,  Kovetz E.~D.,  2022, \mn@doi [Astron. Astrophys. Rev.]
  {10.1007/s00159-022-00143-0}, 30, 5

\bibitem[\protect\citeauthoryear{Borrow, Kannan, Garaldi, Smith, Vogelsberger,
  Pakmor, Springel  \& Hernquist}{Borrow et~al.}{2023}]{Borrow:2022djx}
Borrow J.,  Kannan R.,  Garaldi E.,  Smith A.,  Vogelsberger M.,  Pakmor R.,
  Springel V.,   Hernquist L.,  2023, \mn@doi [Mon. Not. Roy. Astron. Soc.]
  {10.1093/mnras/stad2523}, 525, 5932

\bibitem[\protect\citeauthoryear{{Bouch{\'e}} et~al.,}{{Bouch{\'e}}
  et~al.}{2010}]{bathtub}
{Bouch{\'e}} N.,  et~al., 2010, \mn@doi [\apj] {10.1088/0004-637X/718/2/1001},
  \href {https://ui.adsabs.harvard.edu/abs/2010ApJ...718.1001B} {718, 1001}

\bibitem[\protect\citeauthoryear{{Bouwens} et~al.,}{{Bouwens}
  et~al.}{2014}]{Bouwens2014}
{Bouwens} R.~J.,  et~al., 2014, \mn@doi [\apj] {10.1088/0004-637X/793/2/115},
  \href {https://ui.adsabs.harvard.edu/abs/2014ApJ...793..115B} {793, 115}

\bibitem[\protect\citeauthoryear{{Bouwens} et~al.,}{{Bouwens}
  et~al.}{2016}]{Bouwens16xiion}
{Bouwens} R.~J.,  et~al., 2016, \mn@doi [\apj] {10.3847/1538-4357/833/1/72},
  \href {https://ui.adsabs.harvard.edu/abs/2016ApJ...833...72B} {833, 72}

\bibitem[\protect\citeauthoryear{{Bouwens} et~al.,}{{Bouwens}
  et~al.}{2021}]{Bouwens21}
{Bouwens} R.~J.,  et~al., 2021, \mn@doi [\aj] {10.3847/1538-3881/abf83e}, \href
  {https://ui.adsabs.harvard.edu/abs/2021AJ....162...47B} {162, 47}

\bibitem[\protect\citeauthoryear{{Bouwens}, {Illingworth}, {Ellis}, {Oesch}  \&
  {Stefanon}}{{Bouwens} et~al.}{2022}]{Bouwens22}
{Bouwens} R.~J.,  {Illingworth} G.,  {Ellis} R.~S.,  {Oesch} P.,   {Stefanon}
  M.,  2022, \mn@doi [\apj] {10.3847/1538-4357/ac86d1}, \href
  {https://ui.adsabs.harvard.edu/abs/2022ApJ...940...55B} {940, 55}

\bibitem[\protect\citeauthoryear{Boylan-Kolchin}{Boylan-Kolchin}{2025}]{Boylan-Kolchin:2024roq}
Boylan-Kolchin M.,  2025, \mn@doi [Mon. Not. Roy. Astron. Soc.]
  {10.1093/mnras/staf471}, 538, 3210

\bibitem[\protect\citeauthoryear{{Boylan-Kolchin}, {Ma}  \&
  {Quataert}}{{Boylan-Kolchin} et~al.}{2008}]{0707.2960}
{Boylan-Kolchin} M.,  {Ma} C.-P.,   {Quataert} E.,  2008, \mn@doi [\mnras]
  {10.1111/j.1365-2966.2007.12530.x}, \href
  {https://ui.adsabs.harvard.edu/abs/2008MNRAS.383...93B} {383, 93}

\bibitem[\protect\citeauthoryear{{Brinchmann}, {Charlot}, {White}, {Tremonti},
  {Kauffmann}, {Heckman}  \& {Brinkmann}}{{Brinchmann}
  et~al.}{2004}]{astro-ph/0311060}
{Brinchmann} J.,  {Charlot} S.,  {White} S.~D.~M.,  {Tremonti} C.,  {Kauffmann}
  G.,  {Heckman} T.,   {Brinkmann} J.,  2004, \mn@doi [\mnras]
  {10.1111/j.1365-2966.2004.07881.x}, \href
  {https://ui.adsabs.harvard.edu/abs/2004MNRAS.351.1151B} {351, 1151}

\bibitem[\protect\citeauthoryear{{Bromm} \& {Yoshida}}{{Bromm} \&
  {Yoshida}}{2011}]{Bromm}
{Bromm} V.,  {Yoshida} N.,  2011, \mn@doi [\araa]
  {10.1146/annurev-astro-081710-102608}, \href
  {https://ui.adsabs.harvard.edu/abs/2011ARA&A..49..373B} {49, 373}

\bibitem[\protect\citeauthoryear{{Bruzual} \& {Charlot}}{{Bruzual} \&
  {Charlot}}{2003}]{BC03}
{Bruzual} G.,  {Charlot} S.,  2003, \mn@doi [\mnras]
  {10.1046/j.1365-8711.2003.06897.x}, \href
  {https://ui.adsabs.harvard.edu/abs/2003MNRAS.344.1000B} {344, 1000}

\bibitem[\protect\citeauthoryear{{Caplar} \& {Tacchella}}{{Caplar} \&
  {Tacchella}}{2019}]{1901.07556}
{Caplar} N.,  {Tacchella} S.,  2019, \mn@doi [\mnras] {10.1093/mnras/stz1449},
  \href {https://ui.adsabs.harvard.edu/abs/2019MNRAS.487.3845C} {487, 3845}

\bibitem[\protect\citeauthoryear{{Carnall}, {McLure}, {Dunlop}  \&
  {Dav{\'e}}}{{Carnall} et~al.}{2018}]{bagpipes}
{Carnall} A.~C.,  {McLure} R.~J.,  {Dunlop} J.~S.,   {Dav{\'e}} R.,  2018,
  \mn@doi [\mnras] {10.1093/mnras/sty2169}, \href
  {https://ui.adsabs.harvard.edu/abs/2018MNRAS.480.4379C} {480, 4379}

\bibitem[\protect\citeauthoryear{{Carnall} et~al.,}{{Carnall}
  et~al.}{2023}]{Carnall2023_SMACS}
{Carnall} A.~C.,  et~al., 2023, \mn@doi [\mnras] {10.1093/mnrasl/slac136},
  \href {https://ui.adsabs.harvard.edu/abs/2023MNRAS.518L..45C} {518, L45}

\bibitem[\protect\citeauthoryear{{Carniani} et~al.,}{{Carniani}
  et~al.}{2018}]{Carniani2018_dust_distr}
{Carniani} S.,  et~al., 2018, \mn@doi [\mnras] {10.1093/mnras/sty1088}, \href
  {https://ui.adsabs.harvard.edu/abs/2018MNRAS.478.1170C} {478, 1170}

\bibitem[\protect\citeauthoryear{Casey et~al.}{Casey
  et~al.}{2014}]{Casey:2014cqa}
Casey C.~M.,  et~al., 2014, \mn@doi [Astrophys. J.]
  {10.1088/0004-637X/796/2/95}, 796, 95

\bibitem[\protect\citeauthoryear{Casey et~al.}{Casey
  et~al.}{2022}]{Casey:2022amu}
Casey C.~M.,  et~al., 2022

\bibitem[\protect\citeauthoryear{{Castellano}, {Fontana}, {Treu}, {Santini},
  {Merlin}  et~al.}{{Castellano} et~al.}{2022}]{Castellano_GLASS_hiz}
{Castellano} M.,  {Fontana} A.,  {Treu} T.,  {Santini} P.,  {Merlin} E.,
  et~al., 2022, \mn@doi [Astrophys. J. Lett.] {10.3847/2041-8213/ac94d0}, \href
  {https://ui.adsabs.harvard.edu/abs/2022ApJ...938L..15C} {938, L15}

\bibitem[\protect\citeauthoryear{{Chemerynska} et~al.,}{{Chemerynska}
  et~al.}{2025}]{Chemerynska25_UVLF}
{Chemerynska} I.,  et~al., 2025, \mn@doi [arXiv e-prints]
  {10.48550/arXiv.2509.24881}, \href
  {https://ui.adsabs.harvard.edu/abs/2025arXiv250924881C} {p. arXiv:2509.24881}

\bibitem[\protect\citeauthoryear{{Chevallard} \& {Charlot}}{{Chevallard} \&
  {Charlot}}{2016}]{Chevallard16_BEAGLE}
{Chevallard} J.,  {Charlot} S.,  2016, \mn@doi [\mnras]
  {10.1093/mnras/stw1756}, \href
  {https://ui.adsabs.harvard.edu/abs/2016MNRAS.462.1415C} {462, 1415}

\bibitem[\protect\citeauthoryear{{Chevance}, {Krumholz}, {McLeod}, {Ostriker},
  {Rosolowsky}  \& {Sternberg}}{{Chevance} et~al.}{2023}]{2203.09570}
{Chevance} M.,  {Krumholz} M.~R.,  {McLeod} A.~F.,  {Ostriker} E.~C.,
  {Rosolowsky} E.~W.,   {Sternberg} A.,  2023, ] {10.48550/arXiv.2203.09570},
  \href {https://ui.adsabs.harvard.edu/abs/2023ASPC..534....1C} {534, 1}

\bibitem[\protect\citeauthoryear{{Chisholm et al.}}{{Chisholm et
  al.}}{2026}]{Chisholm}
{Chisholm et al.} J.,  2026

\bibitem[\protect\citeauthoryear{{Ciesla} et~al.,}{{Ciesla}
  et~al.}{2024}]{Ciesla2024_bursty}
{Ciesla} L.,  et~al., 2024, \mn@doi [\aap] {10.1051/0004-6361/202348091}, \href
  {https://ui.adsabs.harvard.edu/abs/2024A&A...686A.128C} {686, A128}

\bibitem[\protect\citeauthoryear{{Clarke}, {Shapley}, {Sanders}, {Topping},
  {Brammer}, {Bento}, {Reddy}  \& {Kehoe}}{{Clarke} et~al.}{2024}]{Clarke24}
{Clarke} L.,  {Shapley} A.~E.,  {Sanders} R.~L.,  {Topping} M.~W.,  {Brammer}
  G.~B.,  {Bento} T.,  {Reddy} N.~A.,   {Kehoe} E.,  2024, \mn@doi [\apj]
  {10.3847/1538-4357/ad8ba4}, \href
  {https://ui.adsabs.harvard.edu/abs/2024ApJ...977..133C} {977, 133}

\bibitem[\protect\citeauthoryear{{Cochrane} et~al.,}{{Cochrane}
  et~al.}{2019}]{Cochrane}
{Cochrane} R.~K.,  et~al., 2019, \mn@doi [\mnras] {10.1093/mnras/stz1736},
  \href {https://ui.adsabs.harvard.edu/abs/2019MNRAS.488.1779C} {488, 1779}

\bibitem[\protect\citeauthoryear{{Cole} et~al.,}{{Cole}
  et~al.}{2025}]{Cole24_burstSFMS}
{Cole} J.~W.,  et~al., 2025, \mn@doi [\apj] {10.3847/1538-4357/ad9a6a}, \href
  {https://ui.adsabs.harvard.edu/abs/2025ApJ...979..193C} {979, 193}

\bibitem[\protect\citeauthoryear{{Covelo-Paz} et~al.,}{{Covelo-Paz}
  et~al.}{2025}]{CoveloPaz25_haLF}
{Covelo-Paz} A.,  et~al., 2025, \mn@doi [\aap] {10.1051/0004-6361/202452363},
  \href {https://ui.adsabs.harvard.edu/abs/2025A&A...694A.178C} {694, A178}

\bibitem[\protect\citeauthoryear{{Cruz}, {Mu{\~n}oz}, {Sabti}  \&
  {Kamionkowski}}{{Cruz} et~al.}{2025}]{Cruz}
{Cruz} H. A.~G.,  {Mu{\~n}oz} J.~B.,  {Sabti} N.,   {Kamionkowski} M.,  2025,
  \mn@doi [\prd] {10.1103/PhysRevD.111.083503}, \href
  {https://ui.adsabs.harvard.edu/abs/2025PhRvD.111h3503C} {111, 083503}

\bibitem[\protect\citeauthoryear{{Cullen} et~al.,}{{Cullen}
  et~al.}{2023}]{Cullen23_jwst_UVSlopes}
{Cullen} F.,  et~al., 2023, \mn@doi [\mnras] {10.1093/mnras/stad073}, \href
  {https://ui.adsabs.harvard.edu/abs/2023MNRAS.520...14C} {520, 14}

\bibitem[\protect\citeauthoryear{{Curtis-Lake} et~al.,}{{Curtis-Lake}
  et~al.}{2023}]{CurtisLakeJades_xiion}
{Curtis-Lake} E.,  et~al., 2023, \mn@doi [Nature Astronomy]
  {10.1038/s41550-023-01918-w}, \href
  {https://ui.adsabs.harvard.edu/abs/2023NatAs...7..622C} {7, 622}

\bibitem[\protect\citeauthoryear{{Dav{\'e}}, {Finlator}  \&
  {Oppenheimer}}{{Dav{\'e}} et~al.}{2012}]{Dave_12_analytic}
{Dav{\'e}} R.,  {Finlator} K.,   {Oppenheimer} B.~D.,  2012, \mn@doi [\mnras]
  {10.1111/j.1365-2966.2011.20148.x}, \href
  {https://ui.adsabs.harvard.edu/abs/2012MNRAS.421...98D} {421, 98}

\bibitem[\protect\citeauthoryear{{Dekel} \& {Silk}}{{Dekel} \&
  {Silk}}{1986}]{Dekel:1986ehj}
{Dekel} A.,  {Silk} J.,  1986, \mn@doi [\apj] {10.1086/164050}, \href
  {https://ui.adsabs.harvard.edu/abs/1986ApJ...303...39D} {303, 39}

\bibitem[\protect\citeauthoryear{Dekel, Zolotov, Tweed, Cacciato, Ceverino  \&
  Primack}{Dekel et~al.}{2013}]{Dekel:2013uaa}
Dekel A.,  Zolotov A.,  Tweed D.,  Cacciato M.,  Ceverino D.,   Primack J.~R.,
  2013, \mn@doi [Mon. Not. Roy. Astron. Soc.] {10.1093/mnras/stt1338}, 435, 999

\bibitem[\protect\citeauthoryear{{Dekel}, {Sarkar}, {Birnboim}, {Mandelker}  \&
  {Li}}{{Dekel} et~al.}{2023}]{Dekel23_FFB}
{Dekel} A.,  {Sarkar} K.~C.,  {Birnboim} Y.,  {Mandelker} N.,   {Li} Z.,  2023,
  \mn@doi [\mnras] {10.1093/mnras/stad1557}, \href
  {https://ui.adsabs.harvard.edu/abs/2023MNRAS.523.3201D} {523, 3201}

\bibitem[\protect\citeauthoryear{{Di Matteo}, {Springel}  \& {Hernquist}}{{Di
  Matteo} et~al.}{2005}]{DiMatteo:2005ttp}
{Di Matteo} T.,  {Springel} V.,   {Hernquist} L.,  2005, \mn@doi [\nat]
  {10.1038/nature03335}, \href
  {https://ui.adsabs.harvard.edu/abs/2005Natur.433..604D} {433, 604}

\bibitem[\protect\citeauthoryear{{Donnan} et~al.,}{{Donnan}
  et~al.}{2024}]{Donnan24_UVLF}
{Donnan} C.~T.,  et~al., 2024, \mn@doi [arXiv e-prints]
  {10.48550/arXiv.2403.03171}, \href
  {https://ui.adsabs.harvard.edu/abs/2024arXiv240303171D} {p. arXiv:2403.03171}

\bibitem[\protect\citeauthoryear{{Dressler} et~al.,}{{Dressler}
  et~al.}{2024}]{Dressler2024_bursty}
{Dressler} A.,  et~al., 2024, \mn@doi [\apj] {10.3847/1538-4357/ad1923}, \href
  {https://ui.adsabs.harvard.edu/abs/2024ApJ...964..150D} {964, 150}

\bibitem[\protect\citeauthoryear{{Eisenstein} et~al.,}{{Eisenstein}
  et~al.}{2023}]{Eisenstein_JADES}
{Eisenstein} D.~J.,  et~al., 2023, \mn@doi [arXiv e-prints]
  {10.48550/arXiv.2306.02465}, \href
  {https://ui.adsabs.harvard.edu/abs/2023arXiv230602465E} {p. arXiv:2306.02465}

\bibitem[\protect\citeauthoryear{{Eldridge}, {Stanway}, {Xiao}, {McClelland},
  {Taylor}, {Ng}, {Greis}  \& {Bray}}{{Eldridge} et~al.}{2017}]{BPASS}
{Eldridge} J.~J.,  {Stanway} E.~R.,  {Xiao} L.,  {McClelland} L.~A.~S.,
  {Taylor} G.,  {Ng} M.,  {Greis} S.~M.~L.,   {Bray} J.~C.,  2017, \mn@doi
  [\pasa] {10.1017/pasa.2017.51}, \href
  {https://ui.adsabs.harvard.edu/abs/2017PASA...34...58E} {34, e058}

\bibitem[\protect\citeauthoryear{{Emami}, {Siana}, {Weisz}, {Johnson}, {Ma}  \&
  {El-Badry}}{{Emami} et~al.}{2019}]{1809.06380}
{Emami} N.,  {Siana} B.,  {Weisz} D.~R.,  {Johnson} B.~D.,  {Ma} X.,
  {El-Badry} K.,  2019, \mn@doi [\apj] {10.3847/1538-4357/ab211a}, \href
  {https://ui.adsabs.harvard.edu/abs/2019ApJ...881...71E} {881, 71}

\bibitem[\protect\citeauthoryear{{Endsley} et~al.,}{{Endsley}
  et~al.}{2023}]{Endsley23_reionization}
{Endsley} R.,  et~al., 2023, \mn@doi [arXiv e-prints]
  {10.48550/arXiv.2306.05295}, \href
  {https://ui.adsabs.harvard.edu/abs/2023arXiv230605295E} {p. arXiv:2306.05295}

\bibitem[\protect\citeauthoryear{{Endsley}, {Chisholm}, {Stark}, {Topping}  \&
  {Whitler}}{{Endsley} et~al.}{2025}]{Endsley24_bursty}
{Endsley} R.,  {Chisholm} J.,  {Stark} D.~P.,  {Topping} M.~W.,   {Whitler} L.,
   2025, \mn@doi [\apj] {10.3847/1538-4357/addc74}, \href
  {https://ui.adsabs.harvard.edu/abs/2025ApJ...987..189E} {987, 189}

\bibitem[\protect\citeauthoryear{{Faisst}, {Capak}, {Emami}, {Tacchella}  \&
  {Larson}}{{Faisst} et~al.}{2019}]{Faisst2019_Habursty}
{Faisst} A.~L.,  {Capak} P.~L.,  {Emami} N.,  {Tacchella} S.,   {Larson} K.~L.,
   2019, \mn@doi [\apj] {10.3847/1538-4357/ab425b}, \href
  {https://ui.adsabs.harvard.edu/abs/2019ApJ...884..133F} {884, 133}

\bibitem[\protect\citeauthoryear{Faucher-Giguere}{Faucher-Giguere}{2018}]{Faucher-Giguere:2017lgp}
Faucher-Giguere C.~A.,  2018, \mn@doi [Mon. Not. Roy. Astron. Soc.]
  {10.1093/mnras/stx2595}, 473, 3717

\bibitem[\protect\citeauthoryear{{Faucher-Gigu{\`e}re}, {Quataert}  \&
  {Hopkins}}{{Faucher-Gigu{\`e}re} et~al.}{2013}]{Faucher-Giguere:2013nkp}
{Faucher-Gigu{\`e}re} C.-A.,  {Quataert} E.,   {Hopkins} P.~F.,  2013, \mn@doi
  [\mnras] {10.1093/mnras/stt866}, \href
  {https://ui.adsabs.harvard.edu/abs/2013MNRAS.433.1970F} {433, 1970}

\bibitem[\protect\citeauthoryear{Feldmann et~al.}{Feldmann
  et~al.}{2023}]{Feldmann:2022qvd}
Feldmann R.,  et~al., 2023, \mn@doi [Mon. Not. Roy. Astron. Soc.]
  {10.1093/mnras/stad1205}, 522, 3831

\bibitem[\protect\citeauthoryear{{Ferrara}}{{Ferrara}}{2024}]{Ferrara24}
{Ferrara} A.,  2024, \mn@doi [\aap] {10.1051/0004-6361/202348321}, \href
  {https://ui.adsabs.harvard.edu/abs/2024A&A...684A.207F} {684, A207}

\bibitem[\protect\citeauthoryear{Ferrara, Pallottini  \& Dayal}{Ferrara
  et~al.}{2023}]{Ferrara2022}
Ferrara A.,  Pallottini A.,   Dayal P.,  2023, \mn@doi [Mon. Not. Roy. Astron.
  Soc.] {10.1093/mnras/stad1095}, 522, 3986

\bibitem[\protect\citeauthoryear{{Finkelstein} \& {Bagley}}{{Finkelstein} \&
  {Bagley}}{2022}]{FinkelsteinBagley22_UVLF}
{Finkelstein} S.~L.,  {Bagley} M.~B.,  2022, \mn@doi [\apj]
  {10.3847/1538-4357/ac89eb}, \href
  {https://ui.adsabs.harvard.edu/abs/2022ApJ...938...25F} {938, 25}

\bibitem[\protect\citeauthoryear{{Finkelstein} et~al.,}{{Finkelstein}
  et~al.}{2023}]{Finkelstein_CEERS}
{Finkelstein} S.~L.,  et~al., 2023, \mn@doi [\apjl] {10.3847/2041-8213/acade4},
  \href {https://ui.adsabs.harvard.edu/abs/2023ApJ...946L..13F} {946, L13}

\bibitem[\protect\citeauthoryear{{Finkelstein} et~al.,}{{Finkelstein}
  et~al.}{2024}]{Finkelstein_CEERS2}
{Finkelstein} S.~L.,  et~al., 2024, \mn@doi [\apjl] {10.3847/2041-8213/ad4495},
  \href {https://ui.adsabs.harvard.edu/abs/2024ApJ...969L...2F} {969, L2}

\bibitem[\protect\citeauthoryear{{Fisher} et~al.,}{{Fisher}
  et~al.}{2025}]{Fisher25_REBELS}
{Fisher} R.,  et~al., 2025, \mn@doi [arXiv e-prints]
  {10.48550/arXiv.2511.10741}, \href
  {https://ui.adsabs.harvard.edu/abs/2025arXiv251110741F} {p. arXiv:2511.10741}

\bibitem[\protect\citeauthoryear{{Flores Vel{\'a}zquez} et~al.,}{{Flores
  Vel{\'a}zquez} et~al.}{2021}]{Flores2021_indicators}
{Flores Vel{\'a}zquez} J.~A.,  et~al., 2021, \mn@doi [\mnras]
  {10.1093/mnras/staa3893}, \href
  {https://ui.adsabs.harvard.edu/abs/2021MNRAS.501.4812F} {501, 4812}

\bibitem[\protect\citeauthoryear{{Foreman-Mackey}, {Hogg}, {Lang}  \&
  {Goodman}}{{Foreman-Mackey} et~al.}{2013}]{emcee}
{Foreman-Mackey} D.,  {Hogg} D.~W.,  {Lang} D.,   {Goodman} J.,  2013, \mn@doi
  [\pasp] {10.1086/670067}, \href
  {https://ui.adsabs.harvard.edu/abs/2013PASP..125..306F} {125, 306}

\bibitem[\protect\citeauthoryear{{Franco} et~al.,}{{Franco}
  et~al.}{2025}]{Franco_25_COSMOSW}
{Franco} M.,  et~al., 2025, \mn@doi [arXiv e-prints]
  {10.48550/arXiv.2508.04791}, \href
  {https://ui.adsabs.harvard.edu/abs/2025arXiv250804791F} {p. arXiv:2508.04791}

\bibitem[\protect\citeauthoryear{{Fumagalli}, {da Silva}  \&
  {Krumholz}}{{Fumagalli} et~al.}{2011}]{Fumagalli_slug_2011}
{Fumagalli} M.,  {da Silva} R.~L.,   {Krumholz} M.~R.,  2011, \mn@doi [\apjl]
  {10.1088/2041-8205/741/2/L26}, \href
  {https://ui.adsabs.harvard.edu/abs/2011ApJ...741L..26F} {741, L26}

\bibitem[\protect\citeauthoryear{{Furlanetto} \& {Mirocha}}{{Furlanetto} \&
  {Mirocha}}{2022}]{Furlanetto22_feedback}
{Furlanetto} S.~R.,  {Mirocha} J.,  2022, \mn@doi [\mnras]
  {10.1093/mnras/stac310}, \href
  {https://ui.adsabs.harvard.edu/abs/2022MNRAS.511.3895F} {511, 3895}

\bibitem[\protect\citeauthoryear{{Furlanetto}, {Mirocha}, {Mebane}  \&
  {Sun}}{{Furlanetto} et~al.}{2017}]{Furlanetto2017_feedback}
{Furlanetto} S.~R.,  {Mirocha} J.,  {Mebane} R.~H.,   {Sun} G.,  2017, \mn@doi
  [\mnras] {10.1093/mnras/stx2132}, \href
  {https://ui.adsabs.harvard.edu/abs/2017MNRAS.472.1576F} {472, 1576}

\bibitem[\protect\citeauthoryear{{Gelli}, {Mason}  \& {Hayward}}{{Gelli}
  et~al.}{2024}]{Gelli24}
{Gelli} V.,  {Mason} C.,   {Hayward} C.~C.,  2024, \mn@doi [\apj]
  {10.3847/1538-4357/ad7b36}, \href
  {https://ui.adsabs.harvard.edu/abs/2024ApJ...975..192G} {975, 192}

\bibitem[\protect\citeauthoryear{{Giavalisco} \& {Dickinson}}{{Giavalisco} \&
  {Dickinson}}{2001}]{Giavalisco01_clustering}
{Giavalisco} M.,  {Dickinson} M.,  2001, \mn@doi [\apj] {10.1086/319715}, \href
  {https://ui.adsabs.harvard.edu/abs/2001ApJ...550..177G} {550, 177}

\bibitem[\protect\citeauthoryear{{Harikane} et~al.,}{{Harikane}
  et~al.}{2022}]{Harikane}
{Harikane} Y.,  et~al., 2022, \mn@doi [\apjs] {10.3847/1538-4365/ac3dfc}, \href
  {https://ui.adsabs.harvard.edu/abs/2022ApJS..259...20H} {259, 20}

\bibitem[\protect\citeauthoryear{{Harikane}, {Ouchi}, {Oguri}, {Ono},
  {Nakajima}  et~al.}{{Harikane} et~al.}{2023}]{Harikane_UVLFs}
{Harikane} Y.,  {Ouchi} M.,  {Oguri} M.,  {Ono} Y.,  {Nakajima} K.,   et~al.,
  2023, \mn@doi [Astrophys. J., Suppl. Ser.] {10.3847/1538-4365/acaaa9}, \href
  {https://ui.adsabs.harvard.edu/abs/2023ApJS..265....5H} {265, 5}

\bibitem[\protect\citeauthoryear{{Harvey} et~al.,}{{Harvey}
  et~al.}{2025}]{Harvey2025}
{Harvey} T.,  et~al., 2025, \mn@doi [\mnras] {10.1093/mnras/staf1396}, \href
  {https://ui.adsabs.harvard.edu/abs/2025MNRAS.542.2998H} {542, 2998}

\bibitem[\protect\citeauthoryear{{Hegde} \& {Furlanetto}}{{Hegde} \&
  {Furlanetto}}{2025}]{hegde}
{Hegde} S.,  {Furlanetto} S.~R.,  2025, \mn@doi [The Open Journal of
  Astrophysics] {10.33232/001c.145070}, \href
  {https://ui.adsabs.harvard.edu/abs/2025OJAp....8E.147H} {8, 147}

\bibitem[\protect\citeauthoryear{Hegde, Wyatt  \& Furlanetto}{Hegde
  et~al.}{2024}]{Hegde:2024kph}
Hegde S.,  Wyatt M.~M.,   Furlanetto S.~R.,  2024, \mn@doi [JCAP]
  {10.1088/1475-7516/2024/08/025}, 08, 025

\bibitem[\protect\citeauthoryear{{Heintz} et~al.,}{{Heintz}
  et~al.}{2023}]{2212.02890}
{Heintz} K.~E.,  et~al., 2023, \mn@doi [Nature Astronomy]
  {10.1038/s41550-023-02078-7}, \href
  {https://ui.adsabs.harvard.edu/abs/2023NatAs...7.1517H} {7, 1517}

\bibitem[\protect\citeauthoryear{{Hopkins}, {Quataert}  \& {Murray}}{{Hopkins}
  et~al.}{2011}]{Hopkins:2011xm}
{Hopkins} P.~F.,  {Quataert} E.,   {Murray} N.,  2011, \mn@doi [\mnras]
  {10.1111/j.1365-2966.2011.19306.x}, \href
  {https://ui.adsabs.harvard.edu/abs/2011MNRAS.417..950H} {417, 950}

\bibitem[\protect\citeauthoryear{{Hopkins}, {Quataert}  \& {Murray}}{{Hopkins}
  et~al.}{2012}]{1110.4638}
{Hopkins} P.~F.,  {Quataert} E.,   {Murray} N.,  2012, \mn@doi [\mnras]
  {10.1111/j.1365-2966.2012.20593.x}, \href
  {https://ui.adsabs.harvard.edu/abs/2012MNRAS.421.3522H} {421, 3522}

\bibitem[\protect\citeauthoryear{{Hopkins}, {Kere{\v{s}}}, {O{\~n}orbe},
  {Faucher-Gigu{\`e}re}, {Quataert}, {Murray}  \& {Bullock}}{{Hopkins}
  et~al.}{2014}]{Hopkins:2013vha}
{Hopkins} P.~F.,  {Kere{\v{s}}} D.,  {O{\~n}orbe} J.,  {Faucher-Gigu{\`e}re}
  C.-A.,  {Quataert} E.,  {Murray} N.,   {Bullock} J.~S.,  2014, \mn@doi
  [\mnras] {10.1093/mnras/stu1738}, \href
  {https://ui.adsabs.harvard.edu/abs/2014MNRAS.445..581H} {445, 581}

\bibitem[\protect\citeauthoryear{{Hopkins} et~al.,}{{Hopkins}
  et~al.}{2023}]{Hopkins2023_FIRE}
{Hopkins} P.~F.,  et~al., 2023, \mn@doi [\mnras] {10.1093/mnras/stac3489},
  \href {https://ui.adsabs.harvard.edu/abs/2023MNRAS.519.3154H} {519, 3154}

\bibitem[\protect\citeauthoryear{{Hsiao} et~al.,}{{Hsiao}
  et~al.}{2023}]{Hsiao23_xiion}
{Hsiao} T. Y.-Y.,  et~al., 2023, \mn@doi [arXiv e-prints]
  {10.48550/arXiv.2305.03042}, \href
  {https://ui.adsabs.harvard.edu/abs/2023arXiv230503042H} {p. arXiv:2305.03042}

\bibitem[\protect\citeauthoryear{{Hu} et~al.,}{{Hu} et~al.}{2023}]{Hu2023}
{Hu} C.-Y.,  et~al., 2023, \mn@doi [\apj] {10.3847/1538-4357/accf9e}, \href
  {https://ui.adsabs.harvard.edu/abs/2023ApJ...950..132H} {950, 132}

\bibitem[\protect\citeauthoryear{Hutter, Cueto, Dayal, Gottl{\"o}ber, Trebitsch
   \& Yepes}{Hutter et~al.}{2025}]{Hutter:2024cvr}
Hutter A.,  Cueto E.~R.,  Dayal P.,  Gottl{\"o}ber S.,  Trebitsch M.,   Yepes
  G.,  2025, \mn@doi [Astron. Astrophys.] {10.1051/0004-6361/202452460}, 694,
  A254

\bibitem[\protect\citeauthoryear{{Inayoshi}, {Harikane}, {Inoue}, {Li}  \&
  {Ho}}{{Inayoshi} et~al.}{2022}]{Inayoshi22_SF}
{Inayoshi} K.,  {Harikane} Y.,  {Inoue} A.~K.,  {Li} W.,   {Ho} L.~C.,  2022,
  \mn@doi [\apjl] {10.3847/2041-8213/ac9310}, \href
  {https://ui.adsabs.harvard.edu/abs/2022ApJ...938L..10I} {938, L10}

\bibitem[\protect\citeauthoryear{{Iyer} et~al.,}{{Iyer}
  et~al.}{2020}]{2007.07916}
{Iyer} K.~G.,  et~al., 2020, \mn@doi [\mnras] {10.1093/mnras/staa2150}, \href
  {https://ui.adsabs.harvard.edu/abs/2020MNRAS.498..430I} {498, 430}

\bibitem[\protect\citeauthoryear{{Iyer}, {Speagle}, {Caplar}, {Forbes},
  {Gawiser}, {Leja}  \& {Tacchella}}{{Iyer} et~al.}{2024}]{2208.05938}
{Iyer} K.~G.,  {Speagle} J.~S.,  {Caplar} N.,  {Forbes} J.~C.,  {Gawiser} E.,
  {Leja} J.,   {Tacchella} S.,  2024, \mn@doi [\apj]
  {10.3847/1538-4357/acff64}, \href
  {https://ui.adsabs.harvard.edu/abs/2024ApJ...961...53I} {961, 53}

\bibitem[\protect\citeauthoryear{{Jecmen} et~al.,}{{Jecmen}
  et~al.}{2026}]{Jecmen_beta26}
{Jecmen} M.~C.,  et~al., 2026, \mn@doi [arXiv e-prints]
  {10.48550/arXiv.2601.19995}, \href
  {https://ui.adsabs.harvard.edu/abs/2026arXiv260119995J} {p. arXiv:2601.19995}

\bibitem[\protect\citeauthoryear{{Johnson}, {Leja}, {Conroy}  \&
  {Speagle}}{{Johnson} et~al.}{2021}]{2012.01426}
{Johnson} B.~D.,  {Leja} J.,  {Conroy} C.,   {Speagle} J.~S.,  2021, \mn@doi
  [\apjs] {10.3847/1538-4365/abef67}, \href
  {https://ui.adsabs.harvard.edu/abs/2021ApJS..254...22J} {254, 22}

\bibitem[\protect\citeauthoryear{Jones, Oliphant, Peterson  et~al.}{Jones
  et~al.}{2001}]{scipy}
Jones E.,  Oliphant T.,  Peterson P.,   et~al., 2001, {SciPy}: Open source
  scientific tools for {Python}, \url {http://www.scipy.org/}

\bibitem[\protect\citeauthoryear{{Katz} et~al.,}{{Katz}
  et~al.}{2025}]{Katz25_megatron}
{Katz} H.,  et~al., 2025, \mn@doi [arXiv e-prints] {10.48550/arXiv.2510.05201},
  \href {https://ui.adsabs.harvard.edu/abs/2025arXiv251005201K} {p.
  arXiv:2510.05201}

\bibitem[\protect\citeauthoryear{{Kennicutt}}{{Kennicutt}}{1998}]{astro-ph/9807187}
{Kennicutt} Jr. R.~C.,  1998, \mn@doi [\araa] {10.1146/annurev.astro.36.1.189},
  \href {https://ui.adsabs.harvard.edu/abs/1998ARA&A..36..189K} {36, 189}

\bibitem[\protect\citeauthoryear{{Kokorev} et~al.,}{{Kokorev}
  et~al.}{2025a}]{Kokorev25_glimpse}
{Kokorev} V.,  et~al., 2025a, \mn@doi [\apjl] {10.3847/2041-8213/adc458}, \href
  {https://ui.adsabs.harvard.edu/abs/2025ApJ...983L..22K} {983, L22}

\bibitem[\protect\citeauthoryear{{Kokorev} et~al.,}{{Kokorev}
  et~al.}{2025b}]{Kokorev25_capers}
{Kokorev} V.,  et~al., 2025b, \mn@doi [\apjl] {10.3847/2041-8213/ade8f5}, \href
  {https://ui.adsabs.harvard.edu/abs/2025ApJ...988L..10K} {988, L10}

\bibitem[\protect\citeauthoryear{{Korber} et~al.,}{{Korber}
  et~al.}{2025}]{Korber25_HbLF}
{Korber} D.,  et~al., 2025, \mn@doi [arXiv e-prints]
  {10.48550/arXiv.2510.04771}, \href
  {https://ui.adsabs.harvard.edu/abs/2025arXiv251004771K} {p. arXiv:2510.04771}

\bibitem[\protect\citeauthoryear{{Kravtsov} \& {Belokurov}}{{Kravtsov} \&
  {Belokurov}}{2024}]{2405.04578}
{Kravtsov} A.,  {Belokurov} V.,  2024, \mn@doi [arXiv e-prints]
  {10.48550/arXiv.2405.04578}, \href
  {https://ui.adsabs.harvard.edu/abs/2024arXiv240504578K} {p. arXiv:2405.04578}

\bibitem[\protect\citeauthoryear{{Kroupa}}{{Kroupa}}{2001}]{Kroupa2001_IMF}
{Kroupa} P.,  2001, \mn@doi [\mnras] {10.1046/j.1365-8711.2001.04022.x}, \href
  {https://ui.adsabs.harvard.edu/abs/2001MNRAS.322..231K} {322, 231}

\bibitem[\protect\citeauthoryear{{Krumholz}, {Fumagalli}, {da Silva}, {Rendahl}
   \& {Parra}}{{Krumholz} et~al.}{2015}]{krumholz_slug_2015}
{Krumholz} M.~R.,  {Fumagalli} M.,  {da Silva} R.~L.,  {Rendahl} T.,   {Parra}
  J.,  2015, \mn@doi [\mnras] {10.1093/mnras/stv1374}, \href
  {https://ui.adsabs.harvard.edu/abs/2015MNRAS.452.1447K} {452, 1447}

\bibitem[\protect\citeauthoryear{{Leitherer} et~al.,}{{Leitherer}
  et~al.}{1999}]{Leitherer_SB99}
{Leitherer} C.,  et~al., 1999, \mn@doi [\apjs] {10.1086/313233}, \href
  {https://ui.adsabs.harvard.edu/abs/1999ApJS..123....3L} {123, 3}

\bibitem[\protect\citeauthoryear{{Leja}, {Carnall}, {Johnson}, {Conroy}  \&
  {Speagle}}{{Leja} et~al.}{2019}]{Leja:2019}
{Leja} J.,  {Carnall} A.~C.,  {Johnson} B.~D.,  {Conroy} C.,   {Speagle} J.~S.,
   2019, \mn@doi [\apj] {10.3847/1538-4357/ab133c}, \href
  {https://ui.adsabs.harvard.edu/abs/2019ApJ...876....3L} {876, 3}

\bibitem[\protect\citeauthoryear{Lo}{Lo}{2012}]{Lo2012}
Lo C.-F.,  2012, \mn@doi [Journal of Applied Mathematics]
  {10.2139/ssrn.2064829}, 2012

\bibitem[\protect\citeauthoryear{Madau \& Dickinson}{Madau \&
  Dickinson}{2014}]{Madau:2014bja}
Madau P.,  Dickinson M.,  2014, \mn@doi [Ann. Rev. Astron. Astrophys.]
  {10.1146/annurev-astro-081811-125615}, 52, 415

\bibitem[\protect\citeauthoryear{{Man} \& {Belli}}{{Man} \&
  {Belli}}{2018}]{1809.00722}
{Man} A.,  {Belli} S.,  2018, \mn@doi [Nature Astronomy]
  {10.1038/s41550-018-0558-1}, \href
  {https://ui.adsabs.harvard.edu/abs/2018NatAs...2..695M} {2, 695}

\bibitem[\protect\citeauthoryear{{Mason}, {Trenti}  \& {Treu}}{{Mason}
  et~al.}{2015}]{Mason15}
{Mason} C.~A.,  {Trenti} M.,   {Treu} T.,  2015, \mn@doi [\apj]
  {10.1088/0004-637X/813/1/21}, \href
  {https://ui.adsabs.harvard.edu/abs/2015ApJ...813...21M} {813, 21}

\bibitem[\protect\citeauthoryear{{Mason}, {Trenti}  \& {Treu}}{{Mason}
  et~al.}{2023}]{mason23}
{Mason} C.~A.,  {Trenti} M.,   {Treu} T.,  2023, \mn@doi [\mnras]
  {10.1093/mnras/stad035}, \href
  {https://ui.adsabs.harvard.edu/abs/2023MNRAS.521..497M} {521, 497}

\bibitem[\protect\citeauthoryear{{McKee} \& {Ostriker}}{{McKee} \&
  {Ostriker}}{1977}]{McKee1977}
{McKee} C.~F.,  {Ostriker} J.~P.,  1977, \mn@doi [\apj] {10.1086/155667}, \href
  {https://ui.adsabs.harvard.edu/abs/1977ApJ...218..148M} {218, 148}

\bibitem[\protect\citeauthoryear{McKinney, Cooper, Casey, Munoz, Akins,
  Lambrides  \& Long}{McKinney et~al.}{2025}]{McKinney:2025htm}
McKinney J.,  Cooper O.~R.,  Casey C.~M.,  Munoz J.~B.,  Akins H.,  Lambrides
  E.,   Long A.~S.,  2025, \mn@doi [Astrophys. J. Lett.]
  {10.3847/2041-8213/add15d}, 985, L21

\bibitem[\protect\citeauthoryear{{Mehta} et~al.,}{{Mehta}
  et~al.}{2023}]{Mehta23_uvcandels}
{Mehta} V.,  et~al., 2023, \mn@doi [\apj] {10.3847/1538-4357/acd9cf}, \href
  {https://ui.adsabs.harvard.edu/abs/2023ApJ...952..133M} {952, 133}

\bibitem[\protect\citeauthoryear{{Meurer}, {Heckman}  \& {Calzetti}}{{Meurer}
  et~al.}{1999}]{astro-ph/9903054}
{Meurer} G.~R.,  {Heckman} T.~M.,   {Calzetti} D.,  1999, \mn@doi [\apj]
  {10.1086/307523}, \href
  {https://ui.adsabs.harvard.edu/abs/1999ApJ...521...64M} {521, 64}

\bibitem[\protect\citeauthoryear{{Mintz} et~al.,}{{Mintz}
  et~al.}{2025}]{2506.16510}
{Mintz} A.,  et~al., 2025, \mn@doi [arXiv e-prints]
  {10.48550/arXiv.2506.16510}, \href
  {https://ui.adsabs.harvard.edu/abs/2025arXiv250616510M} {p. arXiv:2506.16510}

\bibitem[\protect\citeauthoryear{{Mirocha} \& {Furlanetto}}{{Mirocha} \&
  {Furlanetto}}{2023}]{mirocha23}
{Mirocha} J.,  {Furlanetto} S.~R.,  2023, \mn@doi [\mnras]
  {10.1093/mnras/stac3578}, \href
  {https://ui.adsabs.harvard.edu/abs/2023MNRAS.519..843M} {519, 843}

\bibitem[\protect\citeauthoryear{{Moster}, {Naab}  \& {White}}{{Moster}
  et~al.}{2013}]{Moster13_AM}
{Moster} B.~P.,  {Naab} T.,   {White} S. D.~M.,  2013, \mn@doi [\mnras]
  {10.1093/mnras/sts261}, \href
  {https://ui.adsabs.harvard.edu/abs/2013MNRAS.428.3121M} {428, 3121}

\bibitem[\protect\citeauthoryear{Mu\~noz, Mirocha, Furlanetto  \&
  Sabti}{Mu\~noz et~al.}{2023}]{Munoz:2023cup}
Mu\~noz J.~B.,  Mirocha J.,  Furlanetto S.,   Sabti N.,  2023, \mn@doi [Mon.
  Not. Roy. Astron. Soc.] {10.1093/mnrasl/slad115}, 526, L47

\bibitem[\protect\citeauthoryear{Mu{\~n}oz}{Mu{\~n}oz}{2023}]{Munoz:2023kkg}
Mu{\~n}oz J.~B.,  2023, \mn@doi [Mon. Not. Roy. Astron. Soc.]
  {10.1093/mnras/stad1512}, 523, 2587

\bibitem[\protect\citeauthoryear{Mu{\~n}oz, Mirocha, Chisholm, Furlanetto  \&
  Mason}{Mu{\~n}oz et~al.}{2024}]{Munoz:2024fas}
Mu{\~n}oz J.~B.,  Mirocha J.,  Chisholm J.,  Furlanetto S.~R.,   Mason C.,
  2024, \mn@doi [Mon. Not. Roy. Astron. Soc.] {10.1093/mnrasl/slae086}, 535,
  L37

\bibitem[\protect\citeauthoryear{{Murray}}{{Murray}}{2018}]{Murray_powerbox}
{Murray} S.~G.,  2018, \mn@doi [The Journal of Open Source Software]
  {10.21105/joss.00850}, \href
  {https://ui.adsabs.harvard.edu/abs/2018JOSS....3..850M} {3, 850}

\bibitem[\protect\citeauthoryear{{Murray}, {Quataert}  \& {Thompson}}{{Murray}
  et~al.}{2005}]{Murray:2004dd}
{Murray} N.,  {Quataert} E.,   {Thompson} T.~A.,  2005, \mn@doi [\apj]
  {10.1086/426067}, \href
  {https://ui.adsabs.harvard.edu/abs/2005ApJ...618..569M} {618, 569}

\bibitem[\protect\citeauthoryear{{Naidu} et~al.,}{{Naidu}
  et~al.}{2025}]{Momz40}
{Naidu} R.~P.,  et~al., 2025, \mn@doi [arXiv e-prints]
  {10.48550/arXiv.2505.11263}, \href
  {https://ui.adsabs.harvard.edu/abs/2025arXiv250511263N} {p. arXiv:2505.11263}

\bibitem[\protect\citeauthoryear{{Narayanan}, {Conroy}, {Dav{\'e}}, {Johnson}
  \& {Popping}}{{Narayanan} et~al.}{2018}]{Narayanan_18_dust}
{Narayanan} D.,  {Conroy} C.,  {Dav{\'e}} R.,  {Johnson} B.~D.,   {Popping} G.,
   2018, \mn@doi [\apj] {10.3847/1538-4357/aaed25}, \href
  {https://ui.adsabs.harvard.edu/abs/2018ApJ...869...70N} {869, 70}

\bibitem[\protect\citeauthoryear{Nikoli{\'c}, Mesinger, Davies  \&
  Prelogovi{\'c}}{Nikoli{\'c} et~al.}{2024}]{Nikolic:2024xxo}
Nikoli{\'c} I.,  Mesinger A.,  Davies J.~E.,   Prelogovi{\'c} D.,  2024,
  \mn@doi [Astron. Astrophys.] {10.1051/0004-6361/202451213}, 692, A142

\bibitem[\protect\citeauthoryear{{Oh} \& {Haiman}}{{Oh} \& {Haiman}}{2002}]{Oh}
{Oh} S.~P.,  {Haiman} Z.,  2002, \mn@doi [\apj] {10.1086/339393}, \href
  {https://ui.adsabs.harvard.edu/abs/2002ApJ...569..558O} {569, 558}

\bibitem[\protect\citeauthoryear{{Oke} \& {Gunn}}{{Oke} \&
  {Gunn}}{1983}]{OkeGunn}
{Oke} J.~B.,  {Gunn} J.~E.,  1983, \mn@doi [\apj] {10.1086/160817}, \href
  {https://ui.adsabs.harvard.edu/abs/1983ApJ...266..713O} {266, 713}

\bibitem[\protect\citeauthoryear{{Osterbrock} \& {Ferland}}{{Osterbrock} \&
  {Ferland}}{2006}]{Osterbrokferland2006}
{Osterbrock} D.~E.,  {Ferland} G.~J.,  2006, {Astrophysics of gaseous nebulae
  and active galactic nuclei}

\bibitem[\protect\citeauthoryear{{Overzier} et~al.,}{{Overzier}
  et~al.}{2011}]{Overzier11_dust}
{Overzier} R.~A.,  et~al., 2011, \mn@doi [\apjl] {10.1088/2041-8205/726/1/L7},
  \href {https://ui.adsabs.harvard.edu/abs/2011ApJ...726L...7O} {726, L7}

\bibitem[\protect\citeauthoryear{{Pacifici} et~al.,}{{Pacifici}
  et~al.}{2023}]{2212.01915}
{Pacifici} C.,  et~al., 2023, \mn@doi [\apj] {10.3847/1538-4357/acacff}, \href
  {https://ui.adsabs.harvard.edu/abs/2023ApJ...944..141P} {944, 141}

\bibitem[\protect\citeauthoryear{{Pahl} et~al.,}{{Pahl} et~al.}{2025}]{Pahl}
{Pahl} A.,  et~al., 2025, \mn@doi [\apj] {10.3847/1538-4357/adb1ab}, \href
  {https://ui.adsabs.harvard.edu/abs/2025ApJ...981..134P} {981, 134}

\bibitem[\protect\citeauthoryear{Pallottini \& Ferrara}{Pallottini \&
  Ferrara}{2023}]{Pallottini:2023yqg}
Pallottini A.,  Ferrara A.,  2023, \mn@doi [Astron. Astrophys.]
  {10.1051/0004-6361/202347384}, 677, L4

\bibitem[\protect\citeauthoryear{Papovich, Dickinson  \& Ferguson}{Papovich
  et~al.}{2001}]{Papovich:2001bu}
Papovich C.,  Dickinson M.,   Ferguson H.~C.,  2001, \mn@doi [Astrophys. J.]
  {10.1086/322412}, 559, 620

\bibitem[\protect\citeauthoryear{{Paquereau} et~al.,}{{Paquereau}
  et~al.}{2025}]{Paquereau25_clustering}
{Paquereau} L.,  et~al., 2025, \mn@doi [\aap] {10.1051/0004-6361/202553828},
  \href {https://ui.adsabs.harvard.edu/abs/2025A&A...702A.163P} {702, A163}

\bibitem[\protect\citeauthoryear{{P{\'e}rez-Gonz{\'a}lez}
  et~al.,}{{P{\'e}rez-Gonz{\'a}lez} et~al.}{2025}]{PG25}
{P{\'e}rez-Gonz{\'a}lez} P.~G.,  et~al., 2025, \mn@doi [\apj]
  {10.3847/1538-4357/adf8c9}, \href
  {https://ui.adsabs.harvard.edu/abs/2025ApJ...991..179P} {991, 179}

\bibitem[\protect\citeauthoryear{{Popesso} et~al.,}{{Popesso}
  et~al.}{2023}]{Popesso22}
{Popesso} P.,  et~al., 2023, \mn@doi [\mnras] {10.1093/mnras/stac3214}, \href
  {https://ui.adsabs.harvard.edu/abs/2023MNRAS.519.1526P} {519, 1526}

\bibitem[\protect\citeauthoryear{{Prieto-Lyon} et~al.,}{{Prieto-Lyon}
  et~al.}{2023}]{Prieto}
{Prieto-Lyon} G.,  et~al., 2023, \mn@doi [\aap] {10.1051/0004-6361/202245532},
  \href {https://ui.adsabs.harvard.edu/abs/2023A&A...672A.186P} {672, A186}

\bibitem[\protect\citeauthoryear{{Pritchard} \& {Loeb}}{{Pritchard} \&
  {Loeb}}{2012}]{Pritchard2012_review21cm}
{Pritchard} J.~R.,  {Loeb} A.,  2012, \mn@doi [Reports on Progress in Physics]
  {10.1088/0034-4885/75/8/086901}, \href
  {https://ui.adsabs.harvard.edu/abs/2012RPPh...75h6901P} {75, 086901}

\bibitem[\protect\citeauthoryear{Rieke et~al.}{Rieke
  et~al.}{2023}]{Rieke:2023tks}
Rieke M.~J.,  et~al., 2023, \mn@doi [Astrophys. J. Suppl.]
  {10.3847/1538-4365/acf44d}, 269, 16

\bibitem[\protect\citeauthoryear{{Rinaldi} et~al.,}{{Rinaldi}
  et~al.}{2025}]{Rinaldi24}
{Rinaldi} P.,  et~al., 2025, \mn@doi [\apj] {10.3847/1538-4357/adb309}, \href
  {https://ui.adsabs.harvard.edu/abs/2025ApJ...981..161R} {981, 161}

\bibitem[\protect\citeauthoryear{{Roberts-Borsani} et~al.,}{{Roberts-Borsani}
  et~al.}{2025}]{RobertsBorsani25_burstiness}
{Roberts-Borsani} G.,  et~al., 2025, \mn@doi [arXiv e-prints]
  {10.48550/arXiv.2508.21708}, \href
  {https://ui.adsabs.harvard.edu/abs/2025arXiv250821708R} {p. arXiv:2508.21708}

\bibitem[\protect\citeauthoryear{Robertson, Ellis, Furlanetto  \&
  Dunlop}{Robertson et~al.}{2015}]{Robertson:2015uda}
Robertson B.~E.,  Ellis R.~S.,  Furlanetto S.~R.,   Dunlop J.~S.,  2015,
  \mn@doi [Astrophys. J. Lett.] {10.1088/2041-8205/802/2/L19}, 802, L19

\bibitem[\protect\citeauthoryear{{Robertson} et~al.,}{{Robertson}
  et~al.}{2024}]{GSz14}
{Robertson} B.,  et~al., 2024, \mn@doi [\apj] {10.3847/1538-4357/ad463d}, \href
  {https://ui.adsabs.harvard.edu/abs/2024ApJ...970...31R} {970, 31}

\bibitem[\protect\citeauthoryear{Sabti, Mu{\~n}oz  \& Blas}{Sabti
  et~al.}{2022}]{Sabti:2021xvh}
Sabti N.,  Mu{\~n}oz J.~B.,   Blas D.,  2022, \mn@doi [Phys. Rev. D]
  {10.1103/PhysRevD.105.043518}, 105, 043518

\bibitem[\protect\citeauthoryear{{Samuel et al.}}{{Samuel et
  al.}}{2026}]{Jenna25}
{Samuel et al.} J.,  2026

\bibitem[\protect\citeauthoryear{{Santini} et~al.,}{{Santini}
  et~al.}{2017}]{1706.07059}
{Santini} P.,  et~al., 2017, \mn@doi [\apj] {10.3847/1538-4357/aa8874}, \href
  {https://ui.adsabs.harvard.edu/abs/2017ApJ...847...76S} {847, 76}

\bibitem[\protect\citeauthoryear{{Scarlata}, {Hayes}, {Panagia}, {Mehta},
  {Haardt}  \& {Bagley}}{{Scarlata} et~al.}{2024}]{scarlata2024}
{Scarlata} C.,  {Hayes} M.,  {Panagia} N.,  {Mehta} V.,  {Haardt} F.,
  {Bagley} M.,  2024, \mn@doi [arXiv e-prints] {10.48550/arXiv.2404.09015},
  \href {https://ui.adsabs.harvard.edu/abs/2024arXiv240409015S} {p.
  arXiv:2404.09015}

\bibitem[\protect\citeauthoryear{{Schinnerer} \& {Leroy}}{{Schinnerer} \&
  {Leroy}}{2024}]{2403.19843}
{Schinnerer} E.,  {Leroy} A.~K.,  2024, \mn@doi [\araa]
  {10.1146/annurev-astro-071221-052651}, \href
  {https://ui.adsabs.harvard.edu/abs/2024ARA&A..62..369S} {62, 369}

\bibitem[\protect\citeauthoryear{Shapiro, Iliev  \& Raga}{Shapiro
  et~al.}{2004}]{Shapiro:2003gxa}
Shapiro P.~R.,  Iliev I.~T.,   Raga A.~C.,  2004, \mn@doi [Mon. Not. Roy.
  Astron. Soc.] {10.1111/j.1365-2966.2004.07364.x}, 348, 753

\bibitem[\protect\citeauthoryear{{Shen}, {Vogelsberger}, {Boylan-Kolchin},
  {Tacchella}  \& {Kannan}}{{Shen} et~al.}{2023}]{shen23}
{Shen} X.,  {Vogelsberger} M.,  {Boylan-Kolchin} M.,  {Tacchella} S.,
  {Kannan} R.,  2023, \mn@doi [\mnras] {10.1093/mnras/stad2508}, \href
  {https://ui.adsabs.harvard.edu/abs/2023MNRAS.525.3254S} {525, 3254}

\bibitem[\protect\citeauthoryear{{Shen} et~al.,}{{Shen} et~al.}{2025}]{Shen+25}
{Shen} X.,  et~al., 2025, \mn@doi [arXiv e-prints] {10.48550/arXiv.2503.01949},
  \href {https://ui.adsabs.harvard.edu/abs/2025arXiv250301949S} {p.
  arXiv:2503.01949}

\bibitem[\protect\citeauthoryear{{Shivaei} et~al.,}{{Shivaei}
  et~al.}{2018}]{Shivaie18}
{Shivaei} I.,  et~al., 2018, \mn@doi [\apj] {10.3847/1538-4357/aaad62}, \href
  {https://ui.adsabs.harvard.edu/abs/2018ApJ...855...42S} {855, 42}

\bibitem[\protect\citeauthoryear{{Shuntov} et~al.,}{{Shuntov}
  et~al.}{2025}]{Shuntov}
{Shuntov} M.,  et~al., 2025, \mn@doi [\aap] {10.1051/0004-6361/202554618},
  \href {https://ui.adsabs.harvard.edu/abs/2025A&A...699A.231S} {699, A231}

\bibitem[\protect\citeauthoryear{{Silk}}{{Silk}}{1997}]{Silk:1997xw}
{Silk} J.,  1997, \mn@doi [\apj] {10.1086/304073}, \href
  {https://ui.adsabs.harvard.edu/abs/1997ApJ...481..703S} {481, 703}

\bibitem[\protect\citeauthoryear{{Simmonds} et~al.,}{{Simmonds}
  et~al.}{2024a}]{Simmonds24_xiion}
{Simmonds} C.,  et~al., 2024a, \mn@doi [\mnras] {10.1093/mnras/stad3605}, \href
  {https://ui.adsabs.harvard.edu/abs/2024MNRAS.527.6139S} {527, 6139}

\bibitem[\protect\citeauthoryear{{Simmonds} et~al.,}{{Simmonds}
  et~al.}{2024b}]{simmonds24_masscompl}
{Simmonds} C.,  et~al., 2024b, \mn@doi [\mnras] {10.1093/mnras/stae2537}, \href
  {https://ui.adsabs.harvard.edu/abs/2024MNRAS.535.2998S} {535, 2998}

\bibitem[\protect\citeauthoryear{{Simmonds} et~al.,}{{Simmonds}
  et~al.}{2025}]{2508.04410}
{Simmonds} C.,  et~al., 2025, \mn@doi [arXiv e-prints]
  {10.48550/arXiv.2508.04410}, \href
  {https://ui.adsabs.harvard.edu/abs/2025arXiv250804410S} {p. arXiv:2508.04410}

\bibitem[\protect\citeauthoryear{Sipple \& Lidz}{Sipple \&
  Lidz}{2024}]{Sipple:2023tgt}
Sipple J.,  Lidz A.,  2024, \mn@doi [Astrophys. J.] {10.3847/1538-4357/ad06a7},
  961, 50

\bibitem[\protect\citeauthoryear{{Smit}, {Bouwens}, {Franx}, {Illingworth},
  {Labb{\'e}}, {Oesch}  \& {van Dokkum}}{{Smit} et~al.}{2012}]{1204.3626}
{Smit} R.,  {Bouwens} R.~J.,  {Franx} M.,  {Illingworth} G.~D.,  {Labb{\'e}}
  I.,  {Oesch} P.~A.,   {van Dokkum} P.~G.,  2012, \mn@doi [\apj]
  {10.1088/0004-637X/756/1/14}, \href
  {https://ui.adsabs.harvard.edu/abs/2012ApJ...756...14S} {756, 14}

\bibitem[\protect\citeauthoryear{{Somerville} \& {Dav{\'e}}}{{Somerville} \&
  {Dav{\'e}}}{2015}]{Somerville2014}
{Somerville} R.~S.,  {Dav{\'e}} R.,  2015, \mn@doi [\araa]
  {10.1146/annurev-astro-082812-140951}, \href
  {https://ui.adsabs.harvard.edu/abs/2015ARA&A..53...51S} {53, 51}

\bibitem[\protect\citeauthoryear{Somerville \& Primack}{Somerville \&
  Primack}{1999}]{Somerville:1998bb}
Somerville R.~S.,  Primack J.~R.,  1999, \mn@doi [Mon. Not. Roy. Astron. Soc.]
  {10.1046/j.1365-8711.1999.03032.x}, 310, 1087

\bibitem[\protect\citeauthoryear{{Speagle}, {Steinhardt}, {Capak}  \&
  {Silverman}}{{Speagle} et~al.}{2014}]{1405.2041}
{Speagle} J.~S.,  {Steinhardt} C.~L.,  {Capak} P.~L.,   {Silverman} J.~D.,
  2014, \mn@doi [\apjs] {10.1088/0067-0049/214/2/15}, \href
  {https://ui.adsabs.harvard.edu/abs/2014ApJS..214...15S} {214, 15}

\bibitem[\protect\citeauthoryear{{Stark}, {Topping}, {Endsley}  \&
  {Tang}}{{Stark} et~al.}{2025}]{Stark_JWST_review}
{Stark} D.~P.,  {Topping} M.~W.,  {Endsley} R.,   {Tang} M.,  2025, \mn@doi
  [arXiv e-prints] {10.48550/arXiv.2501.17078}, \href
  {https://ui.adsabs.harvard.edu/abs/2025arXiv250117078S} {p. arXiv:2501.17078}

\bibitem[\protect\citeauthoryear{{Stefanon}, {Bouwens}, {Illingworth},
  {Labb{\'e}}, {Oesch}  \& {Gonzalez}}{{Stefanon}
  et~al.}{2022}]{Stefanon22_xiion}
{Stefanon} M.,  {Bouwens} R.~J.,  {Illingworth} G.~D.,  {Labb{\'e}} I.,
  {Oesch} P.~A.,   {Gonzalez} V.,  2022, \mn@doi [\apj]
  {10.3847/1538-4357/ac7e44}, \href
  {https://ui.adsabs.harvard.edu/abs/2022ApJ...935...94S} {935, 94}

\bibitem[\protect\citeauthoryear{{Sun} \& {Furlanetto}}{{Sun} \&
  {Furlanetto}}{2016}]{SunFurlanetto2016}
{Sun} G.,  {Furlanetto} S.~R.,  2016, \mn@doi [\mnras] {10.1093/mnras/stw980},
  \href {https://ui.adsabs.harvard.edu/abs/2016MNRAS.460..417S} {460, 417}

\bibitem[\protect\citeauthoryear{{Sun}, {Lidz}, {Faisst}  \&
  {Faucher-Gigu{\`e}re}}{{Sun} et~al.}{2023a}]{Sun2023}
{Sun} G.,  {Lidz} A.,  {Faisst} A.~L.,   {Faucher-Gigu{\`e}re} C.-A.,  2023a,
  \mn@doi [\mnras] {10.1093/mnras/stad2000}, \href
  {https://ui.adsabs.harvard.edu/abs/2023MNRAS.524.2395S} {524, 2395}

\bibitem[\protect\citeauthoryear{{Sun}, {Faucher-Gigu{\`e}re}, {Hayward},
  {Shen}, {Wetzel}  \& {Cochrane}}{{Sun} et~al.}{2023b}]{Sun23_FIRE_bursty}
{Sun} G.,  {Faucher-Gigu{\`e}re} C.-A.,  {Hayward} C.~C.,  {Shen} X.,  {Wetzel}
  A.,   {Cochrane} R.~K.,  2023b, \mn@doi [\apjl] {10.3847/2041-8213/acf85a},
  \href {https://ui.adsabs.harvard.edu/abs/2023ApJ...955L..35S} {955, L35}

\bibitem[\protect\citeauthoryear{{Sun}, {Mu{\~n}oz}, {Mirocha}  \&
  {Faucher-Gigu{\`e}re}}{{Sun} et~al.}{2025}]{2410.21409}
{Sun} G.,  {Mu{\~n}oz} J.~B.,  {Mirocha} J.,   {Faucher-Gigu{\`e}re} C.-A.,
  2025, \mn@doi [\jcap] {10.1088/1475-7516/2025/04/034}, \href
  {https://ui.adsabs.harvard.edu/abs/2025JCAP...04..034S} {2025, 034}

\bibitem[\protect\citeauthoryear{{Tacchella}, {Bose}, {Conroy}, {Eisenstein}
  \& {Johnson}}{{Tacchella} et~al.}{2018}]{Tacchella}
{Tacchella} S.,  {Bose} S.,  {Conroy} C.,  {Eisenstein} D.~J.,   {Johnson}
  B.~D.,  2018, \mn@doi [\apj] {10.3847/1538-4357/aae8e0}, \href
  {https://ui.adsabs.harvard.edu/abs/2018ApJ...868...92T} {868, 92}

\bibitem[\protect\citeauthoryear{{Tacchella}, {Forbes}  \&
  {Caplar}}{{Tacchella} et~al.}{2020}]{2006.09382}
{Tacchella} S.,  {Forbes} J.~C.,   {Caplar} N.,  2020, \mn@doi [\mnras]
  {10.1093/mnras/staa1838}, \href
  {https://ui.adsabs.harvard.edu/abs/2020MNRAS.497..698T} {497, 698}

\bibitem[\protect\citeauthoryear{{Tang}, {Stark}, {Mason}, {Gelli}, {Chen}  \&
  {Topping}}{{Tang} et~al.}{2025}]{Tang25_JWST_spectra}
{Tang} M.,  {Stark} D.~P.,  {Mason} C.~A.,  {Gelli} V.,  {Chen} Z.,   {Topping}
  M.~W.,  2025, \mn@doi [arXiv e-prints] {10.48550/arXiv.2507.08245}, \href
  {https://ui.adsabs.harvard.edu/abs/2025arXiv250708245T} {p. arXiv:2507.08245}

\bibitem[\protect\citeauthoryear{{Tinker}, {Robertson}, {Kravtsov}, {Klypin},
  {Warren}, {Yepes}  \& {Gottl{\"o}ber}}{{Tinker} et~al.}{2010}]{Tinker10_bias}
{Tinker} J.~L.,  {Robertson} B.~E.,  {Kravtsov} A.~V.,  {Klypin} A.,  {Warren}
  M.~S.,  {Yepes} G.,   {Gottl{\"o}ber} S.,  2010, \mn@doi [\apj]
  {10.1088/0004-637X/724/2/878}, \href
  {https://ui.adsabs.harvard.edu/abs/2010ApJ...724..878T} {724, 878}

\bibitem[\protect\citeauthoryear{{Topping} et~al.,}{{Topping}
  et~al.}{2024}]{Topping24_UVslopes}
{Topping} M.~W.,  et~al., 2024, \mn@doi [\mnras] {10.1093/mnras/stae800}, \href
  {https://ui.adsabs.harvard.edu/abs/2024MNRAS.529.4087T} {529, 4087}

\bibitem[\protect\citeauthoryear{{Topping} et~al.,}{{Topping}
  et~al.}{2025}]{Topping25}
{Topping} M.~W.,  et~al., 2025, \mn@doi [\apj] {10.3847/1538-4357/ada95c},
  \href {https://ui.adsabs.harvard.edu/abs/2025ApJ...980..225T} {980, 225}

\bibitem[\protect\citeauthoryear{{Trenti}, {Stiavelli}, {Bouwens}, {Oesch},
  {Shull}, {Illingworth}, {Bradley}  \& {Carollo}}{{Trenti}
  et~al.}{2010}]{Trenti10}
{Trenti} M.,  {Stiavelli} M.,  {Bouwens} R.~J.,  {Oesch} P.,  {Shull} J.~M.,
  {Illingworth} G.~D.,  {Bradley} L.~D.,   {Carollo} C.~M.,  2010, \mn@doi
  [\apjl] {10.1088/2041-8205/714/2/L202}, \href
  {https://ui.adsabs.harvard.edu/abs/2010ApJ...714L.202T} {714, L202}

\bibitem[\protect\citeauthoryear{Vale \& Ostriker}{Vale \&
  Ostriker}{2004}]{Vale:2004yt}
Vale A.,  Ostriker J.~P.,  2004, \mn@doi [Mon. Not. Roy. Astron. Soc.]
  {10.1111/j.1365-2966.2004.08059.x}, 353, 189

\bibitem[\protect\citeauthoryear{{Venditti}, {Mu{\~n}oz}, {Bromm}, {Fujimoto},
  {Finkelstein}  \& {Chisholm}}{{Venditti} et~al.}{2025}]{Venditti:2025mgi}
{Venditti} A.,  {Mu{\~n}oz} J.~B.,  {Bromm} V.,  {Fujimoto} S.,  {Finkelstein}
  S.~L.,   {Chisholm} J.,  2025, \mn@doi [\apj] {10.3847/1538-4357/ae0610},
  \href {https://ui.adsabs.harvard.edu/abs/2025ApJ...994...32V} {994, 32}

\bibitem[\protect\citeauthoryear{{Vink}, {de Koter}  \& {Lamers}}{{Vink}
  et~al.}{2001}]{Vink2001_winds}
{Vink} J.~S.,  {de Koter} A.,   {Lamers} H.~J.~G.~L.~M.,  2001, \mn@doi [\aap]
  {10.1051/0004-6361:20010127}, \href
  {https://ui.adsabs.harvard.edu/abs/2001A&A...369..574V} {369, 574}

\bibitem[\protect\citeauthoryear{{Vogelsberger} et~al.,}{{Vogelsberger}
  et~al.}{2020}]{1904.07238}
{Vogelsberger} M.,  et~al., 2020, \mn@doi [\mnras] {10.1093/mnras/staa137},
  \href {https://ui.adsabs.harvard.edu/abs/2020MNRAS.492.5167V} {492, 5167}

\bibitem[\protect\citeauthoryear{{Wang} \& {Lilly}}{{Wang} \&
  {Lilly}}{2020a}]{1912.06523}
{Wang} E.,  {Lilly} S.~J.,  2020a, \mn@doi [\apj] {10.3847/1538-4357/ab7b7d},
  \href {https://ui.adsabs.harvard.edu/abs/2020ApJ...892...87W} {892, 87}

\bibitem[\protect\citeauthoryear{{Wang} \& {Lilly}}{{Wang} \&
  {Lilly}}{2020b}]{2003.02146}
{Wang} E.,  {Lilly} S.~J.,  2020b, \mn@doi [\apj] {10.3847/1538-4357/ab8b5e},
  \href {https://ui.adsabs.harvard.edu/abs/2020ApJ...895...25W} {895, 25}

\bibitem[\protect\citeauthoryear{{Weibel} et~al.,}{{Weibel}
  et~al.}{2025}]{Weibel}
{Weibel} A.,  et~al., 2025, \mn@doi [arXiv e-prints]
  {10.48550/arXiv.2507.06292}, \href
  {https://ui.adsabs.harvard.edu/abs/2025arXiv250706292W} {p. arXiv:2507.06292}

\bibitem[\protect\citeauthoryear{{Weisz} et~al.,}{{Weisz}
  et~al.}{2012}]{Weisz12}
{Weisz} D.~R.,  et~al., 2012, \mn@doi [\apj] {10.1088/0004-637X/744/1/44},
  \href {https://ui.adsabs.harvard.edu/abs/2012ApJ...744...44W} {744, 44}

\bibitem[\protect\citeauthoryear{{Whitler}, {Endsley}, {Stark}, {Topping},
  {Chen}  \& {Charlot}}{{Whitler} et~al.}{2023}]{Whitler2022:JWST_Ages}
{Whitler} L.,  {Endsley} R.,  {Stark} D.~P.,  {Topping} M.,  {Chen} Z.,
  {Charlot} S.,  2023, \mn@doi [\mnras] {10.1093/mnras/stac3535}, \href
  {https://ui.adsabs.harvard.edu/abs/2023MNRAS.519..157W} {519, 157}

\bibitem[\protect\citeauthoryear{{Whitler} et~al.,}{{Whitler}
  et~al.}{2025}]{Whitler25_UVLF}
{Whitler} L.,  et~al., 2025, \mn@doi [\apj] {10.3847/1538-4357/adfddc}, \href
  {https://ui.adsabs.harvard.edu/abs/2025ApJ...992...63W} {992, 63}

\bibitem[\protect\citeauthoryear{Xavier, Abdalla  \& Joachimi}{Xavier
  et~al.}{2016}]{Xavier:2016elr}
Xavier H.~S.,  Abdalla F.~B.,   Joachimi B.,  2016, \mn@doi [Mon. Not. Roy.
  Astron. Soc.] {10.1093/mnras/stw874}, 459, 3693

\bibitem[\protect\citeauthoryear{{Yung}, {Somerville}, {Finkelstein}, {Wilkins}
   \& {Gardner}}{{Yung} et~al.}{2024a}]{Yung24_brightUV}
{Yung} L.~Y.~A.,  {Somerville} R.~S.,  {Finkelstein} S.~L.,  {Wilkins} S.~M.,
  {Gardner} J.~P.,  2024a, \mn@doi [\mnras] {10.1093/mnras/stad3484}, \href
  {https://ui.adsabs.harvard.edu/abs/2024MNRAS.527.5929Y} {527, 5929}

\bibitem[\protect\citeauthoryear{{Yung}, {Somerville}, {Nguyen}, {Behroozi},
  {Modi}  \& {Gardner}}{{Yung} et~al.}{2024b}]{Yung}
{Yung} L.~Y.~A.,  {Somerville} R.~S.,  {Nguyen} T.,  {Behroozi} P.,  {Modi} C.,
    {Gardner} J.~P.,  2024b, \mn@doi [\mnras] {10.1093/mnras/stae1188}, \href
  {https://ui.adsabs.harvard.edu/abs/2024MNRAS.530.4868Y} {530, 4868}

\bibitem[\protect\citeauthoryear{{da Silva}, {Fumagalli}  \& {Krumholz}}{{da
  Silva} et~al.}{2012}]{slug}
{da Silva} R.~L.,  {Fumagalli} M.,   {Krumholz} M.,  2012, \mn@doi [\apj]
  {10.1088/0004-637X/745/2/145}, \href
  {https://ui.adsabs.harvard.edu/abs/2012ApJ...745..145D} {745, 145}

\bibitem[\protect\citeauthoryear{{van der Walt}, {Colbert}  \&
  {Varoquaux}}{{van der Walt} et~al.}{2011}]{numpy}
{van der Walt} S.,  {Colbert} S.~C.,   {Varoquaux} G.,  2011, \mn@doi
  [Computing in Science and Engineering] {10.1109/MCSE.2011.37}, \href
  {https://ui.adsabs.harvard.edu/abs/2011CSE....13b..22V} {13, 22}

\makeatother
\end{thebibliography}
\bibliographystyle{mnras}

\onecolumn
\appendix

\section{The SFR power spectrum in hydro sims}
\label{App:hydro}

\begin{figure}
    \centering
    \includegraphics[width=0.50\linewidth]{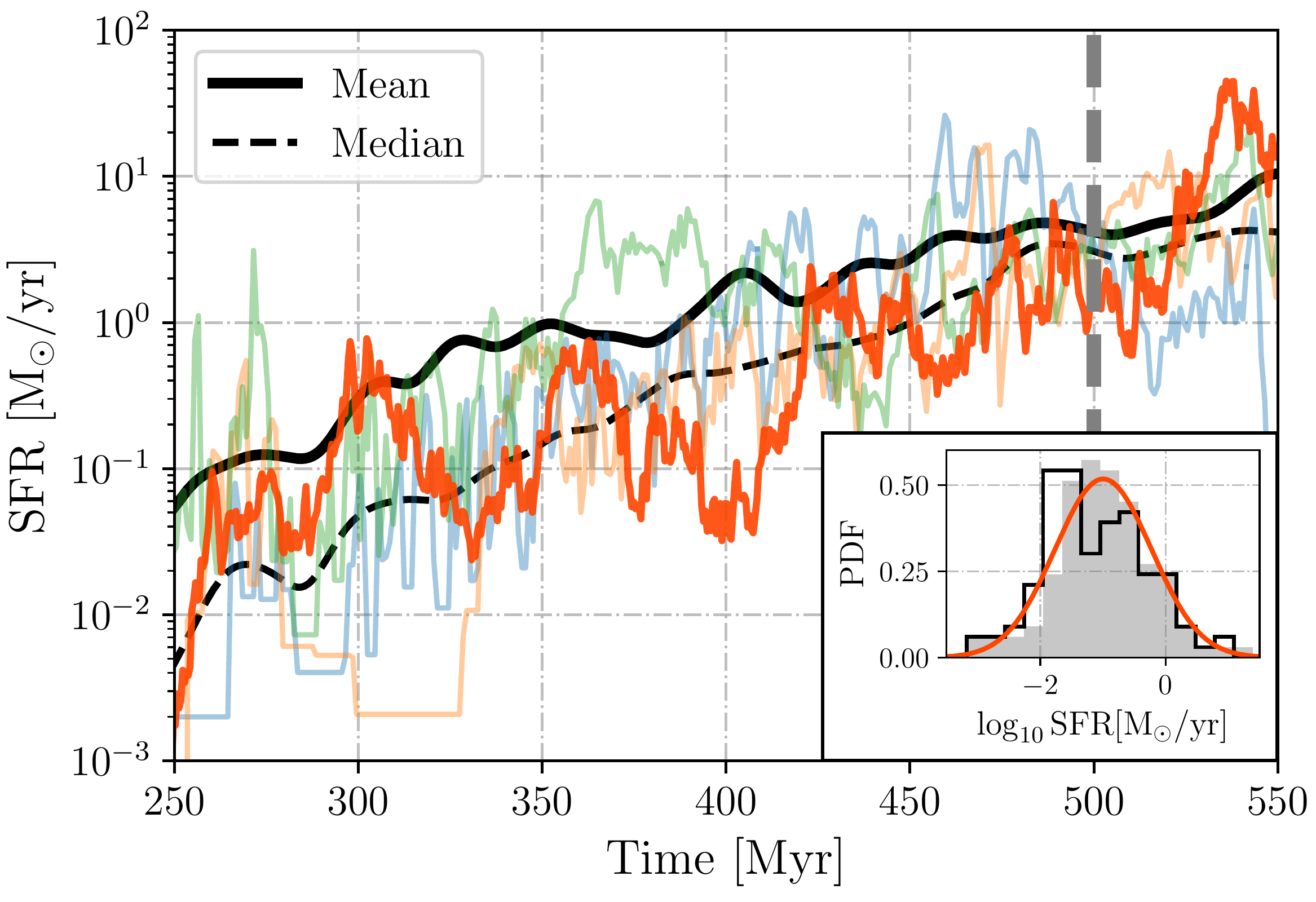}
    \includegraphics[width=0.476\linewidth]{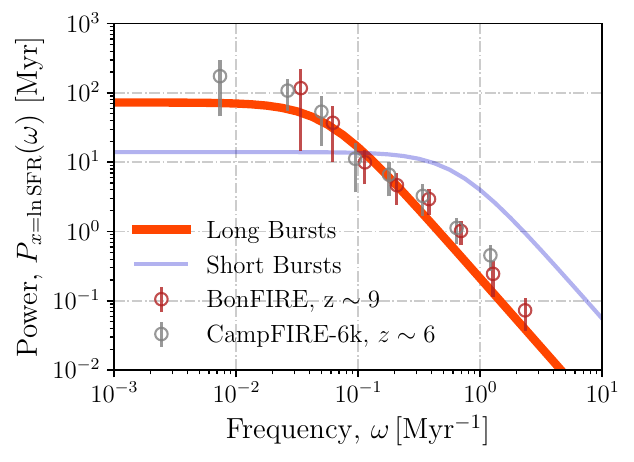}
    \caption{Comparison of our analytic PS model against FIRE simulations (in all cases for galaxies hosted in halos of $M_h=10^{10.5}\,\Msun$). {\bf Left:} Star formation histories drawn from the FIRE simulation (light lines), as well as their mean (thick black) and median (thin black dashed). The thick red line is a synthetic SFR drawn from our PS model (with long bursts, around the median from the simulation). 
    The inset shows the PDF of log SFR at $t=500$ Myr for all galaxies (black), as well as its smoothed version over 5 Myr (gray), which is approximately Gaussian (red line). 
    {\bf Right:} We show the PS for the BonFIRE galaxies (at $z\sim 9$, reaching $t=550$ Myr) and the zoom-in CampFIRE-6k ($z=6$, or $t=900$ Myr). They agree with each other, showing little $z$ evolution. The example short-bursts model (blue) with $\tauPSD=2$ Myr overpredicts the high-$\omega$ power and underpredicts at low $\omega$ compared to the simulations. The long-bursts model (red, with $\tauPSD=20$ Myr), modeled after our best-fit in the main text, agrees well with the FIRE-3 power spectrum.  
    }
    \label{fig:appendix_SFH_FIRE}
\end{figure}

In order to validate that our PS formalism can generate realistic SFHs, we compare against samples drawn from the BonFIRE simulation suite~\citep{Jenna25}. 
BonFIRE is a hydrodynamic cosmological simulation of a large volume ($L=41.2$ cMpc) run at a mass resolution of $m_{\rm baryon}\approx5\times10^3\,\Msun$ to $z\sim9$.
BonFIRE is equipped with fully multiphase gas, sub-parsec resolution in dense gas, and updated FIRE-3 physics including improved low-metallicity gas cooling and a Pop~III model \citep{Hopkins2023_FIRE}.
Due to its finite volume, BonFIRE only has $N=14$ galaxies with $M_h\sim 10^{10.5}\,\Msun$ at $z=9$, and CampFIRE-6k (which is a subregion of the BonFIRE volume re-simulated at higher resolution to later times) has $N=5$ at $z=6$. We show the SFHs of a sample of these galaxies in the left panel of Fig.~\ref{fig:appendix_SFH_FIRE}, along with the mean and median. As we saw in the main text, burstiness increases the mean far above the median, in our formalism this is encapsulated in the fact that a lognormal variable has $\VEV{e^x} > 1$ for $x\sim N(0,\sigma)$.
Along with the example FIRE SFHs we show one of our synthetic SFHs, drawn from the PS model that has long bursts, which is statistically very similar. 
To illustrate this, the right panel of Fig.~\ref{fig:appendix_SFH_FIRE} shows the power spectrum of our long-bursts model along with the one inferred for galaxies hosted in $M_h=10^{10.5}\,\Msun$ halos in BonFIRE (at $z\sim 9$) and CampFIRE ($z\sim 6$). There is remarkable agreement between the two simulations, despite the different coverage in cosmic time, as well as with our analytic model (which is not calibrated on these simulations). 
For comparison, the model with short bursts ($\tauPSD=2$ Myr) predicts far more power at large frequencies $\omega$, in disagreement with the FIRE simulation physics.
In future work we will extend these studies for smaller halos, where due to mass resolution the SFR reaches zero, in which case the log diverges.

While these tests show the reasonableness of our PS model, as well quantitatively probe its amplitude and timescale, they do not prove that the SFR is a lognormal field. A full Kolmogorov–Smirnov test is beyond the scope of this work. However, in the inset of the left panel of Fig.~\ref{fig:appendix_SFH_FIRE} we show the PDF for log(SFR) of galaxies hosted in halos of $M_h=10^{10.5}\,\Msun$, both at the native 1Myr resolution and smoothed over 5 Myr. In both cases they closely resemble a Gaussian, which validates our log-normality assumption.

\section{The power spectrum of SFR}
\label{app:powerspectrum_y}

\begin{figure}
    \centering
    \includegraphics[width=0.5\linewidth]{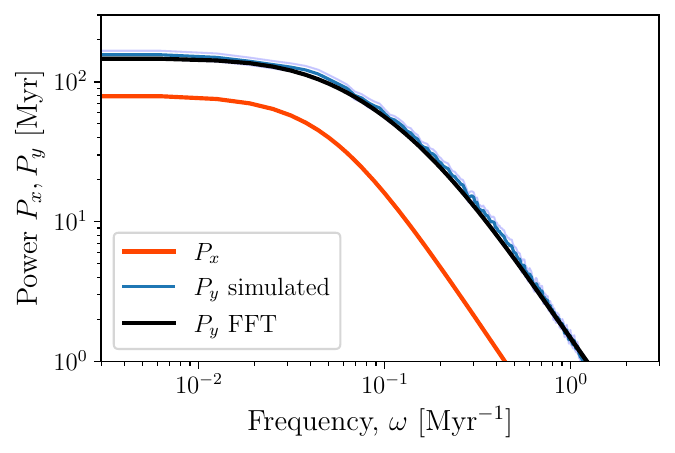}
    \caption{The power spectrum of the normalized SFR ($y=\dot M_\star / \overline{\dot M_\star}$) is a nonlinear function of that of $x = \ln \dot M_\star$. We use an FFT approach to obtain it, which agrees remarkably well with direct simulations. The power $P_y$ is larger than $P_x$ in amplitude, and also extends to larger frequencies, so it is important to properly compute its full shape to model burstiness. }
    \label{fig:PxPy_vs_omega_appendix}
\end{figure}

In our log-normal model the SFH is decomposed as
\be
\dot M_\star (t) = \overline{\dot M_\star} (t) e^{x(t)},
\ee
which includes an average term ($\overline{\dot M_\star}$) and  a Gaussian random variable $x(t)$ with a well-known 2-point function (correlation function/power spectrum in either real/Fourier space).
In particular, we have taken the simplest example of a damped harmonic oscillator with a known correlation function and power spectrum in Eqs.~(\ref{eq:corrfuncx}, \ref{eq:powspecx}). 
However, when computing luminosities the physical quantity that enters calculations is the SFR itself, not its log. 
We, then, defined the quantity $y=e^x$ in the main text, which is a log-normal variable with a correlation function 
\be
\xi_y(t) = e^{\xi_x(t)} - 1,
\ee
in terms of the input correlation function $\xi_x$ for the Gaussian field $x$. 
Numerically, we start with $\xi_x$ (which is a simple function of $\sigmaPSD$ and $\tauPSD$, see Eq.~\ref{eq:powspecx}), compute $\xi_y$ and FFT to obtain the power spectrum $P_y(\omega)$.
To test this calculation, we compare the power spectrum of $y$ to that of a direct simulation for the fiducial long-bursts model shown in the main text.
Fig.~\ref{fig:PxPy_vs_omega_appendix} shows how both approaches are in remarkable agreement.
The key result is that $P_y \gg P_x$, as $y=e^x$ is a nonlinear function of $x$ so it fluctuates more strongly.
We note, in passing, that for very large values of $\sigmaPSD\gtrsim 5$ numerical issues can arise, as the fluctuations become so strong that they require an arbitrarily high number of samples to capture the true PDF of a distribution. As such, we have limited our parameter space to $\sigmaPSD \lesssim 5$ in this work.

\section{PDF of a sum of lognormals}
\label{app:sum_of_lognormals}

In our PS formalism the (log) SFR fluctuations $x(t_1)$ and $x(t_2)$ are jointly Gaussian, so their sum will be as well.
However, the observables (SFR and luminosities) are proportional to $y=e^x$, as we saw in Eq.~\eqref{eq:SEDintegraldef_y}:
\be
L_\lambda = \int_0^{t_{\rm obs}} d \tage W_\lambda(\tage) y(\tage).
\ee
so they are a (weighted) sum of lognormal variables $y$, which is not necessarily lognormal. 
In fact, there is no closed form for the PDF of a sum of lognormals~\citep{Lo2012}.
This is more of an issue for UV than H$\alpha$, as the H$\alpha$ emission is largely localized to the last $\sim 10$ Myr, so not many bursts contribute, whereas UV has a far longer tail.
Yet, we can numerically approximate it. 
We take a shortcut and separate the UV emission into a short- and a long-timescale component:
\be
L_{\rm UV}(\tage) = L_{\rm UV}^{\rm short}(\tage) +L_{\rm UV}^{\rm long}(\tage)
\ee
by cutting at $\tage = 30$ Myr (or $2\times \tauPSD$, whichever is larger). These two components are approximately uncorrelated and lognormal, so we get their PDFs $\mathcal P_{\rm short/long}$ independently and obtain the PDF of the sum $L_{\rm UV}$ through convolution, since:
\be
\mathcal P (L_{\rm UV}) = \int_0^{L_{\rm UV}} dL_1  \mathcal P_{\rm short}(L_1) \mathcal P_{\rm long}(L_{\rm UV} - L_1).
\ee
In practice we do this convolution through an FFT for numerical expediency. 
One could keep slicing up $L_{\rm UV}$ into smaller bins for more precision, but that involves more correlations between the log-normal variables, and further FFTs, and as we will see comparing against a direct simulatons this suffices.

\section{Comparison to a direct simulation}
\label{app:comparison_simulation}

Through the main text we use our efficient analytic method to obtain all the statistics, including $\mathcal \mathcal P(\MUV|M_h)$ and $\mathcal \mathcal P(\MUV, L_{\Ha})|M_h)$, which are used for the UVLF and the $\Ha$/UV ratios, respectively. 
In this Appendix we cross check these results with the outputs of a direct simulation.
We will do so in two steps, first we will compare the PDFs of the luminosity at different wavelengths for halos of a certain mass (that is, $\mathcal P(L_\lambda| M_h)$). Then we will compare to the observed quantities (luminosity functions and H$\alpha$/UV ratios at fixed $M_{\rm UV}$). 
Throughout this section we will generate SFHs for galaxies from a lognormal model with a known input power spectrum, and obtain their SEDs by directly integrating SFH with the Green's function. Then we will bin the galaxies to obtain PDFs.
In this comparison we set a simpler analytical SFH:
\be
\dot M_\star(t) = \overline{\dot M_\star} (t)e^{x(t)} \qquad {\rm where} \qquad
\overline{\dot M_\star}(t) = \overline{M_\star}(t) e^{\alpha t} /f(\alpha), 
\ee
between $t=0$ and the age of the universe $t_{\rm univ}$ at the input redshift $z$.
We set a growth index $\alpha=10^{-2}/$ Myr to match a exponentially growing model at $z=6-7$, and $f(\alpha) = \alpha/(e^{\alpha t_{\rm univ}} - 1)$ normalizes this function. 
Then, we draw $x(t)$ for each galaxy from $0-t_{\rm univ}$, with a power spectrum $P_x(\omega)$ evaluated over frequencies $\omega = 2\pi/t_{\rm univ}$ to $\pi/t_{\rm res}$, where $t_{\rm res}=1$ Myr is our temporal resolution.
Notice that not all galaxies will have the same $M_\star$, and in fact their mean $\VEV{M_\star} > \overline{M_\star}$ due to burstiness. 
In this example we set $M_h=10^{10}\,\Msun$ and $\overline{M_\star} = 10^7\,\Msun$, tough later we will sum over the halo mass function. For concreteness we compare simulation and analytics for the long-bursts fiducial case, with $\sigmaPSD = 2$ and $\tauPSD = 20$ Myr.

\subsection{Separate PDFs for Ha and UV}

\begin{figure}
    \centering
    \includegraphics[height=5cm]{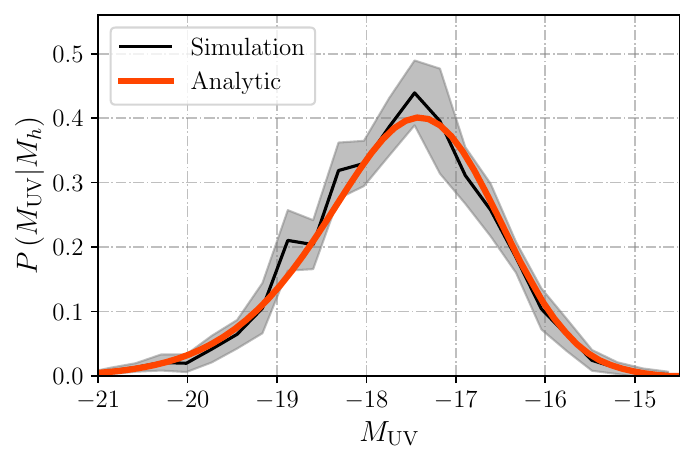}
    \includegraphics[height=5cm]{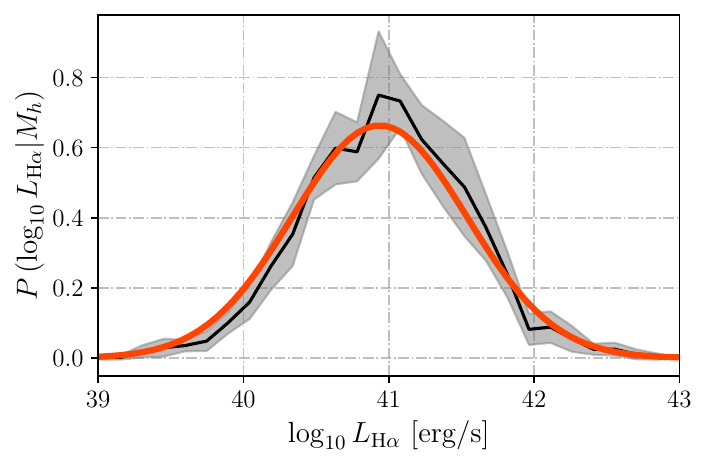}
    \caption{Simulated (black) and analytic PDFs (red) for $\MUV$ (left) and  $\log_{10} L_{\rm H\alpha}$ (right)  a fixed halo mass $M_h=10^{10}\,\Msun$.
    The gray shaded region shows the 1$\sigma$ errors from a bootstrap analysis with $N=10^3$ simulated galaxies at this halo mass.
    The overlap between the PDFs shows that our analytic model well reproduces simulated SEDs. 
    }
    \label{fig:app_PDFs_MUV_LHa}
\end{figure}

Fig.~\ref{fig:app_PDFs_MUV_LHa} shows the PDFs of the $\MUV$ and $\log_{10} L_{\rm H\alpha}$ of galaxies residing in halos of fixed $M_h$, where the near-lognormal nature of the observables is clear.
The PDFs resulting from the simulation and our analytic log-normal approximation are in excellent agreement. 
Specifically, for $\log_{10} \LHa$ the simulation obtains a mean of $40.97$ and an rms 0.51, whereas the analytic calculation predicts 40.98 and 0.52; in agreement to within 0.01 dex. 
For $\MUV$ the PDF is heavy tailed towards the bright end due to burstiness (see Appendix~\ref{app:sum_of_lognormals} for how we compute the PDF). Even in this case the two methods agree: the simulations predict a mean and an rms of $-17.71$ and 0.83, whereas integrating the analytic PDF we obtain $-17.67$ and 0.89, in agreement within 0.05-0.1 mags, which suffices for our purposes (especially considering systematic uncertainties in observations and stellar population synthesis models).

\subsection{H$\alpha$ to UV ratios}
\label{app:eta_PDF}

Moving beyond LFs, in order to obtain the PDF of $\Ha$/UV we need to know how the two observables correlate.
If $\MUV$ and $\log_{10} L_\Ha$ were Gaussianly distributed, even if correlated, this would be trivial. As discussed above, they are only approximately so, and this produces a large difference. Let us discuss in detail how we analytically compute $\Ha$/UV ratios.

The observable we focus on this work is $\etaHaUV \equiv \log_{10}(L_{\rm Ha} /L_{\rm UV} )$.
It is not enough to calculate the PDF $\mathcal P(\etaHaUV | M_h)$ at fixed $M_h$, since we want to condition on $\MUV$ to compare against observations, which are $\MUV$-limited and often binned in this quantity.
That is, we need $\mathcal P(\etaHaUV | M_h, \MUV)$, which we can rewrite as
\be
\mathcal P(\etaHaUV | M_h, M_{\rm UV}) = \mathcal P(\log_{10} L_{\rm Ha} - \log_{10} L_{\rm UV} | M_h, \log_{10} L_{\rm UV}) = \mathcal P(\log_{10} L_{\rm Ha}  |  M_h, \log_{10} L_{\rm UV})
\ee
for the appropriate value of $L_{\rm UV}$.
It is convenient to recast this PDF through Bayes' theorem as
\be
\mathcal P(\log_{10} L_{\rm Ha}  |  M_h, \log_{10} L_{\rm UV}) = \mathcal P(\log_{10} L_{\rm UV}  |  M_h, \log_{10} L_{\rm Ha}) \dfrac{\mathcal P(\log_{10} L_{\rm Ha} | M_h)}{\mathcal P(\log_{10} L_{\rm UV} | M_h)}.
\ee
We know $\mathcal P(\log_{10} L_{\rm UV} | M_h)$ and $\mathcal P(\log_{10} L_{\rm Ha} | M_h)$ from the UV and Ha LF calculations (they are approximated as lognormals). We are just missing $\mathcal P(\log_{10} L_{\rm UV}  |  M_h, \log_{10} L_{\rm Ha})$. For that we will use the physical insight that the UV light is integrated over longer timescales than $\Ha$. As such, we decompose the UV luminosity into two components:
\be
L_{\rm UV} = A L_{\Ha} + L_{\rm extra}, 
\ee
one that is proportional to the $\Ha$ and captures short timescales, with a constant rescaling $A$, and one due to the ``extra'' SFR at long timescales. 
This extra UV light has a Green's function 
\be
G_{\rm extra}(t) = G_{\rm UV} (t) - A G_\alpha(t),
\ee
which allows us to compute its statistics as we did for $L_{\rm UV}$ and $L_\Ha$ in the main text. 
Importantly, it is approximately lognormal and uncorrelated to $L_{\Ha}$, so in order to characterize it we just need its mean and variance. 
We can, then, use the PDF for $\log_{10} L_{\rm UV}$,$\log_{10} L_{\Ha}$, and this ``extra'' component to obtain
\be
\mathcal P(\log_{10} L_{\rm UV}  |  M_h, \log_{10} L_{\rm Ha}) = \mathcal P(\log_{10} L_{\rm extra} = \log_{10} (L_{\rm UV} - A L_{\rm Ha}) |  M_h) \left |\dfrac{d \log_{10}L_{\rm extra} }{d \log_{10}L_{\rm UV}  } \right |
\ee
where the last term is just a lognormal evaluated at a specific value times the Jacobian of the transformation.

While this may appear a convoluted way to compute $\mathcal P(\etaHaUV | M_h, M_{\rm UV})$, it is fully analytic (and therefore fast) and it agrees well with direct simulations.
As an example, Fig.~\ref{fig:simcomparison_PDFs_ratio} shows the distribution of $\etaHaUV$ from our simulation of galaxies with $M_h=10^{10}\,\Msun$ divided in bins of $\MUV$. The analytic calculation matches the simulation fairly well. 
Note, for bright objects ($\MUV \ll \overline \MUV$) the PDF of $\Ha$/UV becomes overly peaked in our analytic model, when compared to simulations. That is because a residual correlation between $L_{\rm extra}$ and $L_{\rm H\alpha}$, since these two variables are decorrelated only over their whole ranges, not at each specific value.
Fortunately, we find that for all bright galaxies residing in haos of a mass ($\MUV \leq \overline \MUV$) the PDF is very similar to that of $\MUV = \overline \MUV$, so we will set it to that value (in Fig.~\ref{fig:simcomparison_PDFs_ratio} we set it to the value at $\MUV=-18.2$ which is slightly brighter).  
Code-wise this is done by setting $L_{\rm UV} = {\rm min}(L_{\rm UV} , \overline L_{\rm UV})$ (though note that when adding up the $P$'s there's an additional $\mathcal P(L_{\rm UV})$ which we do not modify, as that is the actual PDF of galaxies having $L_{\rm UV}$). 
In future work we will revisit this assumption.

Note that we cannot select galaxies based on $M_h$, so in order to compare predictions to observations we compute
\be
\mathcal P(\etaHaUV | \MUV) = \int dM_h \dfrac{dn}{dM_h} \, \mathcal P(\etaHaUV | \MUV, M_h), 
\ee
given the conditional PDF $\mathcal P(\etaHaUV | \MUV, M_h)=\mathcal P( \log_{10} L_\Ha | \MUV, M_h)$ computed above. 

\subsection{Integrating over the HMF}

In addition to the PDF at a fixed $\MUV$ and $M_h$, we want to integrate over all $M_h$ to obtain the observables, as we cannot condition data on $M_h$. 
This is not an obvious variable transformation, as ``on-mode'' galaxies will be high on both UV and H$\alpha$, so the correlation between the two luminosities ought to be modeled carefully.
If not modeled correctly, we could infer distributions for $\etaHaUV$ that are either too broad or too narrow. 
For this exercise we simulate SFHs for an entire halo mass function of halos, and give them (average) stellar masses by a simple relation $\overline M_\star(M_h) = \tilde f_\star \times M_h$, with
\be
\tilde f_{\star}(M_h) = \frac{\tilde \epsilon_{f_\star}}{(M_h / M_{c,f_{\star}})^{\beta_{f_{\star}}} + (M_h / M_{c,f_{\star}})^{-\alpha_{f_{\star}}}}
\ee
with $\tilde \epsilon_{f_\star}=0.037$, $M_{c,f_{\star}}=1.4\times 10^{11}$, $\beta_{f_{\star}}=0.5$, and $\alpha_{f_{\star}}=0.8$, to resemble our best-fit average SFH integrated over time. 
We set this both in the simulation and in our code.
That allows us to rescale the SFHs we simulated above to other $M_h$. We also allow $\sigmaPSD$ to vary with halo mass as in the main text ($d\sigmaPSD/d\log_{10}M_h = -0.5$) by simply broadening our simulation outputs around their mean for each $M_h$. 
While this is simply a toy model to translate simulations at a fixed $M_h$ into a galaxy population, it suffices for our cross-checking purposes, as we can replicate its assumptions within our analytic code. In future work we will refine its assumptions.

\begin{figure}
    \centering
    \includegraphics[width=0.49\linewidth]{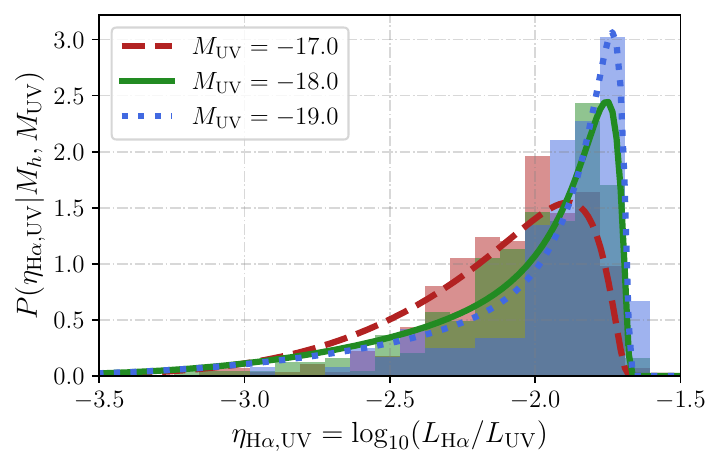}
    \includegraphics[width=0.47\linewidth]{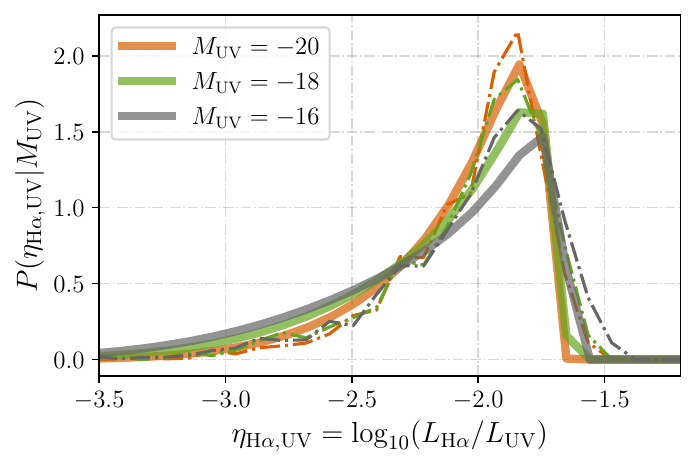}
    \caption{Distributions of $\etaHaUV$ from simulation and our analytic approach. 
    The {\bf left} panel shows the PDF of $\etaHaUV$ for galaxies hosted in halos of fixed $M_h=10^{10}\Msun$ at several $M_{\rm UV}$. Histograms show the output of a simulation and lines are our analytic calculation. 
    These PDFs are then integrated along the HMF to obtain the PDF conditioned only on $\MUV$ (as it will be observed) on the  {\bf right}. Dot-dashed lines are the simulated histograms, and solid are the analytic result (which is shifted by $\etaHaUV=-0.1$ to match the mean, such a shift can be absorbed in the $\AmpHa$ free parameter). 
    } 
\label{fig:simcomparison_PDFs_ratio}
\end{figure}

Given this rescaling we can find the PDFs $\mathcal P(\etaHaUV | \MUV, M_h)$ for any $M_h$, and then average over halo masses as we did for the LFs in the main text, i.e.,
\be
\mathcal P(\etaHaUV | \MUV) = \mathcal N \int dM_h \dfrac{dn}{dM_h}\mathcal P(\etaHaUV | \MUV, M_h),
\ee
where $\mathcal N$ is a normalization that integrates the PDF to unity. 
In this simulation we integrate over 30 halo masses, logspaced in $10^7-10^{14}\,\Msun$. 
The right panel of Fig.~\ref{fig:simcomparison_PDFs_ratio} shows the $\mathcal P(\etaHaUV | \MUV)$ from our simulation, to be compared with the result from our analytic calculation. 
These two approaches agree well, building confidence in our efficient, analytic modeling of burstiness.  

\subsection{Star forming main sequence}
\label{app:SFMS}

A well-studied tracer of burstiness is the star-forming main sequence (SFMS), which parametrizes the $\dot M_\star-M_\star$ relation, including its scattering. We use the simulation procedure described above to produce SFHs. This toy model allows us to reconstruct a relation between $M_\star$ and SFR by binning the simulated population of galaxies.
We show the $z\sim 6$ SFMS estimated from this procedure in Fig.~\ref{fig:balmerbreaks}, along with the photometric+spectroscopic measurements from \citep[][see also \citealt{Popesso22} for brighter galaxies]{Clarke24}, in both cases using $\Ha$ as a tracer of SFR with the same conversion of $ \log_{10} \dot M_\star [{\Msun/\rm yr}] = \log_{10} L_\Ha [{\rm  erg/s}] -41.37$~\citep{astro-ph/9807187}. 
The estimated median SFMS from our toy model agrees well with the observations, though at higher $M_\star$ the prediction is steeper than observed.
This result ought to be interpreted with care, due to the toy nature of the model employed to derive the SFMS here, and we leave for future work including $M_\star$ as an output of our analytic model.
Yet, our prediction that burstiness increases towards smaller $M_h$ naturally makes the SFMS broader towards smaller $M_\star$ as well.

\subsection{Balmer Breaks}
\label{app:BalmerBreaks}

\begin{figure}
    \centering
    \includegraphics[width = 0.50\linewidth]
    {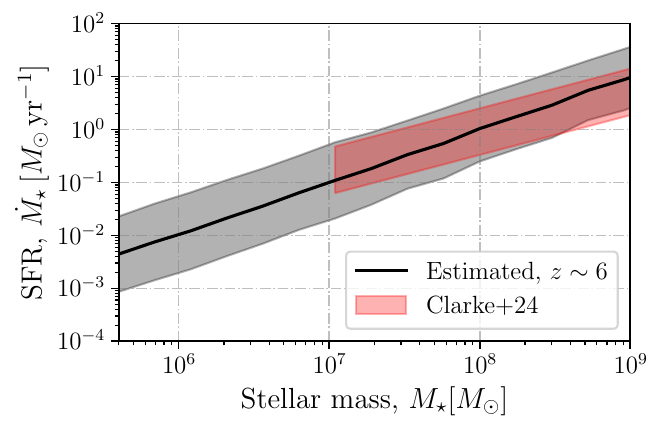}
    \includegraphics[width=0.47\linewidth]{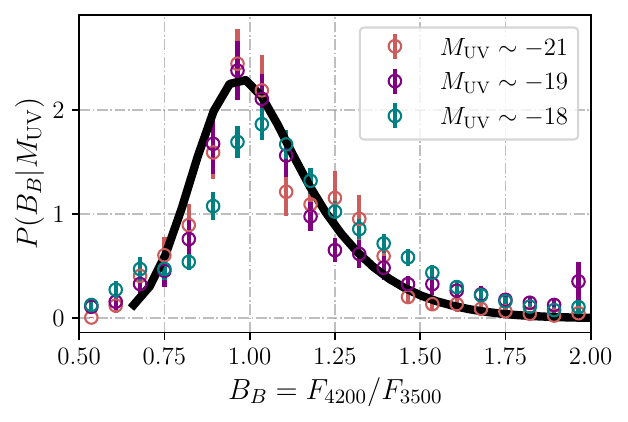}
    \caption{Toy simulation outputs beyond $\Ha$/UV ratios.
    {\bf Left:}
     Predicted star-forming main sequence (black), as well as the NIRCAM+NIRSpec measurements of \citet[red]{Clarke24}, in both cases using $\Ha$ as a tracer of SFR and evaluated at $z\sim 6$. The colored area is the scatter expected from burstiness, which for our model grows towards small stellar masses.
     {\bf Right:}
    Measured (colored points) and predicted (black line) PDFs for Balmer break amplitudes $B_B=F_\lambda(4200)/F_\lambda(3500)$ of UV-selected galaxies at $z\sim 6$, in three bins of $\MUV$. The observations from \citet{Endsley24_bursty} line up nicely with the prediction of the long-bursts model of the main text (which has $\sigmaPSD=2$ and $\tauPSD=20$ Myr), despite not being used for calibrate it. 
    } 
    \label{fig:balmerbreaks}
\end{figure}

There are a plethora of other star-formation tracers beyond $\Ha$ and UV. Some other line fluxes (e.g., Oxygen) require knowledge of the metal content of the galaxies, which we are currently not modeling. However, the flux ratio red to blue-ward of the Balmer transition, or Balmer break amplitude, is an indicator of older, rather than younger stellar populations, so it helps in constraining burstiness along with the UV and $\Ha$ light~\citep[e.g.,][]{Endsley24_bursty}.
We follow \citet{Endsley24_bursty} and define this break as
\be
B_B = F_\lambda (4200{\rm \AA})/F_\lambda (3500 {\rm \AA}).
\ee
Our analytic model is not yet equipped to find distributions of Balmer breaks. However, we can use our toy simulation (rescaling the results at a fixed $M_h$ to others and integrating over a HMF) to approximate it. 
Fig.~\ref{fig:balmerbreaks} shows the theoretical\footnote{Note we smooth the theory curve with the typical error of $B_B$ in the data, which is $\sigma(B_B) = 0.09 B_B + 0.04$. Since we do not correct the flux at 3500 and 4200 the mean values may be offset (there is no $\AmpHa$ factor here), so we manually shift the theory curve by a constant factor of 1.16.} PDF $\mathcal P(B_B | \MUV)$ along with the observations from \citet{Endsley24_bursty}. 
As we saw for the $\Ha$/UV ratios, the PDF is fairly broad, indicating burstiness, and gets broader towards the faint end, albeit mildly.
The predicted distribution of $B_B$ values lines up nicely with the measurements, showcasing a non-Gaussian tail towards high break amplitudes, indicative of off-mode galaxies. We emphasize that we have not calibrated our model to these data.  We leave for future work improving the theoretical model so as to fit this observable.

\section{Clustering measurements}
\label{app:bias_threshold}

We use the clustering measurements for UV-selected galaxies in \citet{Harikane}, obtained from HST + HSC observations. These biases are reported for a fiducial flat {\it Planck} 2018 cosmology, which we match, and on different $\MUV$ cuts.
For each cut $i$ we translate the bias $b_{\rm meas}$ reported to a binned effective bias as 
\be
b_{\rm eff}(M_{\rm UV, eff}^{(i)}) = (b_{\rm meas} n_{\rm meas})(\MUV < \MUV^{(i+1)}) - (b_{\rm meas} n_{\rm meas})(\MUV < \MUV^{(i)}) 
\ee
where $n_{\rm meas}$ is the reported number density of galaxies above each $\MUV$ cut, and $M_{\rm UV, eff}^{(i)} = (\MUV^{(i+1)} + \MUV^{(i)})/2$ is the bin center (all of which have width 0.5 by design. We then add the errors in quadrature, appropriately weighed by $n_{\rm meas}$. 
This procedure is not exact, as the biases can be correlated between bins and the median $\MUV$ of galaxies in the bin $i$ will not necessarily be the mid-point. 
We show the binned biases we infer at $z\sim 4-6$ and different $\MUV^{(i)}$ in Tab.~\ref{tab:biases}. 
These are the data we use in our likelihoods. 

\begin{table}
\centering
\caption{Biases from HST+HSC observations in \citet{Harikane}, converted to $\MUV$ bins of width 0.5 mag.}
\label{tab:biases}
\begin{tabular}{|c|c|c|}
\hline
$z$ & $M_{\mathrm{UV}}$ & $b_{\rm eff}$ \\
\hline
3.8 & $-22.24$ & $6.47 \pm 0.10$ \\
    & $-21.74$ & $5.09 \pm 0.09$ \\
    & $-21.24$ & $4.12 \pm 0.08$ \\
    & $-20.74$ & $3.59 \pm 0.08$ \\
    & $-20.24$ & $3.02 \pm 0.11$ \\
\hline
4.9 & $-22.67$ & $8.45 \pm 0.10$ \\
    & $-22.17$ & $6.80 \pm 0.10$ \\
    & $-21.67$ & $5.83 \pm 0.10$ \\
    & $-21.17$ & $4.96 \pm 0.10$ \\
    & $-20.67$ & $4.68 \pm 0.12$ \\
\hline
5.9 & $-21.97$ & $8.42 \pm 0.18$ \\
    & $-21.47$ & $7.33 \pm 0.14$ \\
\hline
\end{tabular}
\end{table}

\section{Histogram of $\Ha$/UV observations}
\label{app:HaUVhistograms}

\begin{figure}
    \centering
    \includegraphics[width=0.8\linewidth]{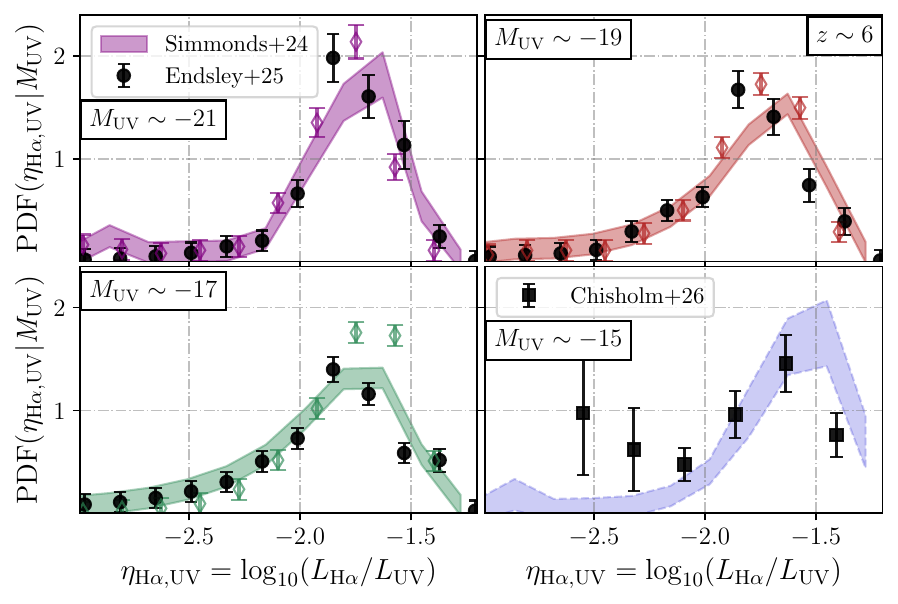}
    \caption{PDFs for the $\etaHaUV$ of galaxies with different $\MUV$ at $z\sim 6$. Black points show the measurements from \citet{Endsley24_bursty} and \citet[][as in Fig.~\ref{fig:HaUVratios_z6}]{Chisholm}. The filled regions correspond to the observations from \citet[][which correspond to slightly different $\MUV$ bins, see Sec.~\ref{sec:compareobservations}]{simmonds24_masscompl} used through the main text, where the faintest (blue) bin was not used due to completeness. Finally, the empty symbols correspond to an alternate sample from \citet[][ with SNR$>5$ and no $M_\star$ cut]{simmonds24_masscompl}, which are shifted by $\Delta \etaHaUV=-0.12$ (corresponding to a different value of the $\AmpHa$ nuisance parameter). 
    The PDFs from all these samples overlap well.}
    \label{fig:appendix_pdf_comparison}
\end{figure}

Galaxy observations often report a series of $\Ha$/UV ratios for Lyman-break galaxies. In this short appendix we outline how we translate these measurements into PDFs for the  $\Ha$/UV ratios from observations, and compare our datasets. 
In all cases we start with a catalog of objects, with measured photo-$z$, $\MUV$, and $\eta_{\rm H\alpha,UV} \equiv \log_{10} L_{\rm H\alpha}/L_{\rm UV}$ all with errorbars (assumed independent as we do not have access to the full covariance matrix for each object).
We bin galaxies in redshift and (observed, i.e., not dust corrected) $\MUV$, as described in the main text.
Then we define bins in $\eta_{\rm H\alpha,UV}$ (which to clarify, is dust corrected) and Monte Carlo to obtain the histogram at each bin. 
For this procedure we sample the data $N=2,000$ times, drawing $\MUV$ and $\eta_{\rm H\alpha,UV}$ within the errorbars. The main source of uncertainty is $\eta_{\rm H\alpha,UV}$, as $\MUV$ errors are much smaller and we take very broad redshift bins (and do not vary $z$). 
From those samples we recover the mean PDF at each $\eta_{\rm H\alpha,UV}$ bin and the covariance matrix $C_{ij}$ between bins, which is approximately diagonal though it has off-diagonal elements towards the ``off-mode'' galaxies ($\etaHaUV\lesssim -2$), as smaller $\Ha$/UV ratios are more difficult to tease out from photometry, but in this first work we ignore them so the error at the bin $i$ is simply $\sqrt{C_{ii}}$.
This is the procedure used to obtain the histograms and their uncertainties in the main text(for $\etaHaUV$ in Figs.~\ref{fig:HaUVratios_z6} and \ref{fig:HaUV_ratios_allz}), as well as for the Balmer breaks in Fig.~\ref{fig:balmerbreaks}).

Fig.~\ref{fig:appendix_pdf_comparison} shows a comparison of the observational PDF $\mathcal P(\etaHaUV|\MUV)$ at $z\sim 6$ computed with different datasets and assumptions. 
The two baseline results used through the main text (from~\citealt{Endsley24_bursty, Chisholm} versus that of \citealt{simmonds24_masscompl}) agree well, despite the different assumptions taken to obtain them. 
In particular, the \citet{simmonds24_masscompl} sample assumes a fairly lax cut of SNR$>3$ and a stellar-mass cut of $M_\star \gtrsim 10^{7.5}\,\Msun$. We generate an alternate catalogue by requiring SNR$>5$ in at least one of the wide JWST bands and removing the stellar-mass cut. The resulting PDFs are shown as empty symbols in Fig.~\ref{fig:appendix_pdf_comparison}, and overlap well both the datasets taken through the main text.

As a further cross-check, we have re-run our entire analysis with this alternate PDF dataset.
We find burstiness paramters: $\sigmaPSD = 1.61\pm 0.11$, $d\sigmaPSD/d\log_{10}M_h = -0.52\pm 0.13$, and $\log_{10}\tauPSD\,[\mathrm{Myr}] = 1.34\pm0.22$, in excellent agreement with the fiducial case used in the main text (but 20\% smaller burstiness amplitude, likely due to the slightly different galaxy population selected through the cuts).
In addition, we have added the faintest bin in the \citet{simmonds24_masscompl} catalog (centered at $\MUV\sim -15$, shown in the lower-right panel of Fig.~\ref{fig:appendix_pdf_comparison}, which is expected to be fairly incomplete). 
In this case we find $\sigmaPSD = 2.035^{+0.068}_{-0.040}$, $d\sigmaPSD/d\log_{10}M_h = -0.45\pm 0.10$, and $\log_{10}\tauPSD\,[\mathrm{Myr}] = 1.33^{+0.21}_{-0.15}$, again in agreement with our main analysis. 
We conclude that differences in how the $\Ha$/UV ratios are extracted from photometry and further processed into PDFs can change specific parameter values at the $10\%$ level but our conclusions remain unchanged.

\begin{figure}
    \centering
    \includegraphics[width=1.04\linewidth]{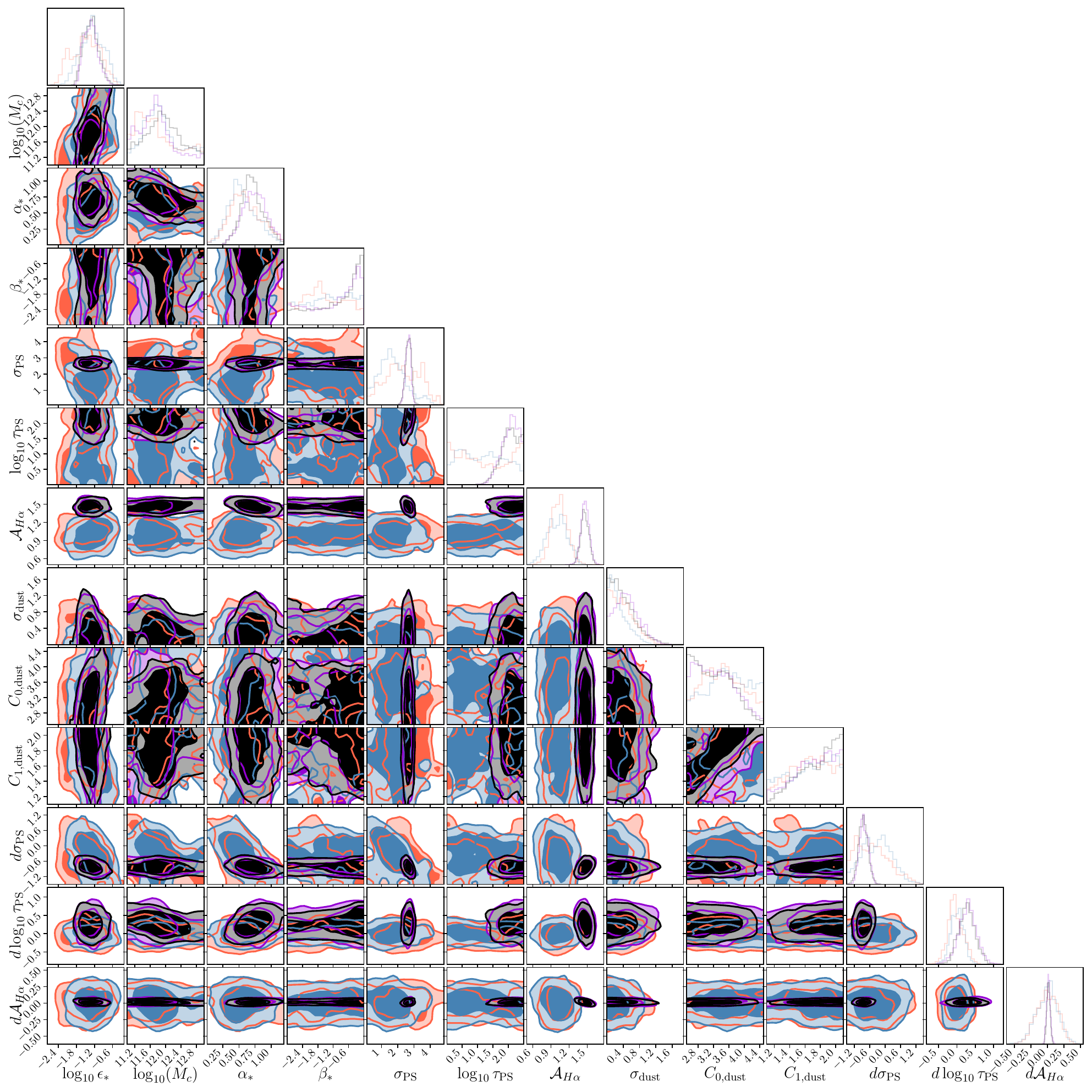}
    \caption{Full corner plot for the parameters at $z\sim 6$, where we show UVLF alone in red, UVLF+clustering in blue, UVLF+H$\alpha$ ratios in purple, and all combined in black.}
    \label{fig:cornerallz6}
\end{figure}

\section{Additional posteriors and cross checks}
\label{app:extraposteriors}

In this appendix we collect the posteriors for parameters varied in for different ancillary analyses.

\subsection{All parameters at redshift $z\sim 6$}

First, we show a posterior for all parameters (beyond the burstiness $\sigmaPSD$ and $\tauPSD$ shown in the main text) for our fiducial $z\sim 6$ analysis (in Sec.~\ref{sec:results}). As a reminder to the reader, the parameters are summarized in Table~\ref{tab:parameters}, and include 4 parameters that tie the average SFH of halos, their derivatives against redshift, the two PSD parameters and their derivatives against halo mass, three dust parameters ($C_0$, $C_1$, and the dust variability $\sigma_{\rm dust}$), and two parameters, $\AmpHa$ and its derivative against (log$_{10}$) halo mass, to account for uncertainties in the IMF and metallicity (as well as SPS modeling).

Fig.~\ref{fig:cornerallz6} shows the full corner plot for our $z\sim 6$ analysis. We do not vary any derivatives against redshift, as we are fitting a single $z$. 
Some key insights are:

$\bullet \AmpHa \approx 1.5$, instead of unity, showing that the data prefer brighter H$\alpha$ at fixed UV on average than predicted by our default IMF and metallicity in the \citetalias{BC03} SPS model.

$\bullet C_0\approx 3.6$ is a bit lower than the typically assumed value of $C_0=4.4$, showcasing slightly less dust attenuation than typically assumed (though the uncertainty covers most of the prior range), which could indicate lower dust production or different attenuation laws in high redshift galaxies. 

$\bullet \sigma_{\rm dust}\lesssim 1$ shows that some level of dust variability is allowed, but not highly preferred.

$\bullet d\tauPSD$ is consistent with zero, and is dominated by the Gaussian prior we set.

\subsection{Fitting different datasets at $z\sim 6$}

\begin{figure}
    \centering
    \includegraphics[width = 0.31\linewidth]{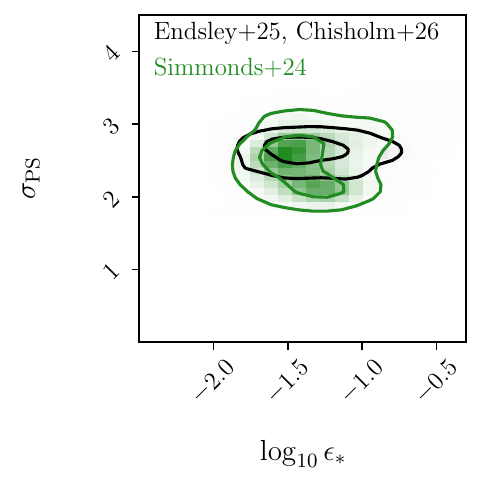}
    \includegraphics[width = 0.32\linewidth]{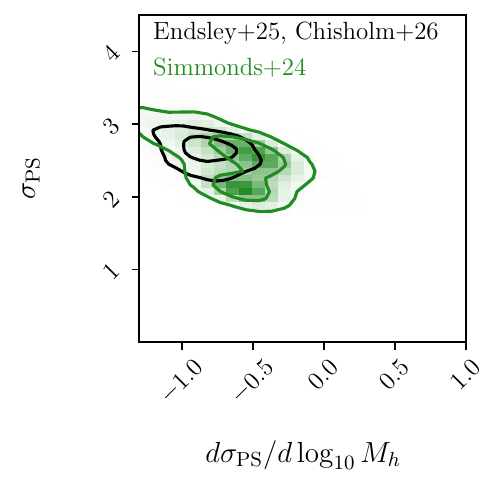}
    \includegraphics[width = 0.32\linewidth]{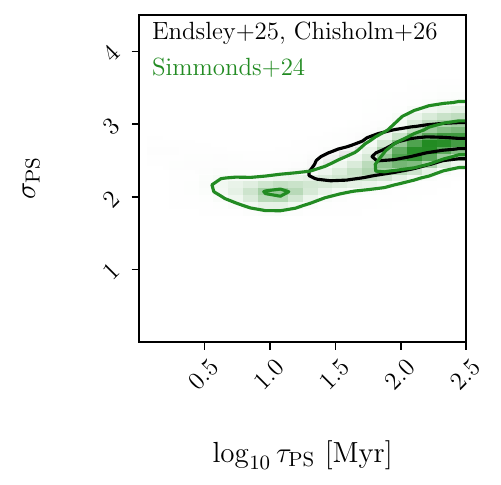}
    
    \caption{We test that the $\Ha$/UV ratios used do not alter our conclusions. Each panel shows posteriors for the burstiness PS amplitude $\sigmaPSD$ against a different parameter. We fit to the UVLF and $\Ha$/UV ratios at $z\sim 6$ from \citet[][black]{Chisholm,Endsley24_bursty} vs \citet[][green]{simmonds24_masscompl}, which are obtained with different data and assumptions on the SED fitting of galaxies. 
    We find great agreement between the parameters derived from both datasets, albeit slightly stronger burstiness and mass dependence in the former. 
    }
    \label{fig:cornersigmaz6_bothdata}
\end{figure}

Through the main text we have used the  H$\alpha$/UV ratios from ~\citet{Endsley24_bursty, Chisholm} as well as \citet{simmonds24_masscompl}.
These two datasets overlap at $z\sim 6$, so here we directly test the burstiness parameters inferred by assuming either dataset, to estimate the uncertainty introduced by the different SED-fitting procedures and ensure convergence. Fig.~\ref{fig:cornersigmaz6_bothdata} shows the posteriors for the PS amplitude $\sigmaPSD$ against the SFE amplitude $\epsilon_\star$, its mass derivative $d\sigmaPSD/d\log_{10}M_h$, and the PS timescale $\tauPSD$ . 
One set of countours is obtained from using the $\Ha$/UV ratios from \citet[][the former derived from a two-component burst SFH model, and the latter directly inferred from photometric excesses]{Endsley24_bursty,Chisholm}, and the other from \citep[][derived through a Bayesian fit of the star-formation history]{simmonds24_masscompl}.
In both cases we also fit to the UVLFs at $z\sim 6$.
We find broadly similar results, despite the diverging datasets used to obtain them. 
The first dataset appears to prefer slightly more burstiness, and a steeper slope against halo mass, but overall the results are in agreement, showing that the prior built in by SED fitting is not negligible, but does not dominate our results.

\subsection{Comparing $z\sim 4, 5,$ and $6$}

\begin{figure}
    \centering
    \includegraphics[width = 0.31\linewidth]{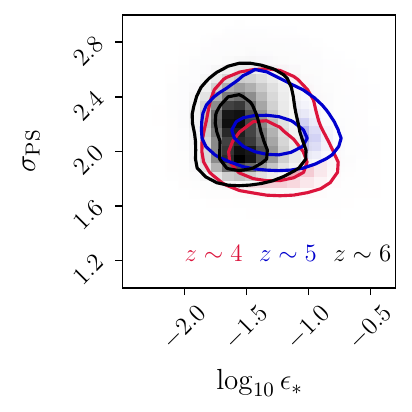}
    \includegraphics[width = 0.32\linewidth]{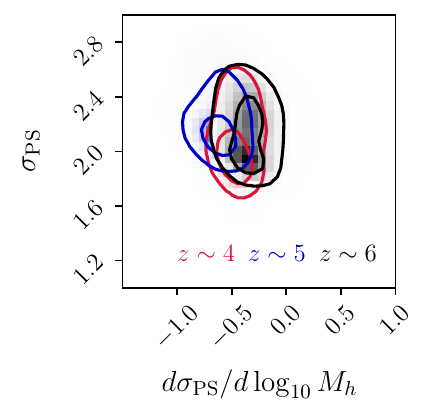}
    \caption{We compare independent fits to the UVLF+$\Ha$/UV ratios of \citet{simmonds24_masscompl} at $z\sim4, 5$, and $6$, finding consistent results in the three redshifts with no discernible evolution.
    }
\label{fig:corner_appendix_z456}
\end{figure}

Fig.~\ref{fig:posteriossigmatau_z456} of the main text showed that the burstiness parameters inferred at different $z$ are similar. 
Here we show that this conclusion extends beyond $\sigmaPSD$ vs $\tauPSD$. Fig.~\ref{fig:corner_appendix_z456} shows posteriors for $\sigmaPSD$ against its mass derivative  and SFE, where it is clear that all three redshifts largely overlap, which indicates little explicit evolution in the burstiness through $z\sim 4-6$.

\begin{figure}
    \centering
    \includegraphics[width = 0.31\linewidth]{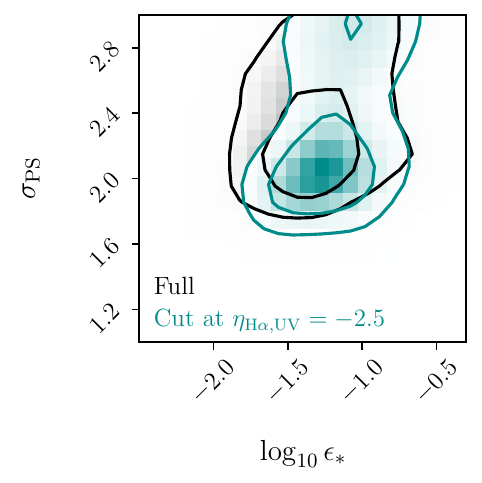}
    \includegraphics[width = 0.32\linewidth]{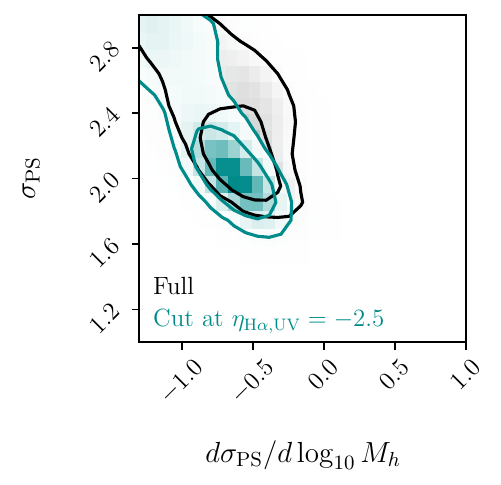}
    \includegraphics[width = 0.32\linewidth]{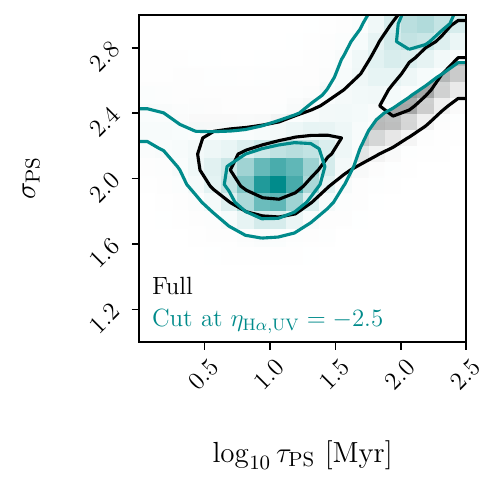}
    \caption{Same as Fig.~\ref{fig:cornersigmaz6_bothdata} but comparing the effect of cutting out all galaxies with $\Ha$/UV ratios $\etaHaUV\leq -2.5$, assuming those faint-$\Ha$ values are harder to observe in photometry. We find larger uncertainties in that case, but the same conclusions. 
    }
\label{fig:corner_cutatm2.5}
\end{figure}

\subsection{Cutting log $\etaHaUV$ values in inference}

Measuring low $\Ha$ fluxes in photometry can be difficult, even with deep medium bands. As such, the low-$\etaHaUV$ end of the PDFs we use in this work may be subject to systematic uncertainties. We test how much this can affect our conclusions by fitting the $z\sim 6$ UVLF, clustering, and the $\Ha$/UV ratios from \citet{simmonds24_masscompl} but removing all points with $\etaHaUV<-2.5$ (corresponding to $\log_{10} \xiion = 24.7 \, {\rm Hz\,erg^{-1}}$ for the usual Case-B assumption). 
Fig.~\ref{fig:corner_cutatm2.5} shows the posterior for this case along with the baseline case (no cut). The errorbars on the burstiness parameters grow, but they remain consistent with our main analysis, so our conclusions do not sensitivetily depend on the faintest-$\Ha$ emitting galaxies.

\begin{figure}
    \centering
    \includegraphics[width = 0.64\linewidth]{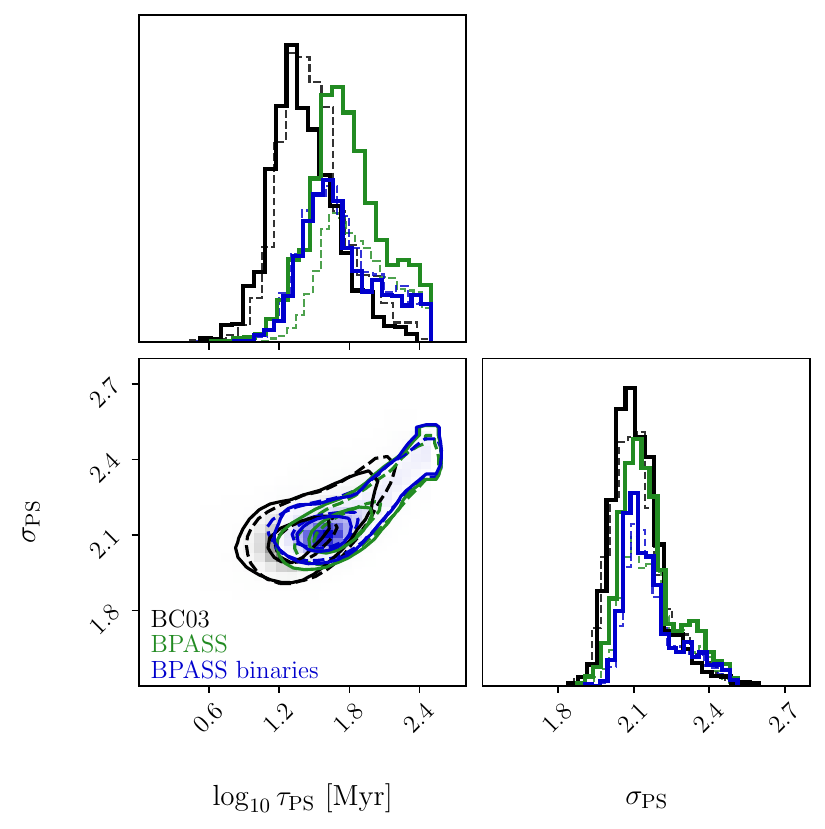}
    \caption{Same as Fig.~\ref{fig:posteriossigmatau_z456} but for the three SPS models considered here, \citetalias{BC03} (black, used in the main text), \citetalias{BPASS} single stars (green) and with binaries (blue). We find very similar burstiness parameters, both with (solid) and without (dashed) including clustering information. 
    }
\label{fig:corner_BPASS}
\end{figure}

\subsection{Testing different SPS models}

One last key assumption we have made through this work is the SPS model of \citetalias{BC03}. This determines how much light is emitted by star bursts of different ages (and thus the Green's functions in Fig.~\ref{fig:windows}), and can vary in other SPS models or when changing stellar parameters such as metallicity. For instance, in Fig.~\ref{fig:windows} we showed the result from using \citetalias{BPASS} with and without binaries, which mainly affects the $\Ha$ emission (by lengthening how long it can be emitted for).

Here we repeat our main analysis of $z\sim 4-6$ data with the \citetalias{BPASS} Green's functions, and compare against our baseline result.
Fig.~\ref{fig:corner_BPASS} shows the posteriors for the two main burstiness parameters, $\sigmaPSD$ and $\tauPSD$. The choice of SPS model affects the burstiness parameters marginally, but we find overall excellent consistency. 
In particular, the amplitudes and timescales are $\sigmaPSD = 2.11^{+0.11}_{-0.08}$ and 
$\log_{10}\tauPSD [\mathrm{Myr}]  = 1.40^{+0.32}_{-0.24}$ for \citetalias{BC03} (as in the main text),
$\sigmaPSD = 2.14^{+0.14}_{-0.08}$ and $\log_{10}\tauPSD\,[\mathrm{Myr}] = 1.75^{+0.36}_{-0.26}$ for BPASS single stars, and
$\sigmaPSD = 2.14^{+0.14}_{-0.06}$ and $\log_{10}\tauPSD\,[\mathrm{Myr}] = 1.66^{+0.43}_{-0.24}$ for BPASS with binaries. 
That is, the burstiness amplitudes agree very precisely and the timescales are slightly longer for the two \citetalias{BPASS} cases than for the fiducial \citetalias{BC03}, though the posteriors are highly overlapping. 
We also find negligible differences in other parameters, such as $\epsilon_\star$ and $d\sigmaPSD/d\log_{10}M_h$.

The reason for the insensitivity to the SPS model chosen is the ``nuisance'' $\AmpHa$ parameter (and its derivative), which rescale the $\Ha$/UV light by a constant, and absorb the bulk of the SPS uncertainty. As such, their values shift between the three cases, where we find $\AmpHa=1.755\pm 0.059$ for \citetalias{BC03}, $\AmpHa= 1.415\pm 0.056$ for \citetalias{BPASS}, and $\AmpHa= 1.259^{+0.052}_{-0.044}$ when adding binaries. The derivatives are closer to each other, with $d\AmpHa/d\log_{10}M_h = -0.034\pm 0.014$, $d\AmpHa/d\log_{10}M_h = -0.039\pm 0.014$, and $d\AmpHa/d\log_{10}M_h = -0.041\pm 0.014$, respectively. 
We conclude that our results do not dependent sensitively on the choice of SPS model, given we marginalize over $\AmpHa$.

\section{H$\alpha$LFs}
\label{app:HaLF}

\begin{figure}
\centering
\includegraphics[width = 0.58\linewidth]{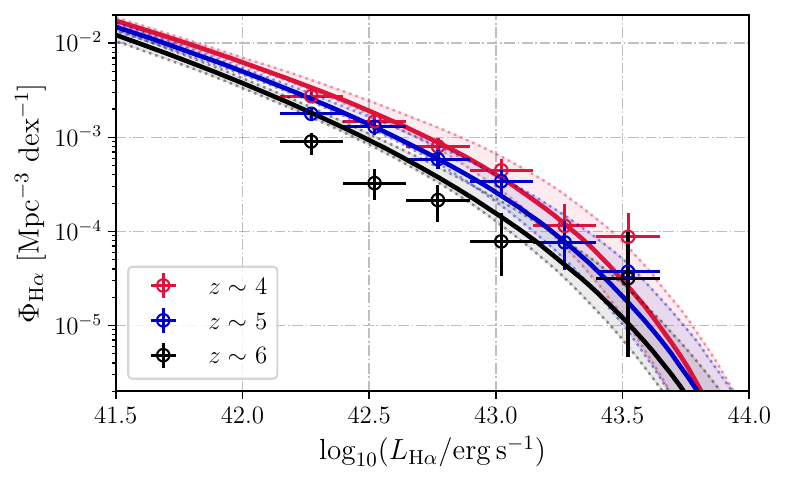}    
\caption{Posterior for the predicted H$\alpha$LF (dust corrected), calibrated from the rest of $z\sim 4-6$ data,
along with the spectroscopic observations from \citet{CoveloPaz25_haLF}. Our model predicts well the $z\sim 4-5$ $\Ha$LFs, though it overpredicts observations at $z\sim 6$.
    }
    \label{fig:HaLFs}
\end{figure}

In addition to UVLFs, clustering, and $\Ha$/UV ratios our model can predict H$\alpha$LFs. For this, we follow our approach for the UVLF, computing
\be
\Phi_{\rm H\alpha}(\log_{10}\LHa) = \int dM_h \dfrac{dn}{dM_h} \mathcal P(\log_{10}\LHa | M_h),
\ee
with the PDF $\mathcal P(\log_{10}\LHa | M_h)$ computed as the $\MUV$ one through this work.
For completeness, we can also model the H$\alpha$ clustering as 
\be
b_{\rm eff} (\LHa) = \Phi_{\rm H\alpha}^{-1}(\LHa) \int dM_h \dfrac{dn}{dM_h} \mathcal P(\log_{10}\LHa |M_h) b(M_h).
\ee
Fig.~\ref{fig:HaLFs} shows our predicted $\Ha$LFs at $z\sim 4-6$ (with 1$\sigma$ uncertainties from fitting the data at those $z$), along with the observations of the (dust-corrected) H$\alpha$LFs from \citet[][see \citealt{Shuntov} for clustering measurements]{CoveloPaz25_haLF}, obtained spectroscopically.
We limit our comparison to $\log_{10}\LHa/{\rm erg\,s^{-1}} > 42.0$, where the data is more complete, and impose a minimum 20\% errorbar on the H$\alpha$LF to account for cosmic variance. 
Our predicted H$\alpha$LFs agree well at $z\sim 4-5$, but have a higher overall amplitude at $z\sim 6$. The uncertainties are fairly large, including on completeness at faint $L_\Ha$ and high $z$, where the NIRCam sensitivity drops rapidly. 
As such,  we conclude that our model reproduces the main features of the $\Ha$LFs, but does not agree with the observed $z\sim6$ faint-end amplitude.

\bsp	
\label{lastpage}
\end{document}